\documentclass[aps,prd,twocolumn,groupedaddress,nofootinbib]{revtex4-1}

\pdfoutput=1
\usepackage{bbold}
\usepackage{mathtools}
\usepackage{ulem}
\usepackage{color}

\usepackage{comment}

\def\laq{~\raise 0.4ex\hbox{$<$}\kern -0.8em\lower 0.62ex\hbox{$\sim$}~}
\def\gaq{~\raise 0.4ex\hbox{$>$}\kern -0.7em\lower 0.62ex\hbox{$\sim$}~}

\def\beq{\begin{equation}}
\def\eeq{\end{equation}}
\def\bea{\begin{eqnarray}}
\def\eea{\end{eqnarray}}
\def\bean{\begin{eqnarray*}}
\def\eean{\end{eqnarray*}}

\newcommand{\class}{{\sc class}}

\newcommand{\HH}{\mathcal{H} }

\def \pa {\partial}
\def \ra {\rightarrow}

\def \la {\lambda}

\def \ga {\gamma}

\def \om {\omega}
\def \Om {\Omega}

\def \U{\Upsilon}

\def\th#1#2{\theta^{#2(#1)}}

\def \s#1#2#3{\left( s^#1_#2 \right)^{(#3)}}

\def\th#1#2{\theta^{(#1)#2}}

\def\laq{~\raise 0.4ex\hbox{$<$}\kern -0.8em\lower 0.62ex\hbox{$\sim$}~}
\def\gaq{~\raise 0.4ex\hbox{$>$}\kern -0.7em\lower 0.62ex\hbox{$\sim$}~}

\def\be{\begin{equation}}
\def\ee{\end{equation}}
\def \ga {\gamma}

\def\beq{\begin{equation}}
\def\eeq{\end{equation}}
\def\bea{\begin{eqnarray}}
\def\eea{\end{eqnarray}}

\def \pa {\partial}
\def \ra {\rightarrow}

\def \Rcal {\mathcal{R}}
\def \bell {\boldsymbol{\ell}}
\def \bx {\boldsymbol{x}}
\def \bn {\boldsymbol{n}}

\newcommand{\Ups}{\Upsilon}

\newcommand{\INT}[1]{\int d^2 \ell_{#1} \ }
\newcommand{\NINT}[1]{\int \frac{d^2 \ell_{#1}}{(2\pi)^2}}
\newcommand{\vl}[1]{\boldsymbol{ \ell}_{#1}}

\newcommand{\T}[2]{\theta^{#1(#2)}}

\newcommand{\Acal}{\mathcal A}
\newcommand{\Bcal}{\mathcal B}
\newcommand{\Dcal}{\mathcal D}
\newcommand{\Ecal}{\mathcal E}

\newcommand{\Hcal}{\mathcal H}

\newcommand{\Pcal}{\mathcal P}
\newcommand{\Mcal}{\mathcal M}

\newcommand{\Qcal}{\mathcal Q}
\newcommand{\Ucal}{\mathcal U}

\newcommand{\bk}{{ \bf k }}

\def \bbe {\boldsymbol{e}}
\def \bbep {\boldsymbol{\epsilon}}

\def\laq{~\raise 0.4ex\hbox{$<$}\kern -0.8em\lower 0.62ex\hbox{$\sim$}~}
\def\gaq{~\raise 0.4ex\hbox{$>$}\kern -0.7em\lower 0.62ex\hbox{$\sim$}~}

\def\beq{\begin{equation}}
\def\eeq{\end{equation}}
\def\bea{\begin{eqnarray}}
\def\eea{\end{eqnarray}}
\def\bean{\begin{eqnarray*}}
\def\eean{\end{eqnarray*}}

\def \pa {\partial}
\def \ra {\rightarrow}

\def \la {\lambda}

\def \ga {\gamma}

\def \om {\omega}
\def \Om {\Omega}

\def \U{\Upsilon}

\def\th#1#2{\theta^{#2(#1)}}

\def\th#1#2{\theta^{#2(#1)}}

\begin{document}

\title{CMB-lensing beyond the leading order: temperature and polarization anisotropies}

\author{Giovanni Marozzi$^{1,2}$,  Giuseppe Fanizza$^3$,  Enea Di Dio$^{4,5,6}$ and Ruth Durrer$^7$,}

\affiliation{$^1$Centro Brasileiro de Pesquisas F\'{\i}sicas, Rua
  Dr. Xavier Sigaud 150, Urca,  CEP 22290-180, Rio de Janeiro, Brazil\\
 $^2$Dipartimento di Fisica, Universit\`a di Pisa and INFN, Sezione di Pisa, Largo Pontecorvo 3, I-56127 Pisa, Italy,  \\
  $^3$Center for Theoretical Astrophysics and Cosmology, Institute for Computational Science, University of Z\"urich, CH-8057 Z\"urich, Switzerland\\
$^4$INAF - Osservatorio Astronomico di Trieste, Via
  G. B. Tiepolo 11, I-34143 Trieste, Italy \\
  $^5$SISSA- International School for Advanced Studies, Via Bonomea 265, 34136 Trieste, Italy \\
  $^6$INFN - National Institute for Nuclear Physics,
via Valerio 2, I-34127 Trieste, Italy \\
$^{7}$Universit\'e de Gen\`eve, D\'epartement de Physique Th\'eorique and CAP,
24 quai Ernest-Ansermet, CH-1211 Gen\`eve 4, Switzerland}

\begin{abstract}
We investigate the weak lensing corrections to the CMB temperature and polarization anisotropies.
We consider all the effects beyond the leading order: post-Born corrections, LSS corrections and, for the 
polarization anisotropies, the correction due to the rotation of the polarization direction between the emission 
at the source and the detection at the observer.
We show that the full next-to-leading order correction to the B-mode polarization is not negligible on small scales and is dominated by the contribution from the rotation, this is a new effect not taken in account in previous works. 
Considering vanishing primordial gravitational waves, the B-mode 
correction due to rotation is comparable to cosmic variance for $\ell \gtrsim 3500$, in contrast to all other spectra where the corrections are always below that threshold for a single multipole. Moreover, the sum of all the effects is larger than cosmic variance at high multipoles, showing that higher-order lensing corrections to B-mode polarization are in principle detectable.
\end{abstract}


\maketitle

\section{Introduction}
\label{Sec1}
\setcounter{equation}{0}

The  temperature and polarization anisotropies
of the cosmic microwave background (CMB) are the most precious cosmological datasets. It is fair to say that virtually all high precision cosmological measurements involve the CMB. The reason for this is twofold: on the one hand there is excellent data available~\cite{Ade:2014afa,Naess:2014wtr,Crites:2014prc,Story:2014hni,Adam:2015rua,Planck:2015xua,Ade:2015zua,Keisler:2015hfa} and on the other hand CMB fluctuations are theoretically well understood and can be calculated perturbatively.  The CMB success story is by no means over, we expect more precision data to arrive especially for polarization and reconstruction of the cosmic lens map~\cite{Abazajian:2016yjj,DiValentino:2016foa}.

As it is well known, CMB fluctuations are lensed by foreground large scale structure (LSS) and this effect is rather large (up to 10\% and more) on small scales~\cite{Seljak:1995ve,Lewis:2006fu,RuthBook}.
Therefore the question is justified whether higher order contributions to lensing might be relevant. We naively expect them to be of the order of the square of the first order contribution, hence 1\% and therefore it is necessary to include them as numerical CMB calculations~\cite{Lewis:1999bs,Lesgourgues:2011re,Blas:2011rf,Lesgourgues:2013bra} aim at a precision of 0.1\%. On the other hand, present CMB codes do take into account some of the non-linearities by summing up a series of 'ladder diagrams' into an exponential~\cite{Lewis:2006fu,RuthBook}.  It is easy to check that including these non-linearities is requested to achieve the precision goal.

The question which we address in this paper is: what about the other non-linearities which are not included in this sum? Might they also be relevant? These are mainly contributions coming from the fact that the deflection angle of the photons at higher order can no longer be computed assuming the photons move along their unperturbed path, but the perturbation of the photon path has to be taken into account. These are the so-called 
`post-Born corrections'. 
We have already studied this problem for the temperature anisotropies in a previous paper~\cite{Marozzi:2016uob}. The present paper is a follow up on that work. We complete the previous study by calculating also the effects on polarization. Furthermore, here we  treat also the non-linearities of the matter distribution perturbatively. This is  more consistent than just using a Halofit model \cite{Smith:2002dz,Takahashi:2012em}, as it allows us to correctly take into account the higher order statistics (3- and 4-point functions) assuming Gaussian first order perturbations.
We neglect the radial displacement corrections induced by the time delay effect (which indeed is not a lensing contribution). As shown in \cite{Hu:2001yq}, these corrections are at most of the order $\mathcal{O} \left( 10^{-4} \right)$, apart for the temperature-E-mode cross correlation power spectrum for which can reach the order of $\mathcal{O} \left( 10^{-3} \right)$.
We do, however, take into account all effects of second and third order lensing. This includes also the induced vector and tensor modes. These modes are especially important for B-polarisation as they effectively rotate the photon polarisation.

In addition to our work, there have been three other publications on this topic~\cite{Hagstotz:2014qea,Pratten:2016dsm,Lewis:2016tuj}.  In the first paper, an important cancellation which reduces the final result by more than an order of magnitude has been missed. In~\cite{Pratten:2016dsm} our so called 'third group' terms, which vanish when assuming Gaussian statistics and are very relevant for the final result,
are not included.
In the most recent publication~\cite{Lewis:2016tuj} these terms are included, but the rotation of the polarization which is induced by second order lensing is not considered. We discuss it here for the first time and we actually find that it is the dominant correction for $B$-polarization.

In this paper we present the methodology of our calculations and numerical  results for the corrections of CMB 
temperature and polarization anisotropies by next-to-leading order lensing. In an accompanying letter~\cite{Marozzi:2016und} we
discuss the relevance of our findings for future CMB  experiments.

The paper is organized as follows. In the next section we summarize the small deflection angle approximation for CMB lensing beyond linear order, and present the expressions for the deflection angle up to third order.
In Sect.~\ref{Sec3} we translate the results into harmonic space, '$\vl{}$ space'. We also compare the expressions for temperature anisotropies with the corresponding terms for the polarization spectra at all orders in perturbation theory.
In Sect.~\ref{Sec4} we  briefly recollect  the results 
 for the post-Born corrections to the lensed power spectrum of the CMB temperature anisotropies first given in~\cite{Marozzi:2016uob} considering also the non-Gaussian nature of the 
deflection angle at higher order.
In Sect.~\ref{Sec5} we evaluate the contributions from higher orders in the gravitational potential (or equivalently in the matter density) to  corrections of the lensed power spectrum of the CMB temperature and polarization anisotropies. Following \cite{Pratten:2016dsm,Lewis:2016tuj} we call them
`LSS corrections'.
In Sect.~\ref{Sec6} we derive the last missing contribution coming from the fact that   
parallel transported polarization  direction changes along the path of the photon from the source to the observer. This contribution which turns out to be very substantial has been missed in previous work.
Our results are summarized in Sect.~\ref{Sec7}, where we evaluate  the different contributions numerically  considering  a Halofit matter power spectrum.  In Sect.~\ref{Sec8} we conclude.
Several technical aspects and calculations are presented in four appendices.


\section{Weak lensing corrections beyond leading order in real space}
\label{Sec2}
\setcounter{equation}{0}
We want to determine the effect of lensing on the CMB temperature and polarization anisotropies beyond the well studied leading order from first order perturbation theory~\cite{Lewis:2006fu,RuthBook}. 

Following the derivation of the post-Born correction to temperature anisotropies in \cite{Marozzi:2016uob}, we  first generalize the results of~\cite{Lewis:2006fu,RuthBook}  writing
the following relation between the lensed and unlensed temperature anisotropies $\Mcal$ and 
polarization tensor $\Pcal_{mn}$ of the photon field
valid up to fourth order in the deflection angles $\T{a}{i}$ (the superscript $(i)$ denotes the order).
\bea
\label{eq:expansion}
\tilde\Mcal(x^a)&\equiv&\Mcal\left(x^a+\delta \theta^{a}\right)\simeq\Mcal(x^a)+\sum_{i=1}^4\T{b}{i}\nabla_b\Mcal(x^a)
\nonumber \\
&+&
\frac{1}{2}\sum_{i+j\le 4}\T{b}{i}\T{c}{j}\nabla_{b}\nabla_c\Mcal(x^a)\nonumber\\
&+&\frac{1}{6}\sum_{i+j+k\le 4}\T{b}{i}\T{c}{j}\T{d}{k}\nabla_b\nabla_c\nabla_d\Mcal(x^a) 
\nonumber \\
&+&\frac{1}{24} \T{b}{1}\T{c}{1}\T{d}{1}\T{e}{1}\nabla_b\nabla_c\nabla_d\nabla_e\Mcal(x^a)\,,
\label{IniExpT}
\eea
\bea
\label{eq:expansion}
\tilde\Pcal_{mn}(x^a)&\equiv&\Pcal_{mn}\left(x^a+\delta \theta^{a}\right)
\nonumber \\
&\simeq&\Pcal_{mn}(x^a)+\sum_{i=1}^4\T{b}{i}\nabla_b\Pcal_{mn}(x^a)\nonumber\\
&+&\frac{1}{2}\sum_{i+j\le 4}\T{b}{i}\T{c}{j}\nabla_{b}\nabla_c\Pcal_{mn}(x^a)
\nonumber \\
&+&\frac{1}{6}\sum_{i+j+k\le 4}\T{b}{i}\T{c}{j}\T{d}{k}\nabla_b\nabla_c\nabla_d\Pcal_{mn}(x^a)\nonumber\\
&+&\frac{1}{24} \T{b}{1}\T{c}{1}\T{d}{1}\T{e}{1}\nabla_b\nabla_c\nabla_d\nabla_e\Pcal_{mn}(x^a)\,. \nonumber \\
\label{IniExp}
\eea
A consistent treatment of the polarization in the form of $\Pcal_{mn}$ or, using the Stokes parameters $\Qcal$ and $\Ucal$, in the form of 
$\Pcal=\Qcal+i \Ucal$ and $\bar{\Pcal}=\Qcal-i \Ucal$ has to consider that the polarization tensor is 
parallel-transported along the perturbed photon geodesics. Neglecting this effect (we shall add it at a second stage in Sect.~\ref{Sec6}) we can substitute $\Pcal_{mn}$ with $\Pcal$ and $\bar{\Pcal}$.  An over-bar denotes complex conjugation.

Following  \cite{Marozzi:2016uob}, we can then write
\bea
\tilde\Mcal(x^a)&\simeq&
\Acal^{(0)}(x^a)+\sum_{i=1}^4\Acal^{(i)}(x^a)
+\sum_{\substack{ i+j\le 4 \\ 1\le i\le j}}\Acal^{(ij)}(x^a)
\nonumber \\
&+&\sum_{\substack{i+j+k\le 4 \\ 1\le i\le j\le k}} \Acal^{(ijk)}(x^a)
+ \Acal^{(1111)}(x^a)
\, , \\
\label{eq:expansion-compact}
\tilde\Pcal(x^a)&\simeq&
\Dcal^{(0)}(x^a)+
\sum_{i=1}^4\Dcal^{(i)}(x^a)
+\sum_{\substack{ i+j\le 4 \\ 1\le i\le j}}\Dcal^{(ij)}(x^a)
\nonumber \\
&+&\sum_{\substack{i+j+k\le 4 \\ 1\le i\le j\le k}} \Dcal^{(ijk)}(x^a)
+ \Dcal^{(1111)}(x^a)
\,, \ 
\label{eq:expansion-compact_x}
\eea
where
\bea
&&\hspace{-0.8cm}\Acal^{(i_1 i_2....i_n)}(x^a)=
\nonumber \\
&&\!\!\!\!\!\!\!\!\!\!\!\!
=\frac{{\rm Perm}(i_1 i_2....i_n)}{n!}\T{b}{i_1}\T{c}{i_2}.....\nabla_{b}\nabla_c.......\Mcal(x^a)\,,
\\
&& \hspace{-0.8cm} \Dcal^{(i_1 i_2....i_n)}(x^a) =
\nonumber \\
&& \!\!\!\!\!\!\!\!\!\!\!\!
=\frac{{\rm Perm}(i_1 i_2....i_n)}{n!}\T{b}{i_1}\T{c}{i_2}.....\nabla_{b}\nabla_c.......\Pcal(x^a)\,,
\eea
where 
$\Acal^{(0)}(x^a)\equiv \Mcal(x^a)$, 
$\Dcal^{(0)}(x^a)\equiv \Pcal(x^a)$ and
${\rm Perm}(i_1 i_2....i_n)$ denotes the number of permutation of the set $(i_1 i_2....i_n)$.

We introduce also the Weyl potential
\be
\Phi_W = \frac{1}{2}\left(\Phi+\Psi\right)\,
\ee
in terms of the Bardeen potentials $\Phi$ and $\Psi$.
The lensing potential $\psi$ to the last scattering surface
is then determined by
\bea
\psi({\bf n},z_s) &=& \frac{-2}{\eta_o-\eta_s} \int_{\eta_s}^{\eta_o} d\eta \frac{\eta-\eta_s}{\eta_o-\eta} \Phi_W((\eta-\eta_o){\bf n},\eta)
\nonumber \\
&=&-2\int_0^{r_s} dr' \frac{r_s-r'}{r_s r'}\Phi_W(-r'{\bf n},\eta_o-r') \,, 
\label{Lenspot}
\eea
where $\bf{n}$ is the direction of photon propagation, $\eta$ denotes conformal time and $r$ the comoving distance, $r=\eta_o-\eta$, where $\eta_o$ stands for present time. The index $_s$ indicates the corresponding quantity evaluated at the last scattering surface.
The first order deflection angle is simply the gradient of the lensing potential~\cite{Bartelmann:1999yn,RuthBook}. Beyond the linear order, we need to account also for the lensing of the direction $\bf n$ on the path of the photon. 
Then one obtains the following expressions for the deflection angle up to third perturbative order~\cite{Fanizza:2015swa}
\bea
\T{a}{1} &=&
 -2\int_{0}^{r_s}\!dr' \frac{r_s-r'}{r_s\,r'}\nabla^a\Phi_W(r') \,,
\label{TheOrd1}
\\
\T{a}{2}&=&
-2\int_{0}^{r_s}\!dr'\frac{r_s-r'}{r_s\,r'} \left[\nabla^a\Phi_W^{(2)}(r')
\right.
\nonumber \\
&& \left. 
+\nabla_b\nabla^a\Phi_W(r')\T{b}{1}(r')\right]\,,
\label{TheOrd2LSS}
\\
\T{a}{3}\!&=&\!
-2\int_{0}^{r_s}\!dr'\frac{r_s-r'}{r_s\,r'}
\!\left[\nabla^a\Phi_W^{(3)}(r')
\right.
\nonumber \\
&& \left. 
+\nabla_b\nabla^a\Phi_W(r')\T{b}{2}(r')
+
\nabla_b\nabla^a\Phi_W^{(2)}(r')\T{b}{1}(r')\! \right.
\nonumber \\
& & \left.  +
\frac{1}{2}\nabla_b\nabla_c\nabla^a\Phi_W(r')\T{b}{1}(r')\T{c}{1}(r') \right]\,.
\label{LSS-TheOrd3}
\eea
Latin letters $a,b,c,d$ run over the two directions on the sphere.
In Eqs.~(\ref{TheOrd1}-\ref{LSS-TheOrd3}) we consider the terms with the maximal number of  transverse derivatives, including the ones that come from expanding the Weyl potential, $\Phi_W$, to higher order. Note that $\T{a}{2}$ as well as $\T{a}{3}$ are not purely scalar perturbations, they also contain vector contributions as, for example, the curl of $\nabla_b\nabla^a\Phi_W\T{b}{1}$ does not vanish. But for our purpose a decomposition of the higher order deflection angle into scalar and vector parts are of no particular use.
On the other hand, let us point out that we have neglected the second order vector and tensor perturbations of the metric appearing as a  consequence of the nonlinear coupling among scalar, vector and tensor in the Einstein equation. These corrections are subleading with respect to the ones discussed here.

Let us also remind that the Taylor expansion in Eqs.~(\ref{IniExpT}) and (\ref{IniExp}) holds in the approximation of small deflection angles, i.e.~when the deflection angle is much smaller than the angular separations related to a given $C_\ell$. This is valid for  an angular separation of about $4.5$ arc minutes  which corresponds to $\ell \lesssim 2500$ (see \cite{Seljak:1995ve,RuthBook,Lewis:2006fu}). 
In this work, we adopt the small deflection angle approximation for the second and third order deflection angles only, which are much smaller than this value,
as a consequence 
our results are valid to much higher $\ell$s and we can safely present them up to $\ell =3500$.

\section{Weak lensing corrections of the power spectra}
\label{Sec3}
\setcounter{equation}{0}

We evaluate the lensing correction to the angular power spectra $C^{\Mcal}_\ell$, $C^{\Ecal \Mcal}_\ell$,  $C^{\Ecal}_\ell$
and $C^{\Bcal}_\ell$ in the flat sky limit. In this approximation, see e.g.~\cite{RuthBook}, we replace the combination $(\ell,m)$ with a 2-dimensional vector $\vl{}$.
Therefore, the angular position is then the 2-dimensional Fourier transform of the position in $\vl{}$ space at 
redshift $z$.
For a generic variable $Y(z,{\bf x})$ we have
\be
Y(z,{\bf x})=\frac{1}{2\pi}\INT{}Y(z,\vl{})e^{-i\vl{}\cdot{\bf x}}\,,
\label{DeflspaceB}
\ee
and 
\be
\langle Y(z_1,\vl{})\bar{Y}(z_2,\vl{}\,') \rangle=\delta\left( \vl{}-\vl{}\,' \right)C_\ell^{Y}(z_1,z_2)\,,
\label{GSpet}
\ee
while for polarization we have ($\varphi_\ell$ denotes the polar angle in $\vl{}$-space)
\be
\Pcal(z,{\bf x})=-\frac{1}{2 \pi} \INT{} \left[ \Ecal(z,\vl{}) +i \Bcal(z,\vl{}) \right] e^{-2 i \varphi_\ell} 
e^{-i\vl{}\cdot{\bf x}}\,,
\label{Deflspace}
\ee
with
\bea
\langle \Ecal(z_s,\vl{})\bar\Mcal(z_s,\vl{}\,') \rangle&=&\delta\left( \vl{}-\vl{}\,' \right)C_\ell^{\Ecal \Mcal}(z_s)\,,\nonumber\\
\langle \Ecal(z_s,\vl{})\bar\Ecal(z_s,\vl{}\,') \rangle&=&\delta\left( \vl{}-\vl{}\,' \right)C_\ell^{\Ecal}(z_s)\,,\nonumber\\
\langle \Bcal(z_s,\vl{})\bar\Bcal(z_s,\vl{}\,') \rangle&=&\delta\left( \vl{}-\vl{}\,' \right)C_\ell^\Bcal(z_s)\,,\nonumber\\
\langle \Bcal(z_s,\vl{})\bar\Mcal(z_s,\vl{}\,') \rangle&=& 0\,,\nonumber\\
\langle \Bcal(z_s,\vl{})\bar\Ecal(z_s,\vl{}\,') \rangle&=&  0\, .
\label{SpectraAll}
\eea

We follow the notation of~\cite{DiDio:2013bqa, DiDio:2014lka} to determine the angular power spectra defined above and we introduce the (3-dimensional) initial curvature power spectrum
\be
\langle R_{\rm in} \left( {\bf k} \right) \bar R_{\rm in} \left( {\bf k}' \right) \rangle = \delta_D \left( {\bf k} - {\bf k}' \right) P_R \left(k \right)\,.
\ee
(In both 2- and 3-dimensional Fourier transforms we adopt the unitary Fourier transform normalization, so there are no factors of $2\pi$ in this formula as well as in Eqs. (\ref{GSpet}) and (\ref{SpectraAll}.)

For a given linear perturbation variable $A$ we define its transfer function $T_A(z,k)$ normalized to the initial curvature perturbation by
\be
A \left( z, {\bf k} \right) =T_A(z,k)R_{\rm in}({\bf k}) \, ,
\ee
and an angular power spectrum will be then determined by
\bea\label{e:cAB}
&&\hspace{-0.8cm}C^{AB}_\ell \left( z_1, z_2 \right) = 4 \pi \int \frac{dk}{k} \mathcal{P}_R (k) \Delta^A_\ell (z_1,k)\Delta^B_\ell (z_2,k) 
\nonumber \\
&=& \frac{2}{\pi} \int dk k^2 P_R (k)  \Delta^A_\ell (z_1,k)\Delta^B_\ell (z_2,k) \,,
\eea
where $\mathcal{P}_R (k) = \frac{k^3}{2 \pi^2} P_R (k)$ is the dimensionless primordial power spectrum, and 
$\Delta^A_\ell \left( z, k \right)$ denotes the transfer function in angular and redshift space for the variable $A$. 
For instance, by considering $A=B=\Phi_W$ and $A=B=\psi$ we obtain that (setting $C_\ell^{\Psi_W}(z,z')\equiv C_\ell^W(z,z')$)
\begin{widetext}
\bea
\!\!\!\!\!\!\!\!\!\!C_\ell^W(z,z')&=&\frac{1}{2 \pi}\int dk\,k^2\,P_R(k)
\left[ T_{\Psi+\Phi}(k,z)j_\ell\left((k r\right) \right]
\left[ T_{\Psi+\Phi}(k,z')j_\ell\left(k r'\right) \right]\,,
\\
C_\ell^\psi(z,z')&=&\frac{2}{\pi}\int dk\,k^2\,P_R(k)
\left[ \int_{0}^{r}d r_1\frac{r-r_1}{r r_1}
T_{\Psi+\Phi}(k,z_1)j_\ell\left( k r_1 \right) \right]  \left[\int_{0}^{r'}d r_2 \frac{ r' -r_2 }{r' r_2}
T_{\Psi+\Phi}(k,z_2)j_\ell\left(  k r_2 \right) \right] \,,
\label{Cpsi}
\eea
\end{widetext}
where $j_\ell$ denotes a spherical Bessel function of order $\ell$. As before, $r\equiv \eta_o -\eta$ is the comoving distance to redshift $z$, and analogously $r', r_1,r_2$ denote the distances to redshifts $z',~z_1,~z_2$. Above and hereafter, we define $z=z(r)$, $z'=z(r')$, etc..

Hereafter,
in order to numerically evaluate the next-to-leading order lensing contributions to the CMB temperature and polarization anisotropies,
we will apply the Limber approximation~\cite{Limber:1954zz,Kaiser:1991qi,LoVerde:2008re}.
We remark that this approximation works very well for CMB lensing. Indeed, CMB lensing is appreciable only for 
$\ell>100$, where 
the Limber approximation is very close to the exact solution.
  
Following~\cite{Bernardeau:2011tc}, the Limber approximation can be written as
\bea \label{eq:Limber}
&& \hspace{-2.5cm}\frac{2}{\pi}\int dk\,k^2\,f(k)
j_\ell\left(k x_1\right) j_\ell\left(k x_2\right)
\simeq 
\nonumber \\
\hspace{0.8cm}
&\simeq&
\frac{\delta_D(x_1-x_2)}{x_1^2} f\left(\frac{\ell+1/2}{x_1}\right)\,,
\eea
where $f(k)$ should be a smooth, not strongly oscillating function of $k$ which decreases sufficiently rapidly for $k\ra\infty$ (more precisely, $f(k)$ has to decrease faster than $1/k$ for $k>\ell/x$).
Using this approximation, one  can then obtain the Limber-approximated $C_\ell^W$ and $C_\ell^\psi$
(see \cite{Marozzi:2016uob} for details).

Starting with the definitions~(\ref{DeflspaceB}) and (\ref{Deflspace}), we can transform Eqs.~(\ref{eq:expansion-compact}) 
and (\ref{eq:expansion-compact_x})
into $\vl{}$ space where they become (see \cite{Marozzi:2016uob} for details) 
\bea
\tilde\Mcal(z_s,\vl{})&\simeq&
\Acal^{(0)}(\vl{})+\sum_{i=1}^4\Acal^{(i)}(\vl{})
+\sum_{\substack{i+j\le 4 \\ 1\le i\le j}}\Acal^{(ij)}(\vl{})
\nonumber \\
&+&\sum_{\substack{i+j+k\le 4 \\ 1\le i\le j\le k}} \Acal^{(ijk)}(\vl{})
+\Acal^{(1111)}(\vl{})\,,
\label{eq:expansion-compact-l}
\eea
\bea
\tilde\Pcal(z_s,\vl{})&\simeq&
\Dcal^{(0)}(\vl{})+\sum_{i=1}^4\Dcal^{(i)}(\vl{})
+\sum_{\substack{i+j\le 4 \\ 1\le i\le j}}\Dcal^{(ij)}(\vl{})
\nonumber \\
&+&\sum_{\substack{i+j+k\le 4 \\1\le i\le j\le k}} \Dcal^{(ijk)}(\vl{})
+\Dcal^{(1111)}(\vl{})\,,
\label{eq:expansion-compact-l-Pol}
\eea
where we drop the redshift dependence for simplicity on the right hand side, and
we have
\bea
\Dcal^{(0)}(z_s,\vl{})&\equiv&\Pcal(z_s,\vl{})=\frac{1}{2 \pi}\int d^2 {x} \Pcal(z,\bx) e^{i\vl{}\cdot{\bf x}}
\nonumber \\
&=& -\left[\Ecal(z,\vl{})+i \Bcal(z,\vl{})\right] e^{-2 i \varphi_\ell} \,.
\label{Plspace}
\eea

To evaluate the lensing corrections at next-to-leading order we have now to calculate the 
$\vl{}$ space expressions for the  terms $\Acal^{(i....)}$ and $\Dcal^{(i....)}$.
The  expressions for $\Acal^{(i....)}$ considering at  next-to-leading order only the 
post-Born corrections were determined in~\cite{Marozzi:2016uob}.
Starting from these results (see Appendix A of \cite{Marozzi:2016uob}), 
and from the results of
Sect.~\ref{Sec5} for the LSS corrections, one can easily find 
the corresponding expressions for $\Dcal^{(i....)}$ both at leading and next-to-leading order.
They  are obtained from the  $\Acal^{(i....)}$ by the substitution 
 \be 
 \Mcal(z_s,\vl{}) \quad \rightarrow \quad  - \left[\Ecal(z_s,\vl{})+i \Bcal(z_s,\vl{})\right] e^{-2 i \varphi_{\ell}}  \,,
 \label{SubMEB}
 \ee
 performed for any $\Mcal(z_s,\vl{})$ inside the integrals. 
 For completeness, we report them in Appendix \ref{a:Dl}.
  This is very useful as it means, 
comparing Eq. (\ref{eq:expansion-compact-l-Pol}) with Eq. (\ref{eq:expansion-compact-l}) and 
using Eq.~(\ref{SpectraAll}), 
that the lensing corrections at the next-to-leading order of $C^{\Ecal \Mcal}_\ell$,  $C^{\Ecal}_\ell$
and $C^{\Bcal}_\ell$ can be obtained, as the leading lensing corrections (see \cite{Lewis:2006fu,RuthBook}), by using the results for $C^{\Mcal}_\ell$ by a series of simple substitutions (see also
\cite{Lewis:2016tuj}). 
Namely, we find that the corrections to ${C}_\ell^{\Ecal\Mcal}$ are obtained by substituting
\bea
C_{\ell}^{\Mcal}(z_s)& \rightarrow & {C}_\ell^{\Ecal\Mcal}(z_s) \,,
\nonumber \\
\hat{C}_{\ell_1}^{\Mcal}(z_s)& \rightarrow & {C}_{\ell_1}^{\Ecal\Mcal}(z_s) \cos [2(\varphi_{\ell_1}-\varphi_{\ell})]\,,
\label{subET}
\eea
the corrections to ${C}_\ell^{\Ecal}$ by substituting
\bea
C_{\ell}^{\Mcal}(z_s) & \rightarrow &{C}_\ell^{\Ecal}(z_s) \,,
\nonumber \\
\hat{C}_{\ell_1}^{\Mcal}(z_s) & \rightarrow &{C}_{\ell_1}^{\Ecal}(z_s) \cos^2 [2(\varphi_{\ell_1}-\varphi_{\ell})]
\nonumber \\
&&+
{C}_{\ell_1}^{\Bcal}(z_s) \sin^2 [2(\varphi_{\ell_1}-\varphi_{\ell})]\,,
\label{subE}
\eea
and, finally, the corrections to ${C}_\ell^{\Bcal}$ by substituting
\bea
C_{\ell}^{\Mcal}(z_s)& \rightarrow &{C}_\ell^{\Bcal}(z_s) \,,
\nonumber \\
\hat{C}_{\ell_1}^{\Mcal}(z_s)& \rightarrow &{C}_{\ell_1}^{\Ecal}(z_s) \sin^2 [2(\varphi_{\ell_1}-\varphi_{\ell})]
\nonumber \\
&&+
{C}_{\ell_1}^{\Bcal}(z_s) \cos^2 [2(\varphi_{\ell_1}-\varphi_{\ell})]\,,
\label{subB}
\eea
where we use a $\hat{}$ to indicate the $C_{\ell}^{\Mcal}$ that are inside an integral (for completeness, we present more details in Appendix \ref{AppFC}).

At this point, let us briefly recall our approach to obtain the lensing correction to the temperature anisotropies beyond leading order (see \cite{Marozzi:2016uob} for details).
Following \cite{Marozzi:2016uob}, we have that
\be
\langle \tilde{\Mcal}(\vl{})\bar{\tilde{\Mcal}}(\vl{}\,')\rangle =\langle \Acal (\vl{}) \bar{\Acal}(\vl{}\,')\rangle \,,
\ee
where 
\bea
\Acal(\vl{})&=& \Acal^{(0)}(\vl{})+
\sum_{i=1}^4\Acal^{(i)}(\vl{})
+\sum_{\substack{i+j\le 4 \\  1\le i\le j}}\Acal^{(ij)}(\vl{})
\nonumber \\
&& +\sum_{\substack{i+j+k\le 4 \\ 1\le i\le j\le k}} \Acal^{(ijk)}(\vl{})+\Acal^{(1111)}(\vl{})\,.
\label{AcalTotal}
\eea
We now introduce  $C_\ell^{(i\ldots ,\, j\ldots)}$ defined by
\bea
\delta\left( \vl{}-\vl{}\,' \right) C_\ell^{(ij\ldots,ij\ldots)} &=&\langle \Acal^{(ij\ldots)} (\vl{}) 
\bar{\Acal}^{(ij\ldots)}(\vl{}\,')\rangle\,,
\nonumber \\
\delta\left( \vl{}-\vl{}\,' \right) C_\ell^{(ij\ldots,i'j'\ldots)} &=&
\langle \Acal^{(ij\ldots)} (\vl{}) 
\bar{\Acal}^{(i'j'\ldots)}(\vl{}\,')\rangle
\nonumber \\
&+&\langle \Acal^{(i'j'\ldots)} (\vl{}) 
\bar{\Acal}^{(ij\ldots)}(\vl{}\,')\rangle\,,\quad\quad
\label{HigherOrderCll}
\eea
where the last definition applies when the coefficients $(ij\ldots)$ and $(i'j'\ldots)$ are not identical. 
The delta Dirac function $\delta\left( \vl{}-\vl{}\,' \right)$ is a consequence of statistical isotropy.
By omitting terms of higher  than  fourth order in the Weyl potential and 
terms that vanish as a consequence of  Wick's theorem (odd number of Weyl potentials), we obtain
\bea 
\tilde{C}^{\Mcal}_\ell&=& C^{{\Mcal}}_\ell+C_\ell^{(0,2)}+C_\ell^{(0,11)}+C_\ell^{(1, 1)}+C_\ell^{(0,4)}
+C_\ell^{(0,13)}
\nonumber \\
&&
+C_\ell^{(0,22)}+C_\ell^{(0,112)}+C_\ell^{(0,1111)}
+C_\ell^{(1, 3)}+
C_\ell^{(2, 2)}
\nonumber \\
&  &+C_\ell^{(1, 12)}
+C_\ell^{(1, 111)}+C_\ell^{(2, 11)}+C_\ell^{(11, 11)}\,,
\eea
where $C_\ell^{(0,0)}\equiv C^{{\Mcal}}_\ell$ is the unlensed power spectrum.
The terms $C_\ell^{(0,2)}$, $C_\ell^{(0,4)}$ and  $C_\ell^{(0,112)}$, containing an odd number of deflection angles from only one direction, are identically zero  as a consequence of statistical isotropy.  This was shown explicitly for the post-Born part of $C_\ell^{(0,112)}$ in 
\cite{Marozzi:2016uob} and for 
the second order contribution $C_\ell^{(0,2)}$ in~\cite{Bonvin:2015uha}.

Furthermore, making use of the Gaussian statistics of the first order deflection angle, the full correction from first order deflection angles alone, to the unlensed $C_\ell^{\Mcal}$, i.e. all the terms above containing only 0's and 1's, can be  fully re-summed \cite{Seljak:1995ve,RuthBook,Lewis:2006fu}. Denoting this sum by
 $\tilde C^{\Mcal\, (1)}_\ell $ we have
 \begin{widetext}
 \be 
\tilde C^{\Mcal\, (1)}_\ell = \int dr  r J_0 \left( \ell r \right) \int \frac{d^2 \ell'}{ \left( 2 \pi \right)^2} C^{\Mcal}_{\ell'} e^{-i \boldsymbol{\ell}' \cdot {\bf r}} \exp \left[ -\frac{\ell'^2}{2} \left( A_0 \left( 0 \right) - A_0 \left( r \right) + A_2 \left( r \right) \cos \left( 2 \varphi_\ell \right) \right) \right] \,,
\label{ExpFirstOrder}
\ee
\end{widetext}
\twocolumngrid
with
\bea
A_0 \left( r\right) &=& \int \frac{d \ell \ \ell^3}{ 2 \pi } C^\psi_\ell J_0 \left( r \ell \right)\, , 
\nonumber \\
A_2 \left( r\right) &=& \int \frac{d \ell \ \ell^3}{ 2 \pi } C^\psi_\ell J_2 \left( r \ell \right) \, .
\eea
and where $J_0$ and $J_2$ are the Bessel functions of order zero and two. 

We now write 
\be
\tilde C_\ell^{\Mcal} =\tilde C_\ell^{\Mcal\, (1)} + \Delta C_\ell^{(2)} +  \Delta C_\ell^{(3)}\,,
\label{eq:cltot}
\ee
where (neglecting vanishing contributions)
\bea
 \Delta C_\ell^{(2)} &=&C_\ell^{(0,13)} +  C_\ell^{(0,22)} + 
 C_\ell^{(1, 3)} + C_\ell^{(2, 2)}  \,, \label{eq:cl2}\\
 \Delta C_\ell^{(3)} &=& C_\ell^{(1, 12)} + C_\ell^{(2, 11)}  \label{eq:cl3}  \,.
\eea
As already mentioned,  $\tilde C_\ell^{\Mcal\, (1)} $ denotes the well known resummed correction from the first order
deflection angle \cite{Seljak:1995ve,RuthBook,Lewis:2006fu},  which is computed in standard CMB-codes~\cite{Lewis:1999bs,Lesgourgues:2011re}.  $\Delta C_\ell^{(2)}$ and  $\Delta C_\ell^{(3)}$ denote  corrections involving two or three deflection angles respectively, at least one of them beyond the Born approximation or with and higher order Weyl potential.
With a slight abuse of language we call them the Gaussian and non-Gaussian contribution of the 
deflection angle or, as in \cite{Marozzi:2016uob}, the second and third group respectively. Even though the contributions to the second group are not Gaussian, they would be present also if the higher order deflection angles would be Gaussian. Terms of the third group, however,  would vanish for Gaussian higher order deflection angles.
 Note that even though the number of deflection angles is odd in the third group, statistical isotropy does not require it to vanish as (in the correlation function picture) there is in addition the angle between the two directions $\bn_1$ and $\bn_2$ which can be employed to 'pair up' all the angles. If the deflections are all attached to one of these two directions this additional angle is no longer present and a term of the form $C_\ell^{(0,n_1\cdots n_{2j+1})}$ has to vanish due to statistical isotropy, while a term of the form $C_\ell^{(n_1\cdots n_k,n_{k+1}\cdots n_{2j+1})}$ with $k>0$ does not. Here we of course always assume that CMB anisotropies and deflection angles are uncorrelated as the latter come from much lower redshifts.

Furthermore, within the Limber approximation which is very accurate for these small corrections relevant only at high $\ell$ the two contributions $C_\ell^{(0,13)}$  and $C_\ell^{(0,22)}$   coming from  the post-Born part of the deflection angle exactly cancel, $C_\ell^{(0,13)} = - C_\ell^{(0,22)}$. This is no longer so when we consider the LSS contributions to these terms, see Sect.~\ref{Sec5} below.

\section{Post-Born contributions}
\label{Sec4}
\setcounter{equation}{0}

Let us first recall the results for the post-Born lensing corrections obtained in 
\cite{Marozzi:2016uob} for the temperature anisotropies. The results for polarization spectra can then be obtained as illustrated in the previous section.

\subsection{Second group}
\label{4.3}

The second group, where we study the leading post-Born corrections coming from
the deflection angles up to third order when these appear in pairs like 
$\langle \th{2}{a}\th{2}{b} \rangle$ and $\langle \th{1}{a}\th{3}{b} \rangle$, 
is given by 
\begin{widetext}
\bea
C_{\ell,pB}^{(1, 3)} &=& - \NINT{1}\NINT{2}
\left[\left(\vl{}-\vl{1}\right)\cdot\vl{1}\right]^2\,
\left[\left(\vl{}-\vl{1}\right)\cdot\vl{2}\right]^2\,
\hat{C}_{\ell_1}^{\Mcal}(z_s)
\nonumber\\
&&\times
\int_{0}^{r_s}dr'
\frac{\left(r_s-r'\right)^2}{r_s^2\,r'^4}
C_{\ell_2}^\psi(z',z')
P_R\left(\frac{|\vl{}-\vl{1}|+1/2}{r'}\right)
\left[T_{\Psi+\Phi}\left(\frac{|\vl{}-\vl{1}|+1/2}{r'},z'\right)\right]^2\,,\label{eq:secondgroup1}\\
&&
\nonumber
\\
C_{\ell,pB}^{(2, 2)} &=&
 \NINT{1}\NINT{2}\,\left[\left(\vl{}-\vl{1}+\vl{2}\right)\cdot\vl{1}\right]^2\,
\left[\left(\vl{}-\vl{1}+\vl{2}\right)\cdot\vl{2}\right]^2\,
\hat{C}_{\ell_1}^{\Mcal}(z_s)
\nonumber\\
&&\times\int_{0}^{r_S}dr'\frac{\left(r_s-r'\right)^2}{r_s^2\,r'^4}C_{\ell_2}^\psi(z',z')
 P_R\left(\frac{|\vl{}-\vl{1}+\vl{2}|+1/2}{r'}\right)
\left[T_{\Psi+\Phi}\left(\frac{|\vl{}-\vl{1}+\vl{2}|+1/2}{r'},z'\right)\right]^2\,.
\label{eq:secondgroup2}
\eea

\subsection{Third group}
\label{4.4}

The  third group, where we consider terms with three deflection angles which do not vanish due to the non-Gaussian statistic of 
$\theta^{a(2)}$,
is given by 
\bea
C_{\ell,pB}^{(1, 12)} 
&=&
- 2\,\NINT{1}\NINT{2}
\left(\vl{1}\cdot\vl{2}\right)\,
\left[(\vl{}-\vl{1})\cdot\vl{2}\right]\,
\left[\left(\vl{}-\vl{1}\right)\cdot\vl{1}\right]^2\nonumber\\
&&\times\,
\hat{C}_{\ell_1}^{\Mcal}(z_s)
\int_{0}^{r_s}dr'\frac{\left(r_s-r'\right)^2}{r_s^2\,r'^4}\,
P_R\left(\frac{|\vl{}-\vl{1}|+1/2}{r'}\right)\left[T_{\Psi+\Phi}\left(\frac{|\vl{}-\vl{1}|+1/2}{r'},z'\right)\right]^2
C_{\ell_2}^\psi\left( z_s,z' \right)\,,\label{eq:thirdgroup1}
\\ && \nonumber \\
C_{\ell,pB}^{(2, 11)} 
&=&
2\,\NINT{1}\NINT{2}\,
\left(\vl{1}\cdot \vl{2}\right)\,
\left[\left(\vl{}-\vl{1}+\vl{2}\right)\cdot\vl{2}\right]
\left[\left(\vl{}-\vl{1}+\vl{2}\right)\cdot \vl{1}\right]^2\,
\nonumber\\
&& \hspace{-0.5cm}\times\,
\hat{C}_{\ell_1}^{\Mcal}(z_s)
\int_{0}^{r_s}dr'\frac{\left(r_s-r'\right)^2}{r_s^2\,r'^4}
P_R\left(\frac{|\vl{}-\vl{1}+\vl{2}|+1/2}{r'}\right)
\left[T_{\Psi+\Phi}\left(\frac{|\vl{}-\vl{1}+\vl{2}|+1/2}{r'},z'\right)\right]^2
C_{\ell_2}^\psi\left( z_s,z' \right)\,.
\label{eq:thirdgroup2}
\eea
\end{widetext}

Like for the temperature anisotropies (see  \cite{Marozzi:2016uob}), also for the polarization spectra, 
the contributions above, within each group, partially erase each other. 
In the range of integration where $|\vl{}-\vl{1}+\vl{2}|\simeq
|\vl{}-\vl{1}|$ the integrands in Eqs.~(\ref{eq:secondgroup1}) and (\ref{eq:secondgroup2}) (as well as the ones in Eqs.~(\ref{eq:thirdgroup1}) and (\ref{eq:thirdgroup2})) are nearly identical and the corresponding contributions partially cancel (see \cite{Marozzi:2016uob} for details and a physical interpretation).


\section{LSS contributions}
\label{Sec5}
\setcounter{equation}{0}

In this section we determine the next-to-leading order corrections to CMB lensing  coming from higher order corrections of the Weyl potential (the so-called LSS contributions, see also 
\cite{Lewis:2016tuj}).

 We want to determine the LSS contributions to the
 deflection angle up to third order. As one sees from
Eqs.~(\ref{TheOrd2LSS}) and (\ref{LSS-TheOrd3}), this requires 
$\Phi_W^{(2)}$ and $\Phi_W^{(3)}$. We use the Newtonian approximations to $\Phi_W$ which are very accurate on largely sub-horizon scales, $k/\HH\gg 1$, and in a matter dominated regime. They are given by 
(see for example \cite{Bernardeau:2001qr})
 \bea
 \Phi_W^{(2)}(\bk,\eta) &=& -\frac{3\HH^2\Om_m(\eta)}{2k^2}  \delta^{(2)}(\bk,\eta) \,, \label{e:ka2}
 \\ 
 \delta^{(2)}(\bk,\eta)&=&  \frac{1}{\left( 2 \pi \right)^{3/2}} \int d^3k_1 d^3k_2 \delta_D \left( \bk - \bk_1 - \bk_2 \right) 
 \nonumber \\
 && 
\times F_2 \left( \bk_1, \bk_2 \right) \delta \left( \bk_1 ,\eta\right) \delta\left( \bk_2,\eta \right)\, ,\label{e:da2}\\
F_2(\bk_1,\bk_2) &=&\frac{5}{7} + \frac{1}{2} \frac{\bk_1 \cdot \bk_2}{k_1 k_2} \left( \frac{k_1}{k_2} + \frac{k_2}{k_1} \right) + \frac{2}{7} \left(\frac{\bk_1 \cdot \bk_2}{k_1 k_2} \right)^2  \nonumber\,, \\
\label{e:F2}
\eea
and~\cite{Goroff:1986ep,Nielsen:2016ldx}
\bea
&&
 \Phi_W^{(3)}(\bk,\eta) = -\frac{3\HH^2\Om_m(\eta)}{2k^2}  \delta^{(3)}(\bk,\eta) \,, \label{e:ka3}
 \\ 
 && \delta^{(3)}(\bk,\eta) = 
 \nonumber  \\
&& \qquad = \frac{1}{\left( 2 \pi \right)^3} \!\int \!d^3k_1d^3k_2d^3k_3 \delta_D\left( \bk\!-\bk_1\!-\bk_2 -\bk_3\right) 
 \nonumber \\
 &&
 \qquad
  \times
 F_3 \left( \bk_1, \bk_2,\bk_3\right)
  \delta \left( \bk_1 ,\eta\right)\delta \left( \bk_2 ,\eta\right)\delta \left( \bk_3 ,\eta\right), 
 \label{e:da3}\\
 &&
 F_3(\bk_1,\bk_2,\bk_3) =\frac{1}{18} \left\{ G_2(\bk_1,\bk_2)\left[7\alpha(\bk_1+\bk_2,\bk_3)
 \right. \right.
 \nonumber \\
    &&  \left. \left.
    \qquad
 +4\beta(\bk_1+\bk_2,\bk_3) \right]  
 +7\alpha(\bk_1,\bk_2+\bk_3)F_2(\bk_2,\bk_3) \right\},\label{e:F3} \nonumber \\
\eea
with
\be
\alpha(\bk,\bk') = \frac{(\bk+\bk')\cdot\bk}{k^2} \ , 
\qquad
\beta(\bk,\bk') = \frac{(\bk+\bk')^2\bk\cdot\bk'}{2k^2k'^2} \, ,
\ee
\be
G_2(\bk_1,\bk_2) =\frac{3}{7} + \frac{1}{2} \frac{\bk_1 \cdot \bk_2}{k_1 k_2} \left( \frac{k_1}{k_2} + \frac{k_2}{k_1} \right) + \frac{4}{7} \left(\frac{\bk_1 \cdot \bk_2}{k_1 k_2} \right)^2  \,.\label{e:G2}
\ee

We now write explicit formulas for the case of temperature anisotropies, the corresponding expressions for E- and B-modes are obtained  from the temperature results using the substitutions in Eqs.~(\ref{subET})-(\ref{subB}).

\subsection{Second group}
\label{5.2}
Let us  first evaluate the impact of the LSS corrections on our second group. 
As we will show explicitly in the follow, within the Limber approximation the LSS contribution to the second group is 
already included  when we consider an Halofit model in evaluating the leading first order contribution. Namely, it
 is equivalent to take the leading lensing correction, obtained from first order deflection angle, and consider  in the $C_{\ell}^\psi$ the higher order contributions to the gravitational potential (i.e., considering an higher order power spectrum).

To show this  we write the deflection angles up to third order in terms of the 2-dimensional 
Fourier transform of the Weyl potential including also the LSS contributions from $\Phi_W^{(2)}$ and 
$\Phi_W^{(3)}$. In general, an angle $\theta^{a(n)}$ contains a part which depends only on the first order 
Weyl potential and a second part which depends on higher order corrections to the Weyl potential, up to third order these are $\Phi_W^{(2)}$ and 
$\Phi_W^{(3)}$. The first part is the one evaluated in  \cite{Marozzi:2016uob}, let us call it  $\theta_{St}^{a(n)}$,
while we call the second part $\theta_\text{LSS}^{a(n)}$. Up to third order, the second part is given by
\begin{widetext}
\bea
\T{a}{2}_\text{LSS}({\bf x})
&=&\frac{i}{\pi}
\INT{}\int_{0}^{r_s}dr \frac{r_s-r}{r_s\,r}\,\ell^a
\Phi^{(2)}_W(r,\vl{}) e^{-i\vl{}\cdot {\bf x}}\,,
\\ & & \nonumber \\
\T{a}{3}_\text{LSS}({\bf x})
&=&\frac{i}{\pi}
\INT{}\int_{0}^{r_s}dr \frac{r_s-r}{r_s\,r}\,\ell^a
\Phi^{(3)}_W(r',\vl{}) e^{-i\vl{}\cdot {\bf x}}
\nonumber \\ &&
+\frac{i}{\pi^2}\INT{1}\INT{2}\int_{0}^{r_s}dr\frac{r_s-r}{r_s\,r}
\left(\,\ell_1^a\ell_{1b}\Phi_W^{(2)}(r',\vl{1})e^{-i\vl{1}\cdot {\bf x}}\right)\int_{0}^{r}dr' \frac{r-r'}{r\,r'}\,\ell_2^b
\Phi_W(r',\vl{2})e^{-i\vl{2}\cdot {\bf x}} 
\nonumber \\ &&
+\frac{i}{\pi^2}\INT{1}\INT{2}\int_{0}^{r_s}dr\frac{r_s-r}{r_s\,r}
\left(\,\ell_1^a\ell_{1b}\Phi_W(r,\vl{1})e^{-i\vl{1}\cdot {\bf x}}\right)
\int_{0}^{r}dr' \frac{r-r'}{r\,r'}\,\ell_2^b
\Phi_W^{(2)}(r',\vl{2})e^{-i\vl{2}\cdot {\bf x}} \,.
\eea
The LSS corrections to the second group contribute to $
C_\ell^{(0,22)}$, $C_\ell^{(0,13)}$, $C_\ell^{(2,2)}$ and 
$C_\ell^{(1,3)}$. 
To evaluate them we calculate the contribution of 
$\Phi^{(2)}_W$ and $\Phi^{(3)}_W$ to $\Acal^{(2)}(\vl{})$, $\Acal^{(3)}(\vl{})$, $\Acal^{(13)}(\vl{})$ and $\Acal^{(22)}(\vl{})$.
Following \cite{Marozzi:2016uob}, we obtain

\bea
&&\hspace{-0.6cm}\Acal^{(2)}_\text{LSS}(\vl{})=\frac{1}{2\pi}\int d^2 x\,\T{a}{2}_\text{LSS}\nabla_a\Mcal\,e^{i\vl{}\cdot{\bf x}}
\frac{1}{\pi}\INT{2}\left[\left(\vl{}-\vl{2}\right)\cdot\vl{2}\right]
\int_{0}^{r_s}dr \frac{r_s-r}{r_s\,r}\,
\Phi_W^{(2)}(r,\vl{}-\vl{2})\Mcal(r_s,\vl{2})\,,
\label{A2l}
\\ & & \nonumber \\
&&\hspace{-0.6cm}\Acal^{(3)}_\text{LSS}(\vl{})=\frac{1}{2\pi}\int d^2 x\,\T{a}{3}_\text{LSS}\nabla_a\Mcal\,e^{i\vl{}\cdot{\bf x}}\nonumber\\
&=&\frac{1}{\pi}\INT{2}\left[\left(\vl{}-\vl{2}\right)\cdot\vl{2}\right]
\int_{0}^{r_s}dr \frac{r_s-r}{r_s\,r}\,
\Phi_W^{(3)}(r,\vl{}-\vl{2})\Mcal(r_s,\vl{2}) \nonumber
\\
& & -\frac{1}{\pi^2}\INT{2}\INT{3}
\,\left[\left(\vl{}+\vl{2}-\vl{3}\right)\cdot\vl{3}\right]\,
\left[\left(\vl{}+\vl{2}-\vl{3}\right)\cdot\vl{2}\right]\int_{0}^{r_s}dr\frac{r_s-r}{r_s\,r}\nonumber\\
&&\times\int_{0}^{r}dr' \frac{r-r'}{r\,r'}\,
\left[\Phi_W(r,\vl{}+\vl{2}-\vl{3})
\bar\Phi_W^{(2)}(r',\vl{2})
+\Phi_W^{(2)}(r,\vl{}+\vl{2}-\vl{3})
\bar\Phi_W(r',\vl{2})\right]
\Mcal(r_s,\vl{3})\,,
\\
&&\hspace{-0.6cm}\Acal^{(13)}_\text{LSS}(\vl{})=\frac{1}{2 \pi}\int\! d^2 x \! \T{a}{1}\T{b}{3}_\text{LSS}\nabla_a\nabla_b\Mcal e^{i\vl{}\cdot {\bf x}}\nonumber\\
&=&-\frac{1}{\pi^2}\INT{2}\!\INT{3}\!\left[\left(\vl{}+\vl{2}-\vl{3}\right)\cdot \vl{3}\right]\,\left(\vl{2}\cdot \vl{3}\right)
\int_{0}^{r_s}dr \frac{r_s-r}{r_s\,r}\int_{0}^{r_s}dr' \frac{r_s-r'}{r_s\,r'}
\Phi_W(r,\vl{}+\vl{2}-\vl{3})
\bar{\Phi}^{(3)}_W(r',\vl{3})\,\Mcal(r_s,\vl{3})
\nonumber \\
&& 
+\frac{1}{\pi^3}
\INT{2}\INT{3}\INT{4}\left[\left(\vl{}-\vl{2}-\vl{3}-\vl{4}\right)\cdot\vl{4}\right]\,
\left(\vl{4}\cdot\vl{2}\right)\,
\left(\vl{3}\cdot\vl{2}\right)\nonumber\\
&&\hspace{2cm}\times\int_{0}^{r_s}dr \frac{r_s-r}{r_s\,r}\int_{0}^{r_s}dr'\frac{r_s-r'}{r_s\,r'}
\int_{0}^{r'}dr'' \frac{r'-r''}{r'\,r''}\Phi_W(r,\vl{}-\vl{2}-\vl{3}-\vl{4})
\nonumber
\\
&&
\hspace{2cm}\times
\left[\Phi_W(r',\vl{2})\Phi^{(2)}_W(r'',\vl{3})
+\Phi_W^{(2)}(r',\vl{2})\Phi_W(r'',\vl{3})\right]\Mcal(r_s,\vl{4})\,,
\\
&&\hspace{-0.5cm}\Acal^{(22)}_\text{LSS}(\vl{})=\frac{1}{2 \pi}\int d^2 x\frac{1}{2}
\left[\T{a}{2}_\text{LSS}\T{b}{2}_\text{LSS}+2\T{a}{2}\T{b}{2}_\text{LSS}\right]\nabla_a\nabla_b\Mcal e^{i\vl{}\cdot {\bf x}}\nonumber\\
&=&
-\frac{1}{2}\frac{1}{\pi^2}\INT{2}\INT{3}\,\left[\left(\vl{}+\vl{2}-\vl{3}\right)\cdot \vl{3}\right]\,\left(\vl{2}\cdot \vl{3}\right)
\int_{0}^{r_s}dr \frac{r_s-r}{r_s\,r}\int_{0}^{r_s}dr' \frac{r_s-r'}{r_s\,r'}
\Phi_W^{(2)}(r,\vl{}+\vl{2}-\vl{3})
\bar\Phi_W^{(2)}(r',\vl{2})\,\Mcal(r_s,\vl{3})\nonumber \\
&& +
\frac{1}{\pi^3}
\INT{2}\INT{3}\INT{4}\left[\left(\vl{}-\vl{2}-\vl{3}-\vl{4}\right)\cdot\vl{4}\right]\,
\left(\vl{4}\cdot\vl{2}\right)\,
\left(\vl{3}\cdot\vl{2}\right)\nonumber\\
&&\qquad \times\int_{0}^{r_s}dr \frac{r_s-r}{r_s\,r}\int_{0}^{r_s}dr'\frac{r_s-r'}{r_s\,r'}
\int_{0}^{r'}dr'' \frac{r'-r''}{r'\,r''}
\Phi_W^{(2)}(r,\vl{}-\vl{2}-\vl{3}-\vl{4})\Phi_W(r',\vl{2})\Phi_W(r'',\vl{3})\Mcal(r_s,\vl{4})\,.
\eea
\end{widetext}
With these results and using also the $\Acal^{(i....)}(\vl{})$  containing only the first order Weyl potential given in \cite{Marozzi:2016uob}, we can now determine the LSS contribution to the second group by following the procedure outlined in~\cite{Marozzi:2016uob}. 
We first introduce
\bea
& &\!\!\!\!\!\!\!\!\!\!\!\!\langle \Phi_W^{(2)}
(z,\vl{}) \bar{\Phi}_W^{(2)}(z',\vl{}') \rangle=\delta\left( \vl{}-\vl{}\,' \right)C_\ell^{W(22)}(z,z') \,,
\nonumber
\\
& &\!\!\!\!\!\!\!\!\!\!\!\!\langle \Phi_W
(z,\vl{})\bar{\Phi}_W^{(3)}(z',\vl{}') \rangle=\delta\left( \vl{}-\vl{}\,' \right)C_\ell^{W(13)}(z,z') \,,
\eea
and
\begin{widetext}
\bea
C_\ell^{\psi(22)}(z,z')&=& 4 \int_{0}^{r}d r_1\frac{r-r_1}{r r_1}
 \int_{0}^{r'}d r_2 \frac{ r' -r_2 }{r' r_2} C_\ell^{W(22)}(z_1,z_2) \,, \nonumber \\
C_\ell^{\psi(13)}(z,z')&=& 4 \int_{0}^{r}d r_1\frac{r-r_1}{r r_1}
 \int_{0}^{r'}d r_2 \frac{ r' -r_2 }{r' r_2} C_\ell^{W(13)}(z_1,z_2) \,.
\label{CpsiHO}
\eea
With this we obtain
\bea
C_{\ell,LSS}^{(0, 22)}+ C_{\ell,LSS}^{(0, 13)}
&=&
-C_{\ell}^\Mcal(z_s)\NINT{1}\,\left(\vl{1}\cdot \vl{}\right)^2\,
\left[C_{\ell_1}^{\psi(22)}(z_s,z_s)+2 C_{\ell_1}^{\psi(13)}(z_s,z_s)\right]
\nonumber \\
&&
- 16 {C}_{\ell}^{\Mcal}(z_s)\,\NINT{1}\NINT{2}\,
\left[(\vl{2}+\vl{3})\cdot\vl{}\right]\,
\left(\vl{2}\cdot\vl{}\right)
\left(\vl{3}\cdot\vl{}\right)
\nonumber\\
&&\times\,
\int_{0}^{r_s}dr\frac{r_s-r}{r_s\,r}\,
\int_{0}^{r_s}dr'\frac{r_s-r'}{r_s\,r'}\,
\int_{0}^{r'}dr''\frac{r'-r''}{r'\,r''}\,
b_{|\vl{2}+\vl{3}|\ell_2\ell_3}^{\Phi\Phi\Phi^{(2)}}(r,r',r'')\,,
\label{LSS-eq:secondgroup1} 
\\ & & \nonumber \\
C_{\ell,LSS}^{(2, 2)} +C_{\ell,LSS}^{(1, 3)} 
&=&
\NINT{1}\left[\left(\vl{}-\vl{1}\right)\cdot\vl{1}\right]^2\,
\left[C_{|\vl{}-\vl{1}|}^{\psi(22)}(z_s,z_s)+C_{|\vl{}-\vl{1}|}^{\psi(13)}(z_s,z_s)\right]C_{\ell_1}^\Mcal(z_s)
\nonumber \\
&& 
- 16\,\NINT{1}\NINT{2}
\left[\left(\vl{}+\vl{2}-\vl{1}\right)\cdot\vl{2}\right]\,
\left[(\vl{}-\vl{1})\cdot\vl{1}\right]\,
\left[\left(\vl{}+\vl{2}-\vl{1}\right)\cdot\vl{1}\right]\nonumber\\
&&\times\,
{C}_{\ell_1}^{\Mcal}(z_s)
\int_{0}^{r_s}dr\frac{r_s-r}{r_s\,r}\,
\int_{0}^{r_s}dr'\frac{r_s-r'}{r_s\,r'}\,
\int_{0}^{r'}dr''\frac{r'-r''}{r'\,r''}\,
b_{|\vl{}-\vl{1}||\vl{}-\vl{1}+\vl{2}|\ell_2}^{\Phi\Phi\Phi^{(2)}}(r,r',r'')\,,
\label{LSS-eq:secondgroup2}
\eea
where $b_{\ell_1\ell_2\ell_3}^{\Phi\Phi\Phi^{(2)}}$ is a reduced bispectrum and is defined by
\be 
\langle \Phi_W^{(2)}(r_1,\vl{1}) \Phi_W (r_2,\vl{2}) \Phi_W(r_3,\vl{3})
\rangle_c+\text{perm.}=\delta_D(\vl{1}+\vl{2}+\vl{3}) \frac{1}{2 \pi} 
b_{\ell_1\ell_2\ell_3}^{\Phi^{(2)}\Phi\Phi}(r_1,r_2,r_3)\,.
\ee
Following Sec.~3.4 of \cite{DiDio:2015bua} and using the Limber approximation we  obtain the following expression for the reduced bispectrum
\bea
b^{\Phi^{(2)}\Phi\Phi}_{\ell_1 \ell_2 \ell_3 } \left( z_1 , z_2 , z_3 \right) &=&
-\frac{1}{12} \left[\Hcal(\eta_1)^2(\Omega_m(\eta_1)\right]^{-1}
 \frac{\delta_D (r_2 - r_3)\delta_D (r_1 - r_3)}{r_3^2} \nu_2^2 \nu_3^2 
 \frac{1}{(\ell_1+1/2)^2} \nonumber \\
 &&
 P_R(\nu_2) P_R(\nu_3)T^2_{\Phi+\Psi} \left( \nu_2, \eta_3 \right)  T^2_{\Phi+\Psi} \left( \nu_3, \eta_3 \right)
 F_2\left(\frac{\ell_1+1/2}{r_3}, \nu_2, \nu_3\right) 
 + \text{perm.} \,,
   \label{RedBispPhiOrd2Alt}
\eea

where $\nu_i \equiv \frac{\ell_i +1/2}{r_i}$, $r_i=r(z_i)$ as well as $\eta_i=\eta(z_i)$ and we define 
(see \cite{DiDio:2015bua})
\bea
F_2 \left( k_1 , k_2 ,k_3 \right)&=& \frac{5}{7} + \frac{1}{4} \frac{k_1^2-k_2^2-k_3^2}{k_2 k_3} \left( \frac{k_2}{k_3} + \frac{k_3}{k_2} \right)
+ \frac{1}{14} \left(\frac{k_1^2-k_2^2-k_3^2}{k_2 k_3} \right)^2 \, .
\label{F2k1k2k3}
\eea
\end{widetext}
The first contributions to Eqs.~(\ref{LSS-eq:secondgroup1}) and (\ref{LSS-eq:secondgroup2}) take care of when we take into account higher order contributions to the gravitational potential in $C_{\ell}^\psi$  (a higher order power spectrum) and, therefore, it is included when we consider a Halofit model in evaluating the leading first order contribution (in the sense that if we add this contribution to the first order contribution evaluated via Halofit we would effectively do a double counting).
The second terms in Eqs.~(\ref{LSS-eq:secondgroup1}) and (\ref{LSS-eq:secondgroup2}), depend on the reduced bispectrum. In the Limber approximation given in Eq.~(\ref{RedBispPhiOrd2Alt}) these contributions vanish due to the Dirac-delta function, $\delta(r'-r'')$.

\subsection{Third group}
\label{5.1}

 We now evaluate the LSS corrections to our third group. In this group no 3rd order perturbation occur and  it is sufficient to consider the LSS contribution 
 in the deflection angle up to second order.

From the definitions in Eqs. (\ref{HigherOrderCll}) and (\ref{eq:cl3}) 
the LSS contribution to our third group is due to the contribution of 
$\Phi^{(2)}_W$ present in  $\Acal^{(2)}(\vl{})$ and $\Acal^{(12)}(\vl{})$.
The expression for $\Acal_\text{LSS}^{(2)}(\vl{})$ is given in Eq. (\ref{A2l}). 
While, following \cite{Marozzi:2016uob} we  obtain
\begin{widetext}
\onecolumngrid
\bea
\Acal^{(12)}_\text{LSS}(\vl{})&=&\frac{1}{2 \pi}\int d^2 x\T{a}{1}\T{b}{2}_\text{LSS}\nabla_a\nabla_b\Mcal e^{i\vl{}\cdot {\bf x}}\nonumber\\
&=&-\frac{1}{\pi^2}\INT{2}\INT{3}\,\left[\left(\vl{}+\vl{2}-\vl{3}\right)\cdot \vl{3}\right]\,\left(\vl{2}\cdot \vl{3}\right)\int_{0}^{r_s}dr\frac{r_s-r}{r_s\,r}\nonumber\\
&&\times\int_{0}^{r_s}dr' \frac{r_s-r'}{r_s\,r'}
\Phi_W(r,\vl{}+\vl{2}-\vl{3})
\bar{\Phi}^{(2)}_W(r',\vl{2})\,\Mcal(r_s,\vl{3})\,.
\label{A12l}
\eea
Using Eqs. (\ref{A2l}) and (\ref{A12l}), the expression for $\Acal^{(1)}(\vl{})$ and $\Acal^{(11)}(\vl{})$ given in 
\cite{Marozzi:2016uob}, and  Eq.~(\ref{HigherOrderCll}), we then obtain the following  LSS contribution to the third group
\bea
C_{\ell,LSS}^{(1, 12)} +C_{\ell,LSS}^{(2, 11)} 
&=&
- 8\,\NINT{1}\NINT{2}
\left(\vl{1}\cdot\vl{2}\right)\,
\left[(\vl{}-\vl{1})\cdot\vl{1}\right]\,
\left[\left(\vl{}+\vl{2}-\vl{1}\right)\cdot\vl{1}\right]\nonumber\\
&&\times\,
{C}_{\ell_1}^{\Mcal}(z_s)
\int_{0}^{r_s}dr\frac{r_s-r}{r_s\,r}\,
\int_{0}^{r_s}dr'\frac{r_s-r'}{r_s\,r'}\,
\int_{0}^{r_s}dr''\frac{r_s-r''}{r_s\,r''}\,
b_{|\vl{}-\vl{1}||\vl{}-\vl{1}+\vl{2}|\ell_2}^{\Phi^{(2)}\Phi\Phi}(r,r',r'')
\label{LSS-eq:thirdgroup2} \,.
\eea
Note that this result remains finite in the Limber approximation for the reduced bispectrum as there is no factor $r'-r''$ in the integrand. 
Our expression \eqref{LSS-eq:thirdgroup2} for the LSS correction agrees with the corresponding result of Ref.~\cite{Lewis:2016tuj}.
\end{widetext}

\section{Contribution from rotation}
\label{Sec6}
\setcounter{equation}{0}

When considering the next-to-leading order corrections to the CMB polarization, another new effect has to be taken into account: polarization is oriented along a given direction at emission and this direction may rotate along the path of the photon to the observer position due to the presence of structure. Since this has been debated in the literature~\cite{Lewis:2017ans}, we first give a thorough introduction to the physics of the effect before entering into the computation.

\begin{figure}[th]
\centerline{\includegraphics[width=6cm]{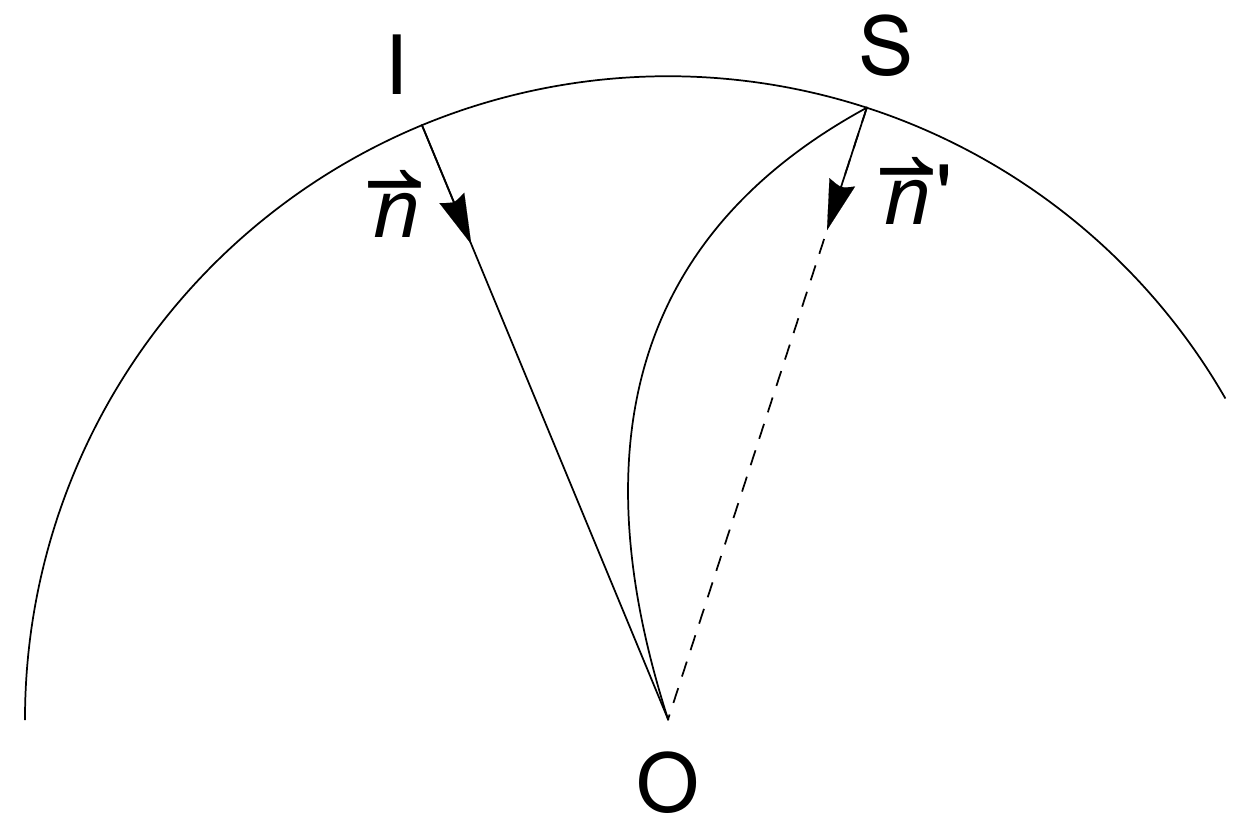}}
\caption{\label{f:direction} The (incoming) source direction $\bn'$ and the image direction $\bn$ are shown. In a generic coordinate 
system $\bn\neq\bn'$ while in GLC angular coordinates follow the photon direction so that $\bn\equiv \bn'$.}
\end{figure}

The problem that appears here is that parallel transport relates the lensed polarisation tensor $\tilde \Pcal_{nm}(\bn)$ with the unlensed polarisation  $\Pcal_{nm}(\bn')$, where $\bn =x^a=(\theta_o^1,\theta_o^2)$ is the direction of the image and 
$\bn'=x^a+\delta \theta^{a}=(\theta_s^1,\theta_s^2)$ is the direction of the source (which is equal to the unlensed position of the image).
To obtain $\tilde \Pcal_{nm}(\bn)$, we have to parallel transport the polarisation from the source position defined by $\bn' \neq \bn$ to the observer, see Fig.~\ref{f:direction}.  However, we must compare $\tilde \Pcal_{nm}(\bn)$ 
with the unlensed polarisation as it
would be observed in the same direction, $\bn$, if no perturbation was present.  The most elegant way to take this subtlety into account is the use of the so-called geodesic light cone (GLC) coordinates~\cite{Gasperini:2011us}. In these coordinates the direction of a photon $(\tilde\theta^1,\tilde\theta^2)$ is constant by definition, $\bn\equiv \bn'$  and we can compare the lensed and unlensed polarisation from the same direction. To find out whether the lensed polarisation is rotated, we therefore just have to study whether the parallel transported Sachs basis is rotated with respect to the directions $(\tilde\theta^1,\tilde\theta^2)$.
We do exactly this in Appendix~\ref{AppA}, where we determine the rotation angle $-\beta$ of the Sachs basis with respect to these directions.

Of course one can also study the problem in Poisson gauge.  A short calculation actually shows that when expressing the polarisation in terms of the directions defined by Poisson gauge, it does not rotate. (This is not exactly true, there actually is a small amount of rotation due to the fact that the photon is not emitted into the direction given by the emission point, $\bn'$, but in a somewhat different direction, see Fig.~\ref{f:direction}. This is discussed in detail in~\cite{Lewis:2017ans}, but since this effect is much smaller than the one discussed here, we neglect it.) In Poisson gauge the directions $\bn$ and $\bn'$ are different and to compare the lensed polarisation seen from direction $\bn$ with the unlensed polarisation from the same direction we have to move the unlensed $\Pcal_{mn}$ from $\bn'$ to $\bn$. In general this is done with the Jacobi map, $(\partial\bn/\partial\bn')$, but since we express the polarisation in terms of an orthonormal basis only the rotation $\om$ of this map contributes. In  Appendix~\ref{AppA}, we show that for scalar perturbations $\beta=\om$ up to second order and one obtains the same result in both ways as it should be.

Therefore, comparing the lensed and the unlensed polarisation {\it from the same direction} $\bn$ doing the calculation in GLC gauge or in Poisson gauge gives the same result. But the rotation of the unlensed $\Pcal_{nm}(\bn')$ into the unlensed result at $\bn$ must be taken into account. This effect has been overlooked in the previous literature~\cite{Pratten:2016dsm,Lewis:2016tuj,Lewis:2017ans} and we show in the following that it is quite substantial.

\begin{widetext}

\begin{figure}[th]
\centerline{\includegraphics[width=12cm]{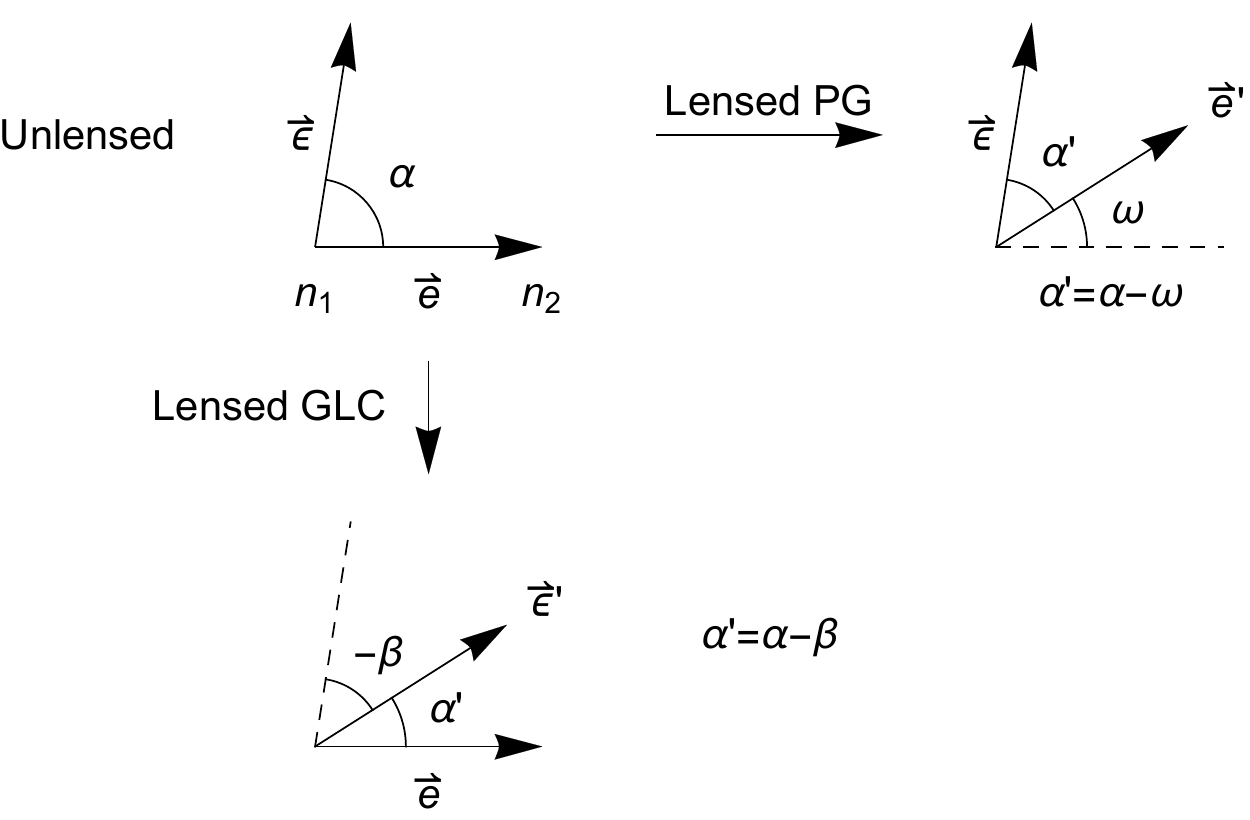}}
\caption{\label{f:bbep} The angle between two close by photons and the direction of polarisation is modified by lensing. Depending on the coordinate system used this is due to the rotation of the connecting vector $\bbe$ or due to the rotation of the polarisation $\bbep$.}
\label{f:bbep}
\end{figure}

\end{widetext}

Another way to understand that $\beta=\omega$ is to consider two nearby photons with connection vector $\bbe$. 
Assume that one of the photons be polarised in direction $\bbep$ enclosing an angle $\alpha$ with $\bbe$. Here, $\bbe$ provides a natural reference direction with respect to which we measure the rotation of polarization. Lensing will change this angle because $\bbe$ and 
$\bbep$ are differently transported (rotated) along their path towards the observer. Indeed, for small separation, $\bbe$ will be Lie transported, like an image, while $\bbep$ will be parallel transported as the Sachs basis, i.e.~the natural basis with respect to which rotation of the image is defined. It is natural to expect that the relative rotation coincides with $\omega$. Indeed in GLC coordinates, since the photon directions are not modified $\bbe$ remains unchanged while the polarisation is rotated by an angle $-\beta$ so that the angle between  $\bbep$ and $\bbe$ becomes $\alpha-\beta$. In Poisson gauge coordinates  $\bbep$ is not modified but the vector connecting the two photons is rotated by $\om=\beta$, hence again $\alpha$ changes into $\alpha-\beta$, see Fig.~\ref{f:bbep}.

To further explain the difference of our result to~\cite{Pratten:2016dsm,Lewis:2016tuj,Lewis:2017ans}, which do not take this rotation into account, let us also mention that when {\it fixing} a coordinate system at the observer, it is the direction of the source of the incoming photons which is rotated w.r.t. this fixed coordinate system by lensing. However, the only directions {\it intrinsic} to the problem are those of incoming photons, and the orientation of the polarisation w.r.t. the one of neighboring incoming photons, as shown in Fig.~\ref{f:bbep}, does rotate due to lensing. In this sense CMB lensing generates frame-dragging on cosmological scales as discussed in~\cite{Marozzi:2016und}.

Note also that this rotation is the only modification of the polarisation tensor which does not involve any derivatives of $\Pcal_{nm}$. So it cannot be confounded with any other term which we have considered before.

Let us now calculate the effects on the polarisation power spectra. We  consider
the rotation angle $\beta$, the effect of this rotation on Eq.~\eqref{eq:expansion-compact_x} is given by a rotation matrix $\Rcal^B_A$ (see Eq.~\eqref{eq:R}) acting on the Sachs basis, as defined in Appendix~\ref{AppA}. To evaluate it, the polarization tensor $\Pcal_{mn}$ is projected on a screen at the observer position given by Eq.~\eqref{eq:RotatedSachs} which 
is rotated by an angle $\beta$ with respect to the screen at the source. 
Because  the screen basis vectors appear twice in the projection of the polarization tensor, a rotation on it will change $\Pcal$ by  $2\beta$. This is simply a consequence of the spin-2 nature of the polarization tensor. Starting from \cite{Kamionkowski:2008fp,Dai:2013nda}
\be
\tilde{\Pcal}^{m n}(x^a) 2 \tilde{s}^{(+)}_m \tilde{s}^{(+)}_n = {\Pcal}^{m n}(x^a + \delta \theta^a)
 2 \tilde{s}^{(+)}_m \tilde{s}^{(+)}_n \,, 
 \ee
 with $\tilde{s}^{(+)}_m(x^a+\delta \theta^a)=e^{-\,i\,\beta}{s}^{(+)}_m(x^a+\delta \theta^a)$ and $s^{(\pm)}_m=\frac{1}{\sqrt{2}}\left(s_m^{1}\pm is_m^{2}\right)$, we obtain\footnote{Note that, to know the rotation $\beta$, the screen basis vector at the source as to be compared with the one at the observer parallel transported to the source following the background geodesic that connects observer and source. Let us point out that this is totally equivalent to what is stated above, the only crucial point is that the two vectors have to be expressed with respect to the same angles when compared.} 
 \be
\tilde\Pcal(x^a)=e^{-2\,i\,\beta}\Pcal(x^a+ \delta \theta^a)\,.
\ee
This rotation has not been included in Refs.~\cite{Pratten:2016dsm} and~\cite{Lewis:2016tuj}.
Note that $\Pcal$ is a scalar with respect to the indices $(m\,n)$ but has helicity $-2$ with respect to the Sachs basis vectors 
$\tilde{s}^{\pm}=\frac{1}{\sqrt{2}}(\tilde{s}^1\pm \tilde{s}^2)$.  Therefore, it does not matter whether we use Poisson gauge or GLC gauge to compute $\Pcal$. As the perturbed Sachs basis is rotated by an angle $\beta$ with respect to the unperturbed one, the invariance of the scalar $\tilde{\Pcal}^{m n}(x^a) 2 \tilde{s}^{(+)}_m \tilde{s}^{(+)}_n$ requests that $\tilde\Pcal$ is rotated by $-2\beta$.  In this work, we have actually used Poisson gauge to compute $\tilde\Pcal$. 

Because we are interested in next-to-leading order corrections, we must in principle take into account the expansion of $\beta$ up to fourth order,  $\beta\simeq\beta^{(0)}+\beta^{(1)}+\beta^{(2)}+\beta^{(3)}+\beta^{(4)}$. 
As explained in \cite{Kamionkowski:2008fp,Dai:2013nda}, 
in their framework 
this angle is also connected to the angle $\omega$ determined by the antisymmetric part of the amplification matrix. Qualitatively, $\omega$ and $\beta$ refer to different physical rotations: the vorticity $\omega$ takes into account the rotation of a bundle of light rays which travel together, whereas $\beta$ is meaningful also just for a single photon. Nevertheless, in Appendix~\ref{AppA} we show that these angles are equal to lowest non-vanishing order
also for scalar fluctuations
and they are both sourced by the curl potential $\Omega$ in the amplification matrix $\Psi^a_b$
(see \cite{Marozzi:2016uob} for definitions). More precisely, 
\beq
\beta^{(2)}=-\frac{1}{2}\Delta\Omega^{(2)}
\label{eq:rotationVSvorticity}
\eeq
which is exactly the vorticity $\omega^{(2)}$. 
In Appendix~\ref{AppA} we calculate $\beta$ from 
 scalar perturbations without reference to  the amplification matrix, by directly solving the parallel transport equation for the Sachs basis, and show the equality $\omega=\beta$ up to second order. Indeed, we find that $\beta^{(0)}$ and $\beta^{(1)}$ are constant along the geodesic, so there is no rotation of polarization between source and observer up to first order. With a global rotation of the Sachs basis we can achieve $\beta^{(0)}=\beta^{(1)}=0$. 
This is perfectly consistent with Eq.~\eqref{eq:rotationVSvorticity} since also $\omega^{(0)}=\omega^{(1)}=0$ for purely scalar first order  perturbations. Then we derive explicitly the non trivial equality $\beta^{(2)}=\omega^{(2)}$ (see Eq.~\eqref{eq:cornerstone} and its derivation in Appendix~\ref{AppA} for details). 

In principle, we should take into account also $\beta^{(3)}$ and $\beta^{(4)}$. However, because of the structure of the rotation, we can neglect all the terms which contain only one angle $\beta^{(i)}$ (this is again a consequence of statistical isotropy). 
The fact that $\beta^{(0)}=\beta^{(1)}=0$ then implies that $\beta^{(3)}$ and $\beta^{(4)}$ can only appear alone in the spectra, hence they do not contribute at next-to-leading order.

Before proceeding with the calculation of the rotated polarisation spectra, let us comment about the nature of the angle $\beta$. At the observer a natural Sachs basis is simply the angular directions 
$\tilde\theta^a =\theta^a_o$. On the path of the photon back to the source this basis is perturbed and at second order it is also rotated by an angle $\beta$. The angle $\beta$ is induced when the photon passes close to a structure but of course does not disappear even if the source and the observer are far away from any structure. Once the Sachs basis is rotated due to the presence of a structure it stays rotated.


In general, the full expansion of the polarization up to 4th order
reads
\begin{widetext}
\bea
&& \hspace{-0.5cm}\tilde{\Pcal}(x^a)=e^{-2 i \left( \beta^{(2)}+\beta^{(3)}+\beta^{(4)}\right)} \Pcal(x^a+\th{1}{a}+\th{2}{a}+\th{3}{a})\nonumber\\
&\simeq&\left[ 1-2i\beta^{(2)}-2i\beta^{(3)}-2i\beta^{(4)}-2\left( \beta^{(2)} \right)^2 \right]\nonumber\\
&&\times\left[ \Dcal^{(0)}(x^a)+
\sum_{i=1}^4\Dcal^{(i)}(x^a)
+\sum_{\substack{i+j\le 4 \\ 1\le i\le j}}\Dcal^{(ij)}(x^a)
+\sum_{\substack{i+j+k\le 4 \\ 1\le i\le j\le k}} \Dcal^{(ijk)}(x^a)\right.
+ \left.\Dcal^{(1111)}(x^a)\right]
\nonumber\\
&\simeq&\Dcal^{(0)}(x^a)+
\sum_{i=1}^4\Dcal^{(i)}(x^a)
+\sum_{\substack{i+j\le 4 \\ 1\le i\le j}}\Dcal^{(ij)}(x^a)
+\sum_{\substack{i+j+k\le 4 \\ 1\le i\le j\le k}} \Dcal^{(ijk)}(x^a)
+ \Dcal^{(1111)}(x^a)
\nonumber \\
& & \hspace{-0.5cm}
-2i\beta^{(2)}\left[ \Dcal^{(0)}(x^a)+
\sum_{i=1}^2\Dcal^{(i)}(x^a)
+\Dcal^{(11)}(x^a)\right]
-2i\beta^{(3)}\left[ \Dcal^{(0)}(x^a)+
\Dcal^{(1)}(x^a)\right]-\left[ 2i\beta^{(4)}+2\left( \beta^{(2)} \right)^2 \right]\Dcal^{(0)}(x^a).
\nonumber \\
&&
\label{eq:fullexpansion}
\eea
According to what we explained above, only two more terms containing $\beta^{(2)}$ contribute, namely
\beq
-2\,i\,\beta^{(2)}\Dcal^{(0)}\qquad\text{and}\qquad-2\left( \beta^{(2)} \right)^2\,\Dcal^{(0)}.
\eeq
Expressing the result for  $\beta^{(2)}$ given in Appendix~\ref{AppA}, in $\vl{}$ space, we obtain\bea
&&\hspace{-0.5cm}\Rcal^{(2)}(\bell)
=-\frac{2i}{2\pi}\int d^2 x\,\beta^{(2)}\Dcal^{(0)}e^{i\bell\cdot\bx}\nonumber\\
&=&-\frac{4i}{(2\pi)^2}\!\int_{0}^{r_s}\!dr \!\frac{r_s-r}{r_s\,r}\!\int_0^{r}dr_1\!\frac{r-r_1}{r\,r_1}
\int d^2\ell_1\,\int d^2\ell_2
\left[\boldsymbol{n}\cdot\left( \boldsymbol{\ell}_2\land\boldsymbol{\ell}_1 \right)\left(\boldsymbol{\ell}_1\cdot\boldsymbol{\ell}_2\right)\right]\,\Phi_W(z,\boldsymbol{\ell}_1)\Phi_W(z_1,\boldsymbol{\ell}_2)\,\Dcal^{(0)}(\bell-\bell_1-\bell_2) \,, \nonumber
\\
&&  \\ 
&&\hspace{-0.5cm}\Rcal^{(22)}(\bell)=-\frac{2}{2\pi}\int d^2x\left( \beta^{(2)} \right)^2 \Dcal^{(0)} e^{i \bell \cdot \bx}\nonumber\\
&=&-\frac{8}{(2\pi)^4} \int_{0}^{r_s}dr\,\frac{r_s-r}{r_s\,r}\int_0^{r}dr_1\,\frac{r-r_1}{r\,r_1}
\int_{0}^{r_s}dr_2\,\frac{r_s-r_2}{r_s\,r_2}\int_{0}^{r_2}dr_3\,\frac{r_2-r_3}{r_2\,r_3}\,
\nonumber\\
&&\times
\int d^2\ell_2\int d^2\ell_3\,\int d^2\ell_4\int d^2\ell_5
\left[\bn\cdot\left( \bell_2\land\bell_1 \right)\,\left(\bell_1 \cdot\bell_2\right)\right]\,
\left[\bn\cdot\left( \bell_4\land\bell_3 \right)\,\left(\bell_3\cdot\bell_4\right)\right]
\nonumber\\
&&
\hspace{4cm}\times\,
\Phi_W(z,\bell_1)\Phi_W(z_1,\bell_2)\Phi_W(z_2,\bell_3)\Phi_W(z_3,\bell_4)\Dcal^{(0)}(z_s,\bell-\bell_1-\bell_2-\bell_3-\bell_4).
\eea
\end{widetext}
Here, as in Appendix~\ref{AppA}, $\bn$ is the unit vector normal to the $\bell$-plane.
Using these expansions, we can now evaluate the contribution of $\beta^{(2)}$ to polarization. The new non-vanishing terms are  (see Appendix \ref{AppFC} for similar calculation for post-Born and LSS contributions)
\bea
\delta(\bell-\bell')\Delta\left( C^\Ecal_\ell+C^\Bcal_\ell \right)^{(22,0)}&=&\langle \Rcal^{(22)}(\bell)\bar\Dcal^{(0)}(\bell') \rangle \,, \nonumber\\
\delta(\bell-\bell')\Delta\left( C^\Ecal_\ell+C^\Bcal_\ell \right)^{(2,2)}&=&\langle \Rcal^{(2)}(\bell)\bar\Rcal^{(2)}(\bell') \rangle\,,\nonumber\\
e^{-4i\phi_\ell}\,\delta(\bell+\bell')\Delta\left( C^\Ecal_\ell-C^\Bcal_\ell \right)^{(22,0)}&=&\langle \Rcal^{(22)}(\bell)\Dcal^{(0)}(\bell') \rangle\,, \nonumber\\
e^{-4i\phi_\ell}\,\delta(\bell+\bell')\Delta\left( C^\Ecal_\ell-C^\Bcal_\ell \right)^{(2,2)}&=&\langle \Rcal^{(2)}(\bell)\Rcal^{(2)}(\bell') \rangle\,,\nonumber\\
-e^{-2i\phi_\ell}\,\delta(\bell-\bell')\Delta C^{\Ecal\Mcal(22,0)}_\ell&=&\langle \Rcal^{(22)}(\bell)\bar\Acal^{(0)}(\bell') \rangle
\,. \nonumber \\
&&
\label{rotationDelta}
\eea
Inserting our expressions for $\Rcal^{(22)}$, $\Rcal^{(2)}$, $\Dcal^{(0)}$ and $\Acal^{(0)}$ we find 
\begin{widetext}
\bea
\Delta\left( C^\Ecal_\ell+C^\Bcal_\ell \right)^{(22,0)}\!&=&\!
-8 \left[ C^\Ecal_{\ell}(z_s)+C^\Bcal_{\ell}(z_s) \right]\!
\int\! \frac{d^2\ell_1}{(2\pi)^2}\!\int\! \frac{d^2\ell_2}{(2\pi)^2}
\left[\bn\cdot\left( \bell_2\land\bell_1 \right)\left(\bell_1\cdot\bell_2\right)\right]^2
\int_{0}^{r_s}dr\,\frac{r_s-r}{r_s\,r}\int_0^{r}dr_1\,\frac{r-r_1}{r\,r_1}
\nonumber\\
&&\times\int_{0}^{r_s}dr_2\,\frac{r_s-r_2}{r_s\,r_2}\int_{0}^{r_2}dr_3\,\frac{r_2-r_3}{r_2\,r_3}
\left[ 
C^W_{\ell_1}(z,z_2)C^W_{\ell_2}(z_1,z_3)-C^W_{\ell_1}(z,z_3)C^W_{\ell_2}(z_1,z_2)\right] \,,
\\
& & \nonumber \\
\Delta\left( C^\Ecal_\ell-C^\Bcal_\ell \right)^{(22,0)}
&=&-8 \left[ C^\Ecal_{\ell}(z_s)-C^\Bcal_{\ell}(z_s) \right]\!
\int\! \frac{d^2\ell_1}{(2\pi)^2}\!\int\! \frac{d^2\ell_2}{(2\pi)^2}
\left[\bn\cdot\left( \bell_2\land\bell_1 \right)\left(\bell_1\cdot\bell_2\right)\right]^2 \int_{0}^{r_s}dr\,\frac{r_s-r}{r_s\,r}\int_0^{r}dr_1\,\frac{r-r_1}{r\,r_1}
\nonumber\\
&&\times
\int_{0}^{r_s}dr_2\,\frac{r_s-r_2}{r_s\,r_2}\int_{0}^{r_2}dr_3\,\frac{r_2-r_3}{r_2\,r_3}
\left[ 
C^W_{\ell_1}(z,z_2)C^W_{\ell_2}(z_1,z_3)-C^W_{\ell_1}(z,z_3)C^W_{\ell_2}(z_1,z_2)\right]\,,
\\
& & \nonumber \\
\Delta\left( C^\Ecal_\ell+C^\Bcal_\ell \right)^{(2,2)}&=&
16\int \frac{d^2\ell_1}{(2\pi)^2}\,\int \frac{d^2\ell_2}{(2\pi)^2}
\,\left[\boldsymbol{n}\cdot\left( \boldsymbol{\ell}_2\land\boldsymbol{\ell}_1 \right)\left(\boldsymbol{\ell}_1\cdot\boldsymbol{\ell}_2\right)\right]^2\int_{0}^{r_s}dr\,\frac{r_s-r}{r_s\,r}\int_0^{r}dr_1\,\frac{r-r_1}{r\,r_1}
\nonumber\\
&&\times
\int_{0}^{r_s}dr_2\,\frac{r_s-r_2}{r_s\,r_2}\int_{0}^{r_2}dr_3\,\frac{r_2-r_3}{r_2\,r_3}
\left[ C^\Ecal_{|\bell-\bell_1-\bell_2|}(z_s)+C^\Bcal_{|\bell-\bell_1-\bell_2|}(z_s) \right]\nonumber\\
&&\times \left[ C^W_{\ell_1}(z,z_2)C^W_{\ell_2}(z_1,z_3)
-C^W_{\ell_1}(z,z_3)C^W_{\ell_2}(z_1,z_2)\right]\,,
\\
\Delta\left( C^\Ecal_\ell-C^\Bcal_\ell \right)^{(2,2)}&=&
-16\int \frac{d^2\ell_1}{(2\pi)^2}\,\int \frac{d^2\ell_2}{(2\pi)^2}
\,\left[\boldsymbol{n}\cdot\left( \boldsymbol{\ell}_2\land\boldsymbol{\ell}_1 \right)\left(\boldsymbol{\ell}_1\cdot\boldsymbol{\ell}_2\right)\right]^2\int_{0}^{r_s}dr\,\frac{r_s-r}{r_s\,r}\int_0^{r}dr_1\,\frac{r-r_1}{r\,r_1}
\nonumber\\
&&\times
\int_{0}^{r_s}dr_2\,\frac{r_s-r_2}{r_s\,r_2}\int_{0}^{r_2}dr_3\,\frac{r_2-r_3}{r_2\,r_3}
\left[ C^\Ecal_{|\bell-\bell_1-\bell_2|}(z_s)-C^\Bcal_{|\bell-\bell_1-\bell_2|}(z_s) \right]\nonumber\\
&&\times \left\{ \cos^2 \left[2 \left( \phi_\ell - \phi_{|\bell-\bell_1-\bell_2|} \right)\right] - \sin^2 \left[ 2 \left( \phi_\ell - \phi_{|\bell-\bell_1-\bell_2|} \right)\right]\right\}
\nonumber\\
&&\times\,
\left[C^W_{\ell1}(z,z_2)C^W_{\ell2}(z_1,z_3)
-C^W_{\ell_1}(z,z_3)C^W_{\ell_2}(z_1,z_2)\right]\,,
\\
\Delta C^{\Ecal\Mcal(22,0)}_\ell&=&
-8\,C^{\Ecal\Mcal}_{\ell}(z_s)
\,\int \frac{d^2\ell_1}{(2\pi)^2}\int \frac{d^2\ell_2}{(2\pi)^2}\,\left[\bn\cdot\left( \bell_2\land\bell_1 \right)\left(\bell_1\cdot\bell_2\right)\right]^2\int_{0}^{r_s}dr\,\frac{r_s-r}{r_s\,r}\int_0^{r}dr_1\,\frac{r-r_1}{r\,r_1}
\nonumber\\
&&\times
\int_{0}^{r_s}dr_2\,\frac{r_s-r_2}{r_s\,r_2}\int_{0}^{r_2}dr_3\,\frac{r_2-r_3}{r_2\,r_3}
\left[ 
C^W_{\ell_1}(z,z_2)C^W_{\ell_2}(z_1,z_3)-C^W_{\ell_1}(z,z_3)C^W_{\ell_2}(z_1,z_2)\right]\,.
\label{DeltaEM220}
\eea

From $\Delta\left( C^\Ecal_\ell\pm C^\Bcal_\ell \right)$, we can easily obtain the corrections to $C^\Ecal_\ell$ and $C^\Bcal_\ell$,
\bea
\Delta C^{\Ecal(22,0)}_\ell&\equiv&\frac{1}{2}\left[ \Delta\left( C^\Ecal_\ell+C^\Bcal_\ell \right)^{(22,0)}+\Delta\left( C^\Ecal_\ell-C^\Bcal_\ell \right)^{(22,0)}\right]\nonumber\\
&=&-8\, C^\Ecal_{\ell}(z_s)\,
\int \frac{d^2\ell_1}{(2\pi)^2}\int \frac{d^2\ell_2}{(2\pi)^2}
\left[\bn\cdot\left( \bell_2\land\bell_1 \right)\left(\bell_1\cdot\bell_2\right)\right]^2\int_{0}^{r_s}dr\,\frac{r_s-r}{r_s\,r}\int_0^{r}dr_1\,\frac{r-r_1}{r\,r_1}
\nonumber\\
&&\times
\int_{0}^{r_s}dr_2\,\frac{r_s-r_2}{r_s\,r_2}\int_{0}^{r_2}dr_3\,\frac{r_2-r_3}{r_2\,r_3}
\left[ 
C^W_{\ell_1}(z,z_2)C^W_{\ell_2}(z_1,z_3)-C^W_{\ell_1}(z,z_3)C^W_{\ell_2}(z_1,z_2)\right]\,,
\label{DeltaE220}
\\
\Delta C^{\Ecal (2,2)}_\ell&\equiv&\frac{1}{2}\left[ \Delta\left( C^\Ecal_\ell+C^\Bcal_\ell \right)^{(2,2)}+\Delta\left( C^\Ecal_\ell-C^\Bcal_\ell \right)^{(2,2)}\right]\nonumber\\
&=&16\int \frac{d^2\ell_1}{(2\pi)^2}\,\int \frac{d^2\ell_2}{(2\pi)^2}
\left[\boldsymbol{n}\cdot\left( \boldsymbol{\ell}_2\land\boldsymbol{\ell}_1 \right)\left(\boldsymbol{\ell}_1\cdot\boldsymbol{\ell}_2\right)\right]^2\,\int_{0}^{r_s}dr\,\frac{r_s-r}{r_s\,r}\int_0^{r}dr_1\,\frac{r-r_1}{r\,r_1}
\nonumber\\
&&\times
\int_{0}^{r_s}dr_2\,\frac{r_s-r_2}{r_s\,r_2}\int_{0}^{r_2}dr_3\,\frac{r_2-r_3}{r_2\,r_3}
\,\left[ C^W_{\ell_1}(z,z_2)C^W_{\ell_2}(z_1,z_3)
-C^W_{\ell_1}(z,z_3)C^W_{\ell_2}(z_1,z_2)\right]\nonumber\\
&&\times \left\{ C^\Ecal_{|\bell-\bell_1-\bell_2|}(z_s)\sin^2\left[2 \left( \phi_\ell - \phi_{|\bell-\bell_1-\bell_2|} \right)\right] \right.
\left. +C^\Bcal_{|\bell-\bell_1-\bell_2|}(z_s)\cos^2\left[2 \left( \phi_\ell - \phi_{|\bell-\bell_1-\bell_2|} \right)\right]\right\}
\label{E22}\,,
\\
&& \nonumber \\ 
\Delta C^{\Bcal(22,0)}_\ell&\equiv&\frac{1}{2}\left[ \Delta\left( C^\Ecal_\ell+C^\Bcal_\ell \right)^{(22,0)}-\Delta\left( C^\Ecal_\ell-C^\Bcal_\ell \right)^{(22,0)}\right]\nonumber\\
&=&-8\, C^\Bcal_{\ell}(z_s)\,
\int \frac{d^2\ell_1}{(2\pi)^2}\int \frac{d^2\ell_2}{(2\pi)^2}
\left[\bn\cdot\left( \bell_2\land\bell_1 \right)\left(\bell_1\cdot\bell_2\right)\right]^2\int_{0}^{r_s}dr\,\frac{r_s-r}{r_s\,r}\int_0^{r}dr_1\,\frac{r-r_1}{r\,r_1}
\nonumber\\
&&\times
\int_{0}^{r_s}dr_2\,\frac{r_s-r_2}{r_s\,r_2}\int_{0}^{r_2}dr_3\,\frac{r_2-r_3}{r_2\,r_3}
\left[ 
C^W_{\ell_1}(z,z_2)C^W_{\ell_2}(z_1,z_3)-C^W_{\ell_1}(z,z_3)C^W_{\ell_2}(z_1,z_2)\right]\,,
\eea
\bea
\Delta C^{\Bcal (2,2)}_\ell&\equiv&\frac{1}{2}\left[ \Delta\left( C^\Ecal_\ell+C^\Bcal_\ell \right)^{(2,2)}-\Delta\left( C^\Ecal_\ell-C^\Bcal_\ell \right)^{(2,2)}\right]\nonumber\\
&=&16\,\int \frac{d^2\ell_1}{(2\pi)^2}\,\int \frac{d^2\ell_2}{(2\pi)^2}
\left[\boldsymbol{n}\cdot\left( \boldsymbol{\ell}_2\land\boldsymbol{\ell}_1 \right)\left(\boldsymbol{\ell}_1\cdot\boldsymbol{\ell}_2\right)\right]^2\,\int_{0}^{r_s}dr\,\frac{r_s-r}{r_s\,r}\int_0^{r}dr_1\,\frac{r-r_1}{r\,r_1}
\nonumber\\
&&\times
\int_{0}^{r_s}dr_2\,\frac{r_s-r_2}{r_s\,r_2}\int_{0}^{r_2}dr_3\,\frac{r_2-r_3}{r_2\,r_3}
\left[ C^W_{\ell_1}(z,z_2)C^W_{\ell_2}(z_1,z_3)
-C^W_{\ell_1}(z,z_3)C^W_{\ell_2}(z_1,z_2)\right]\nonumber\\
&&\times \left\{ C^\Ecal_{|\bell-\bell_1-\bell_2|}(z_s)\cos^2\left[2 \left( \phi_\ell - \phi_{|\bell-\bell_1-\bell_2|} \right)\right]\right.
\left.+C^\Bcal_{|\bell-\bell_1-\bell_2|}(z_s)\sin^2\left[2 \left( \phi_\ell - \phi_{|\bell-\bell_1-\bell_2|} \right)\right] \right\}\,.
\label{B22}
\eea

In a final step we apply the Limber approximation to our integrals. We note that we always encounter the same time integrals, therefore we can evaluate this approximation once and then apply it to all our terms. 
Within the Limber approximation, the $C_\ell$'s for the Weyl potential become

\bea
&&\hspace{-2cm}C^W_{\ell_1}(z,z_2)C^W_{\ell_2}(z_1,z_3) - C^W_{\ell_1}(z,z_3)C^W_{\ell_2}(z_1,z_2)
=\frac{\delta(r_2-r)\delta(r_3-r_1)
-\delta(r_3-r)\delta(r_2-r_1)}{16\,r^2\,r_1^2}
\nonumber\\
&&\times
P_R\left(\frac{\ell_1+1/2}{r}\right)\left[T_{\Phi+\Psi}\left(\frac{\ell_1+1/2}{r},z\right)\right]^2
P_R\left(\frac{\ell_2+1/2}{r_1}\right)\left[T_{\Phi+\Psi}\left(\frac{\ell_2+1/2}{r_1},z_1\right)\right]^2,
\label{limber_weyl}
\eea
so that
\bea
&&\hspace{-2cm}\int_{0}^{r_s}dr\,\frac{r_s-r}{r_s\,r}\int_0^{r}dr_1\,\frac{r-r_1}{r\,r_1}
\int_{0}^{r_s}dr_2\,\frac{r_s-r_2}{r_s\,r_2}\int_{0}^{r_2}dr_3\,\frac{r_2-r_3}{r_2\,r_3}
\left[ C^W_{\ell_1}(z,z_2)C^W_{\ell_2}(z_1,z_3)
-C^W_{\ell_1}(z,z_3)C^W_{\ell_2}(z_1,z_2)\right]\nonumber\\
&=&\frac{1}{16}\int_{0}^{r_s}\frac{dr}{r^2}\int_0^{r}\frac{dr_1}{r_1^2}\,\left(\frac{r-r_1}{r\,r_1}\right)^2\,
\left(\frac{r_s-r}{r_s\,r}\right)^2 P_R\left(\frac{\ell_1+1/2}{r}\right)
 \nonumber\\
&&\times 
P_R\left(\frac{\ell_2+1/2}{r_1}\right)
\left[T_{\Phi+\Psi}\left(\frac{\ell_1+1/2}{r},z\right)\right]^2
\left[T_{\Phi+\Psi}\left(\frac{\ell_2+1/2}{r_1},z_1\right)\right]^2 \,.
\label{limber_weyl2}
\eea
This simplification applies to all the contributions  evaluated above.
\end{widetext}
\section{Numerical Results}
\label{Sec7}
\setcounter{equation}{0}

In this section we present the numerical evaluation of the results given above. 
For the numerical results we consider non-linear (Halofit model~\cite{Smith:2002dz,Takahashi:2012em}) power spectra for the gravitational potential. All the figures have been generated with the following cosmological parameters $h = 0.67$, $\omega_\text{cdm} = 0.12$, $\omega_\text{b} = 0.022$ and vanishing curvature. The primordial curvature power spectrum has the amplitude $A_s = 2.215 \times10^{-9}$, the pivot scale $k_\text{pivot} = 0.05 \ \text{Mpc}^{-1}$, the spectral index $n_s = 0.96$ and no running. The transfer function for the Bardeen potentials, $T_{\Phi+\Psi}$ has been computed with \class{}~\cite{Blas:2011rf}, using Halofit~\cite{Takahashi:2012em}.
In analysing the contribution of $R_{\beta^{(2)}}$ (see below) we compare the non-linear and the linear results. The latter has been obtain with the same cosmological parameters with the linear power spectrum computed with \class{}~\cite{Blas:2011rf}.

First of all, let us note that all the contributions $\Delta C^{X(22,0)}_\ell$ from the rotation of polarization contain the same constant factor multiplying simply the unperturbed spectrum. Let us call it $\Rcal_{\beta^{(2)}}$, so we have that
\bea
\frac{\Delta C^{\Ecal(22,0)}_\ell}{C^\Ecal_\ell}=\frac{\Delta C^{\Bcal(22,0)}_\ell}{C^\Bcal_\ell}
=\frac{\Delta C^{\Ecal\Mcal(22,0)}_\ell}{C^{\Ecal\Mcal}_\ell}=\Rcal_{\beta^{(2)}} \,
\eea
with
\begin{widetext}
\bea
\Rcal_{\beta^{(2)}} &=&-\frac{1}{16}
\int \frac{d \ell_1}{2\pi}\int \frac{d \ell_2}{2\pi}
\left(\ell_1 \ell_2\right)^5
\int_{0}^{r_s}\frac{dr}{ r^2}\int_0^{r}\frac{dr_1}{r_1^2}
\,\left(\frac{r-r_1}{r\,r_1}\right)^2
\,\left(\frac{r_s-r}{r_s\,r}\right)^2P_R\left(\frac{\ell_1+1/2}{r}\right)\nonumber\\
&&\times 
P_R\left(\frac{\ell_2+1/2}{r_1}\right)
\left[T_{\Phi+\Psi}\left(\frac{\ell_1+1/2}{r},z\right)\right]^2
\left[T_{\Phi+\Psi}\left(\frac{\ell_2+1/2}{r_1},z_1\right)\right]^2 \,,
\eea
\end{widetext}
where we have performed the angular integration.
From Eq.~\eqref{beta_square} one infers that $\Rcal_{\beta^{(2)}}$ is proportional to the variance of the rotation angle,
\be
\langle(\beta^{(2)})^2\rangle = -\Rcal_{\beta^{(2)}}/2 \,.
\ee
Using the linear power spectrum~\cite{Blas:2011rf} we obtain $\Rcal^\text{lin}_{\beta^{(2)}}=-7.8\times 10^{-6}$, whereas using Halofit~\cite{Takahashi:2012em} for the matter power spectrum the term becomes  more than one order of magnitude larger, with $\Rcal^\text{Halofit}_{\beta^{(2)}}=-2.5\times 10^{-4}$.
This corresponds to rotation angles of $\sqrt{\langle(\beta^{(2)})^2\rangle}  =6.8'$ and $=38'$ respectively. This is a large effect which cannot be neglected, even though the Halofit approximation may over estimate it (see below). The rotation $\beta^{(2)}$ is due to successive shearing processes along the ray~\cite{Perlick:2004tq}. Parametrically it is of second order in the shear (or the convergence) but since these quantities are second derivatives of the potential they are parametrically of the same order as density fluctuations and can become large, especially on small scales.

The universality of the above coupling and its independence on $\ell$ are due to the fact that, in the related correlators in Eqs.~\eqref{rotationDelta}, no derivatives of $\Pcal$ appear and the two point correlation function of $\beta^{(2)}$ is evaluated at the same direction. On the other hand, Eqs.~\eqref{E22} and \eqref{B22} still have no angular derivatives of $\Pcal$, but they involve the two point correlation function of $\beta^{(2)}$ in two different directions  leading  to a dependence on $\ell$ of the corresponding terms.

\begin{widetext}

\begin{figure}[htb]
\centering
\includegraphics[width=8cm]{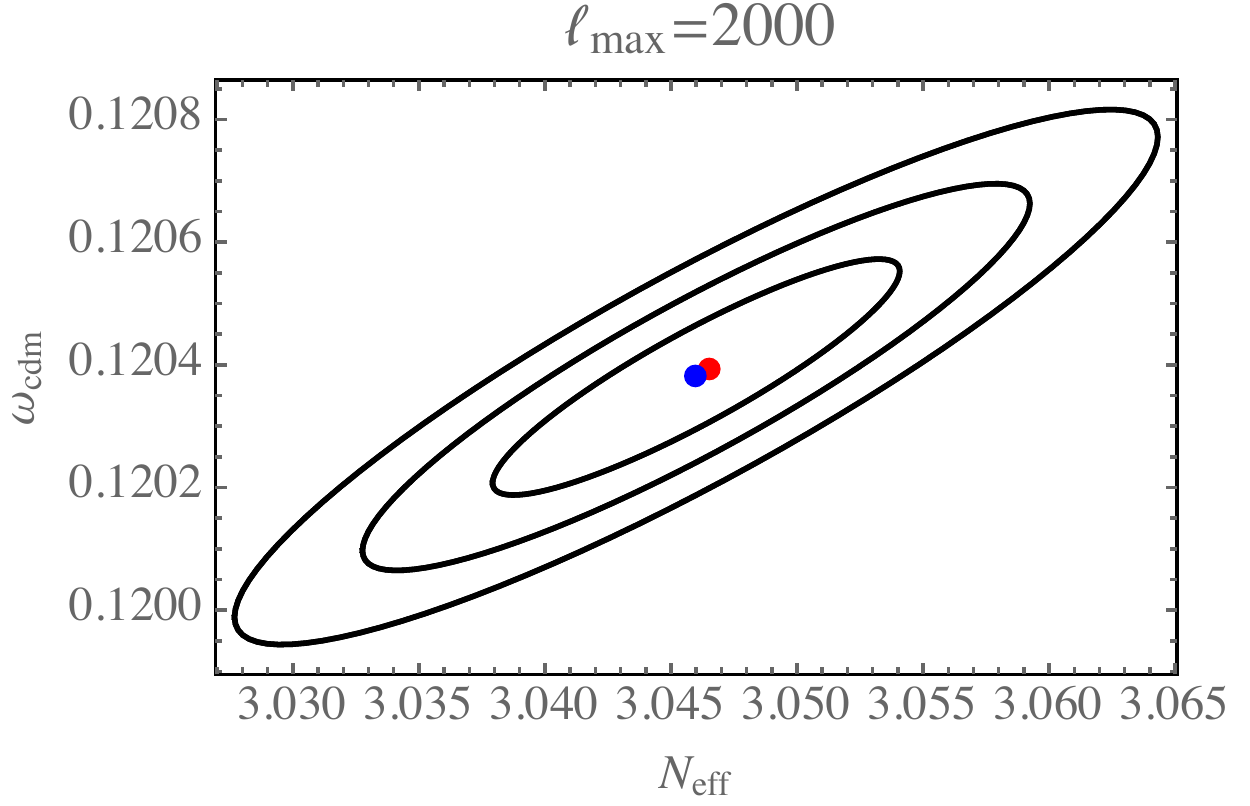}~~~~~~~~~
\includegraphics[width=8cm]{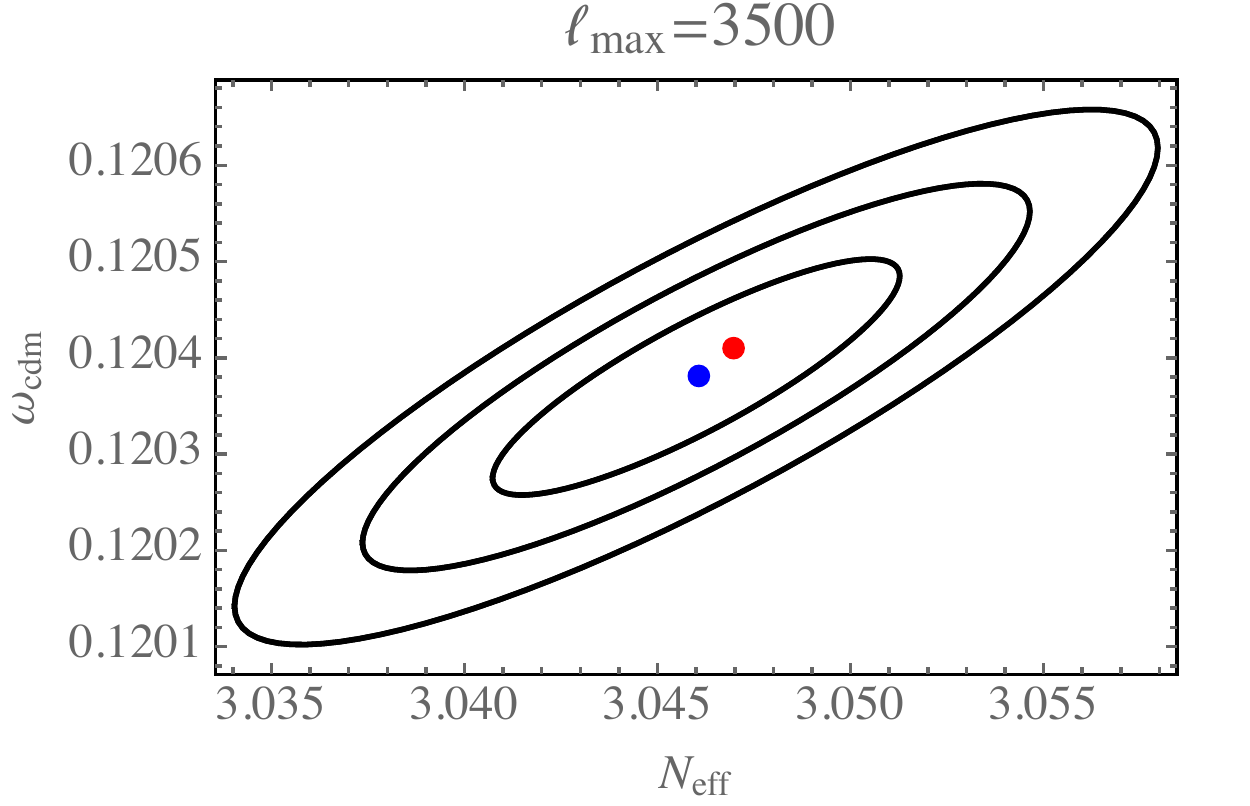}
\centering
\caption{Fisher forecast (see Appendix~\ref{app:Fisher} for details) for a cosmic variance limited survey. 
The blue (red) points show the shift in the best fit parameter for the dark matter density $\om_\text{cdm}=h^2\Om_\text{cdm}$ and the effective number of relativistic species $N_\text{eff}$ induced by the terms in Eqs.~(\ref{DeltaEM220}) and (\ref{DeltaE220}) (we consider vanishing primordial B-modes) using the linear power spectrum (using Halofit). The unshifted best fit value is covered by the blue point. The ellipses denote 1, 2 and 3 sigma contours. The parameters not shown in the panels are fixed to the fiducial cosmology. For both panels we consider B-mode up to $\ell_\text{max}= 1500$ to be consistent with the conservative specifications of CMB-S4~\cite{Abazajian:2016yjj}.}
\label{fig:cosmologicalParameters}
\end{figure}

\end{widetext}

The integrals over $\bell_1$ and $\bell_2$ in $\Rcal_{\beta^{(2)}}$ converge very slowly and are highly UV sensitive.  In particular, a cutoff independent evaluations involves integration domains in $\vl{}$ space where perturbation theory is no longer valid, therefore, also numerical results using Halofit are not reliable. Nevertheless, these corrections just leads to an overall shift
of $\Delta C_\ell/C_\ell$'s and this contribution is negligible in cosmological parameter estimation (see, for instance, Fig.~\ref{fig:cosmologicalParameters}).  For this reason, we do not consider these terms in what follows.

\begin{widetext}

\begin{figure}[ht!]
\centering
\includegraphics[width=8cm]{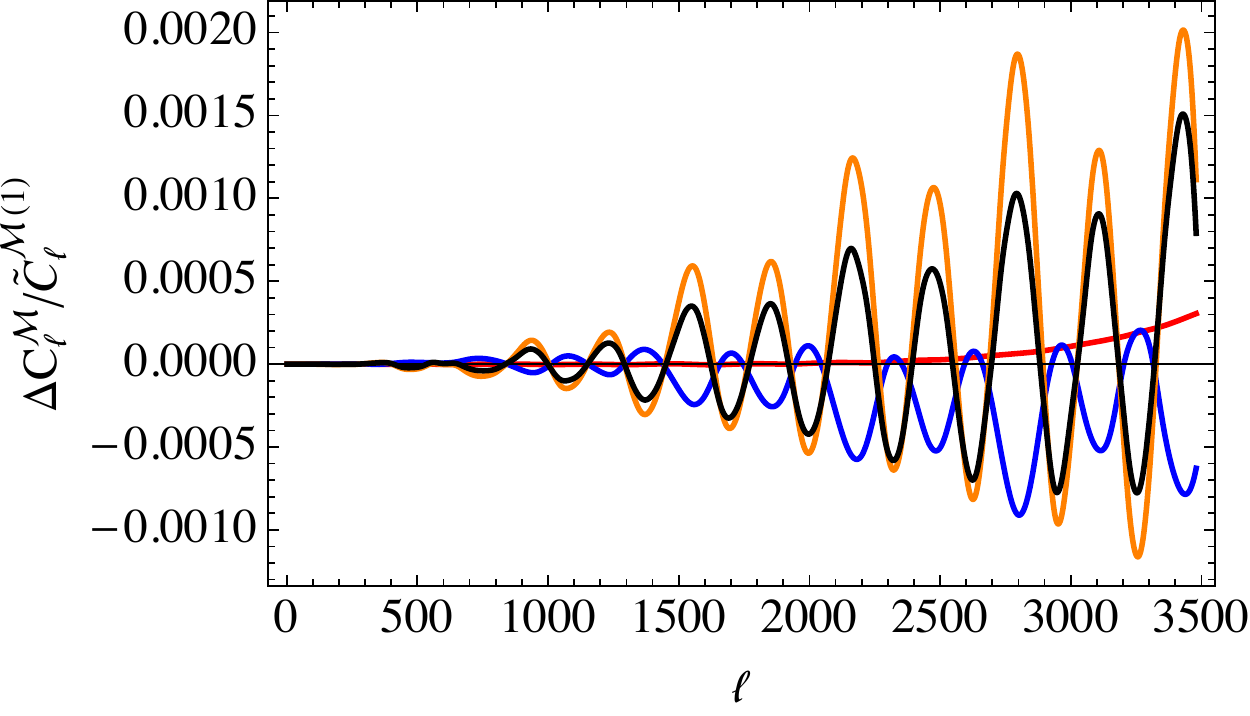}~~~~~~~~~
\includegraphics[width=8cm]{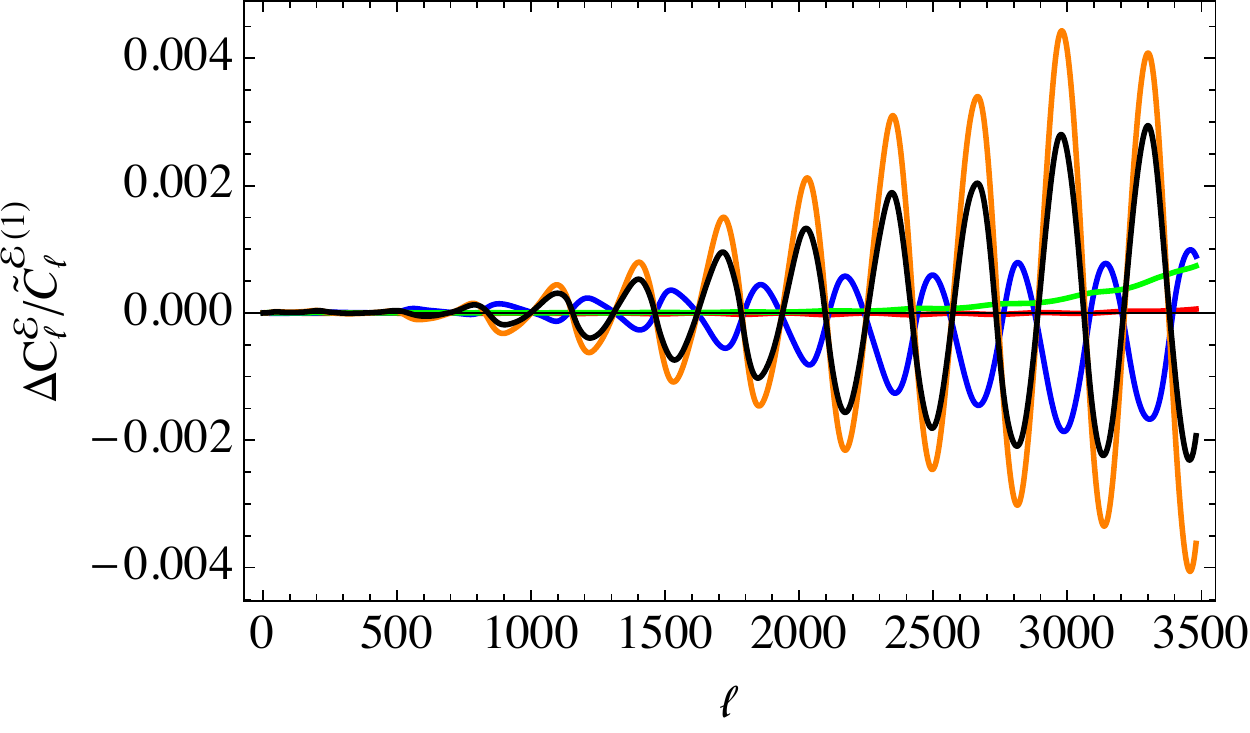}\\
\includegraphics[width=8cm]{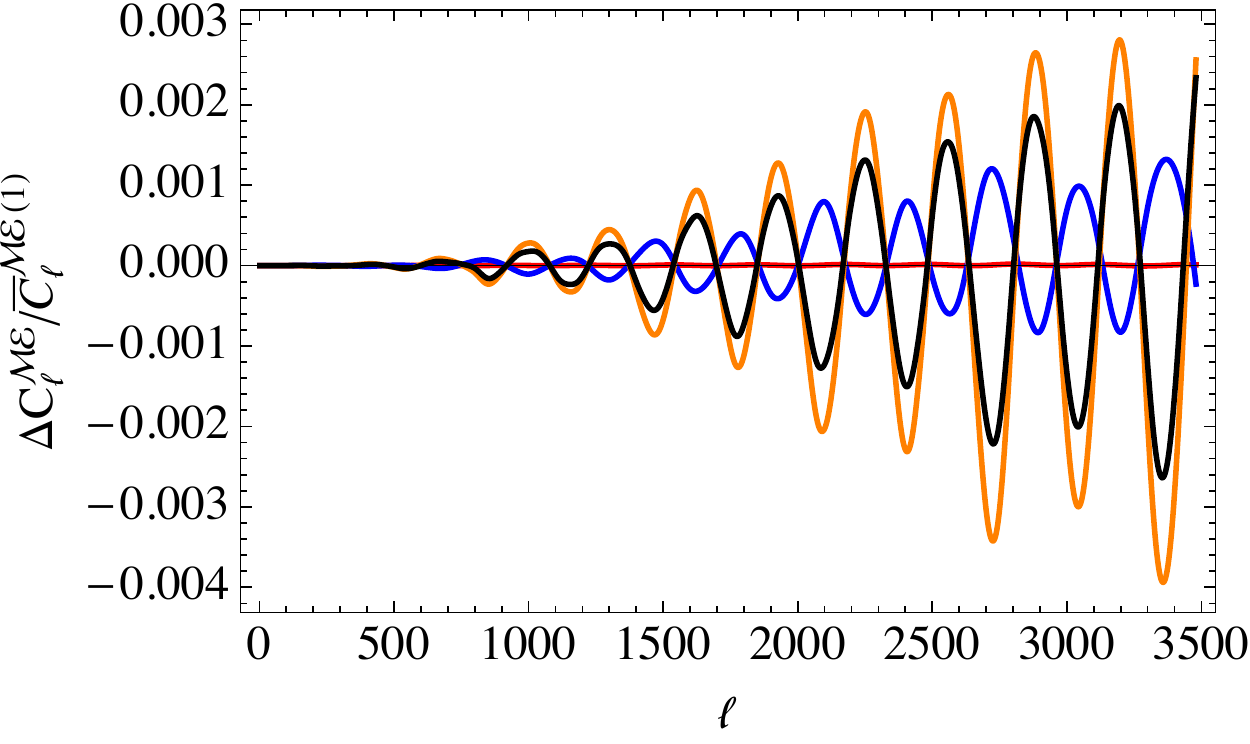}~~~~~~~~~
\includegraphics[width=8cm]{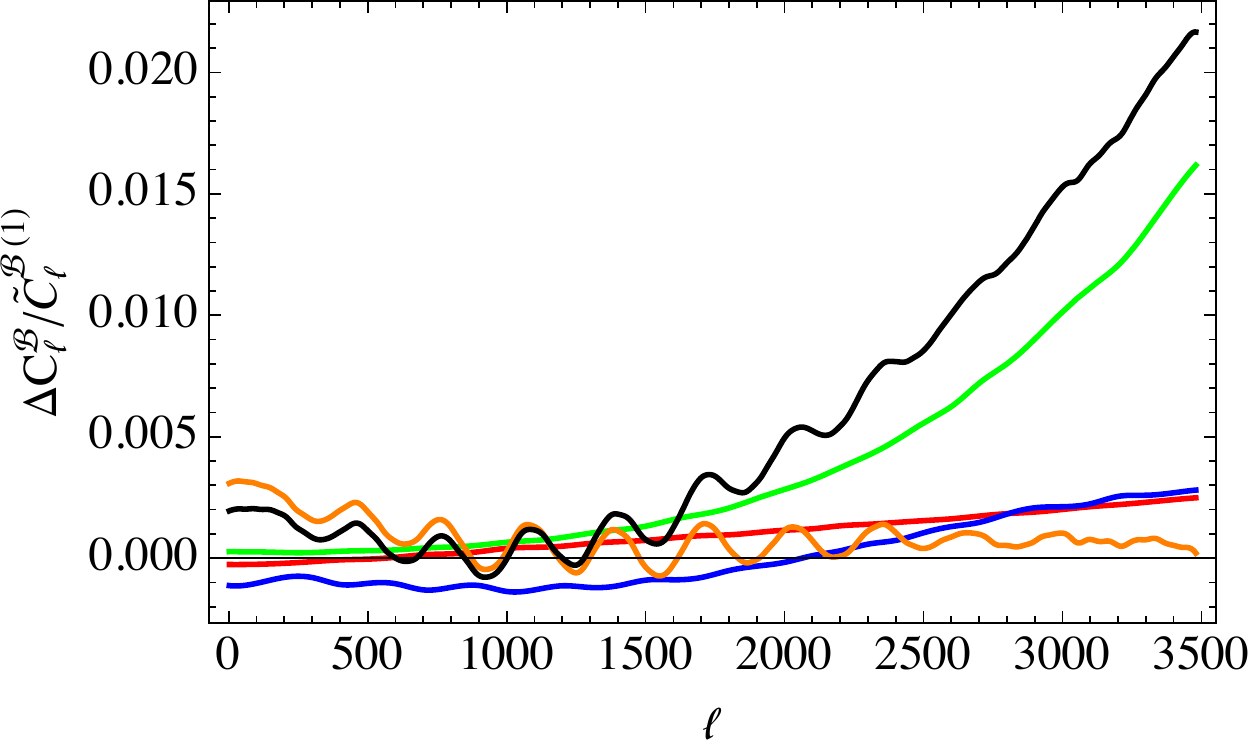}
\caption{Higher order lensing contributions from the post-Born second group (red curves), post-Born third group (blue curves), LSS third group
(orange curves), and rotation angle $\beta^{(2)}$ (green curves,  contributions $(2,2)$). Black curves sum up the total correction. 
We consider  the lensing CMB spectra for temperature (top left-panel), E-modes (top right-panel), cross TE spectra (bottom left-panel), where $\bar{C}_\ell^{\Mcal\Ecal(1)}=\sqrt{\frac{\left(\tilde{C}_\ell^{\Mcal\Ecal (1)}\right)^2+\tilde{C}^{\Mcal (1)}_\ell\,\tilde{C}^{\Ecal (1)}_\ell}{2}}$) and B-modes (bottom right-panel). }
\label{fig:corrections}
\end{figure}

\end{widetext}

In Fig.~\ref{fig:corrections} we compare the different higher order contributions. The 
non-Gaussian (third group) contributions from the post-Born and LSS corrections are relavant for all spectra. They dominate  the temperature (for $\ell<3000$), E-mode and temperature--E-mode cross correlation spectra, whereas they are of the same order of magnitude as the post-Born second group corrections for the B-modes. This post-Born second group is also non-negligible in the temperature spectrum on very small scales ($\ell>3000$).
Moreover, the corrections due to rotation are very important for B-modes in a large range of scales 
(dominant for $\ell>1500$) and give non negligible corrections to E-modes for $\ell>2500$.

\newpage

In Fig.~\ref{fig:CosmicVariance} we present the ratio between these corrections and cosmic variance, $c^X_\ell$, 
$(\sigma^X_\ell)^2$ given by
\bea
\label{sigmaM}
\sigma^\Mcal_\ell&=&\sqrt{\frac{2}{2\ell+1}}\,C_\ell^\Mcal  \,,
\\
\label{sigmaE}
\sigma^\Ecal_\ell&=&\sqrt{\frac{2}{2\ell+1}}\,C_\ell^\Ecal \,,
\\
\label{sigmaME}
\sigma^{\Mcal\Ecal}_\ell&=&\sqrt{\frac{1}{2\ell+1}}\,\sqrt{\left(C_\ell^{\Mcal\Ecal}\right)^2+C^\Mcal_\ell\,C^\Ecal_\ell}  \,,\\
\label{sigmaB}
\sigma^\Bcal_\ell&=&\sqrt{\frac{2}{2\ell+1}}\,\tilde{C}_\ell^{\Bcal(1)}  \,.
\eea
Note that, for B-modes, we have taken into account the first order resummed correction since we consider no primordial gravitational wave, i.e the unlensed spectrum vanishes. 
Therefore, lensed B-modes do not have Gaussian statistics. For this reason its cosmic variance can be significantly  larger than the one from Eq.~\eqref{sigmaB} \cite{Smith:2006nk}.
Considering Gaussian variance also for B-modes, the corrections due to rotation alone are comparable to cosmic variance for $\ell\gtrsim 3500$, in contrast to all other spectra where all the corrections are always below that threshold. 
Moreover, the sum of all the effects can be even larger than cosmic variance at these multipoles, showing that higher-order lensing corrections to B-mode polarization at high multipoles have the best chance to be detectable.

Finally,
in Fig.~\ref{fig:SN} we show the cumulative signal-to-noise ratio defined as
\be
\left( \frac{S}{N} \right)^2 = \sum_{\ell=30}^{\ell_{\rm max}} \left( \frac{\Delta C_\ell}{\sigma_\ell}\right)^2
\ee
where $\sigma_\ell$ are defined like in eqs.~(\ref{sigmaM}-\ref{sigmaB}) but adding a noise contribution to the cosmic variance term, i.e. by replacing $C^X_\ell$ by $C^X_\ell +N^X_\ell$ where
\be
N_\ell= (\Delta X)^2 \exp\left( \frac{\ell \left( \ell+1 \right) \theta^2_{\rm FWHM}}{8 \ln 2} \right)
\ee
and $\Delta X= 1 \ \mu K\times\text{arcmin}$ for temperature, $\Delta X= \sqrt{2} \ \mu K\times\text{arcmin}$ for polarization and an angular resolution of $\theta_{\rm FWHM}=1$ arcmin.
Our results are comparable with Ref.~\cite{Fabbian:2017wfp}. We predict a lower signal-to-noise ratio for the contribution to temperature anisotropies because we limit our analysis to $\ell_{\rm max}=3500$, while they have a smaller  contribution  for E-mode which seems due to non-perturbative effects we do not consider in our approach.


\begin{widetext}

\begin{figure}[ht!]
\centering
\includegraphics[width=8cm]{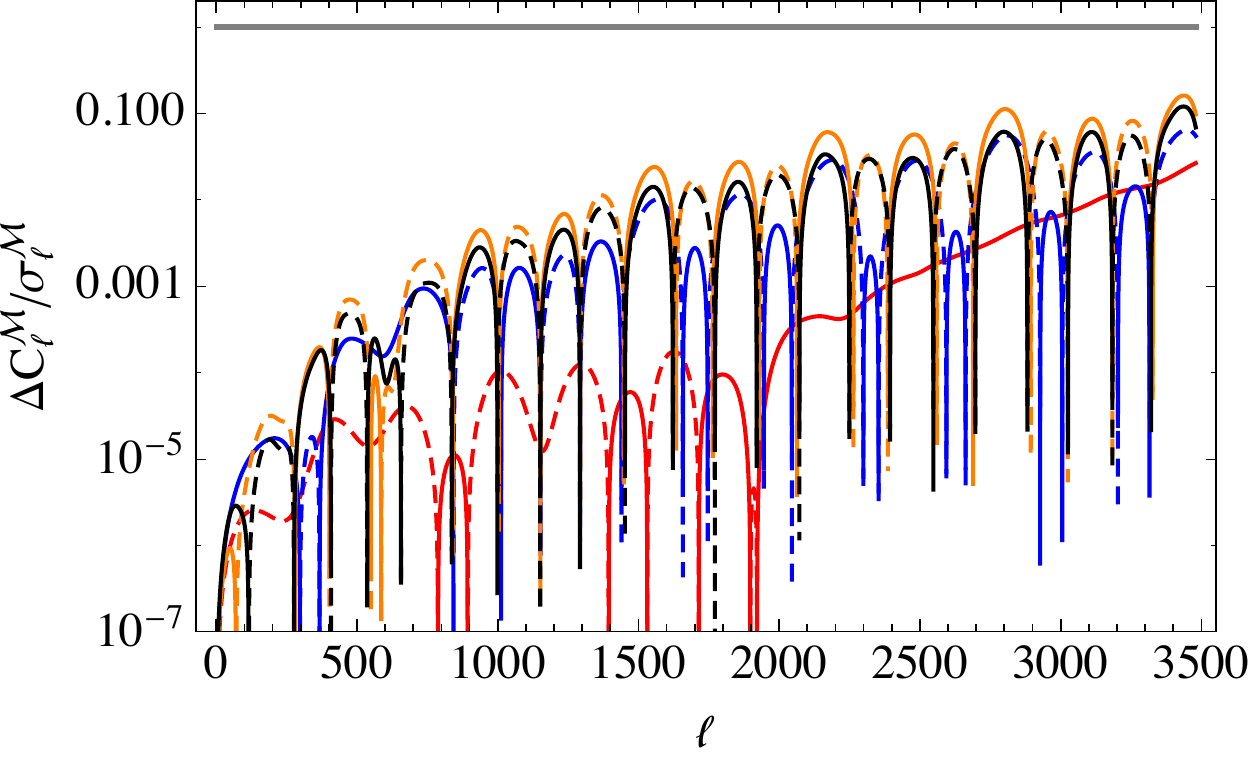}~~~~~~~~~
\includegraphics[width=8cm]{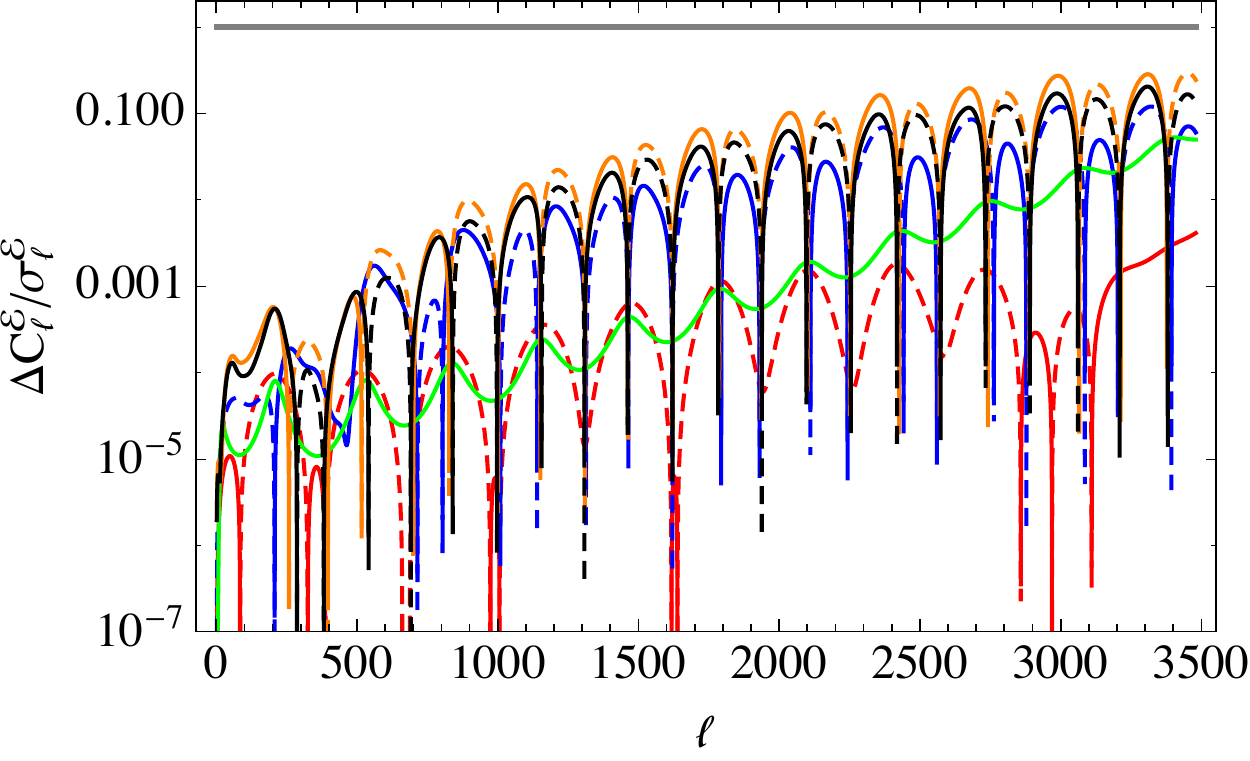}\\ \vspace{3mm}  
\includegraphics[width=8cm]{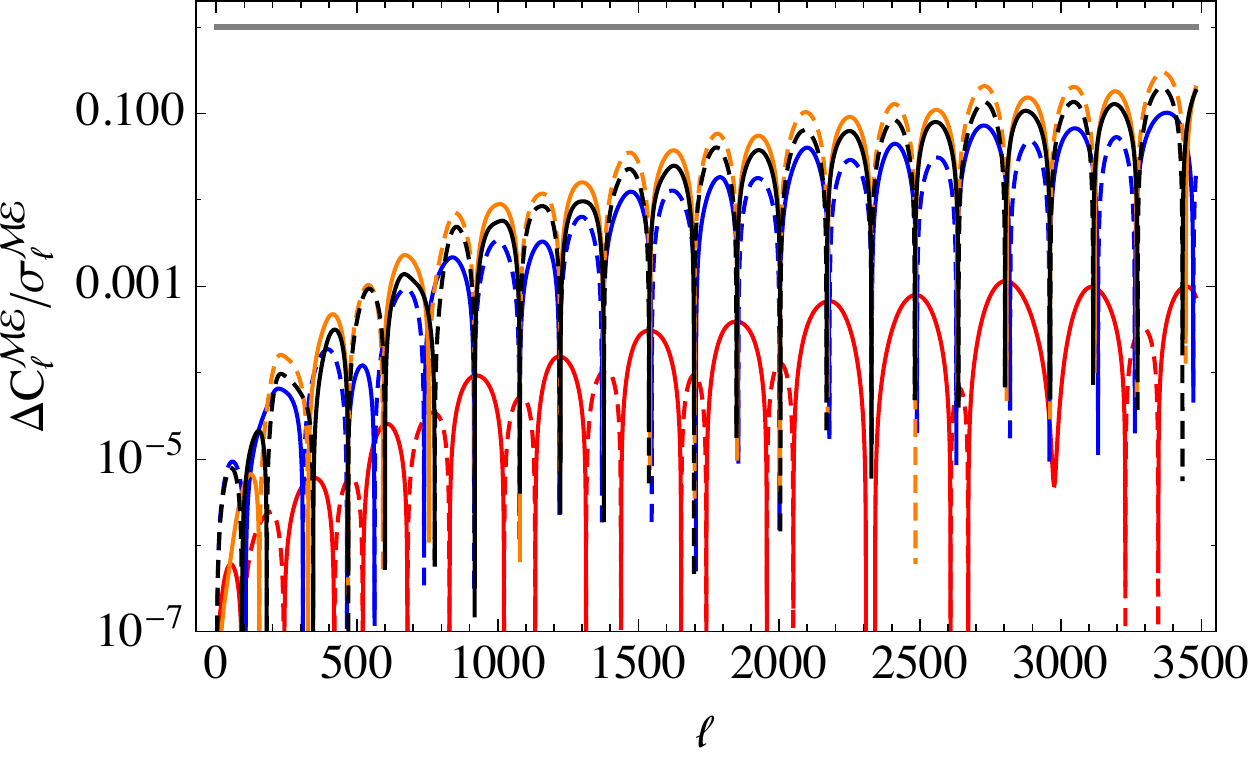}~~~~~~~~~
\includegraphics[width=8cm]{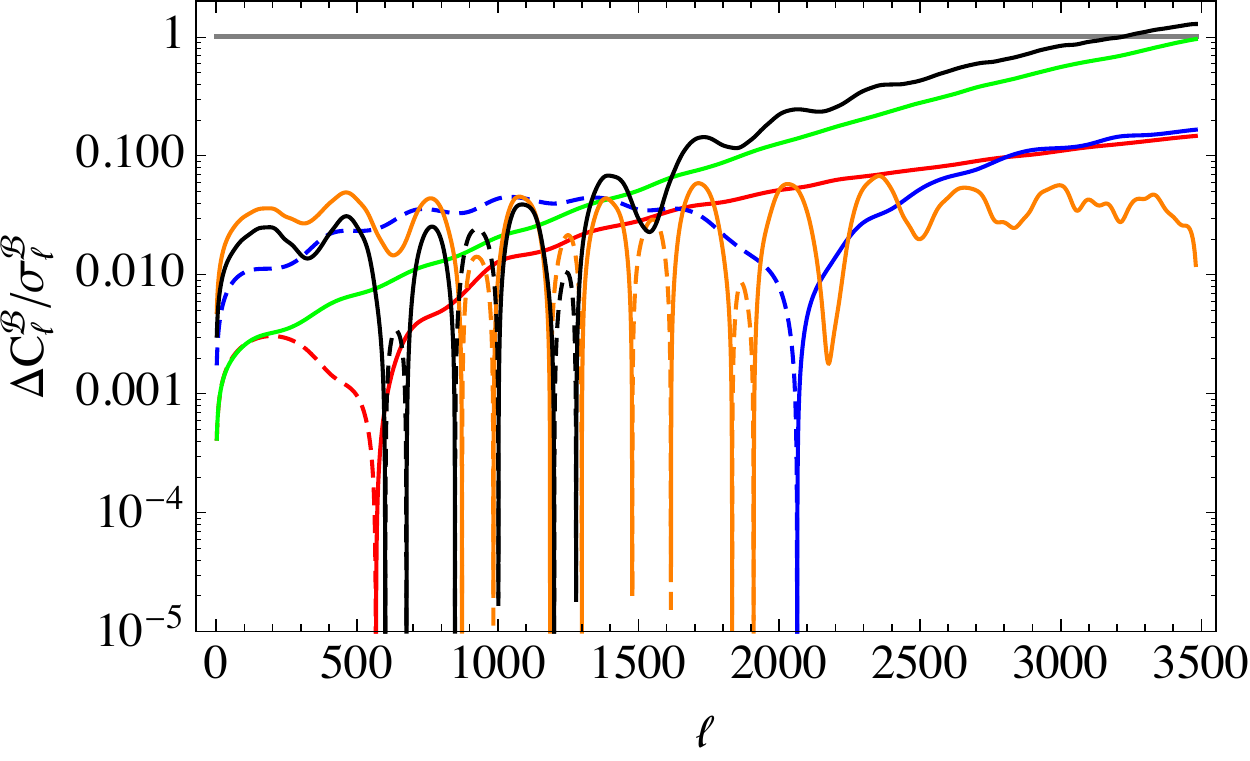}
\caption{Comparison between next-to-leading order corrections and cosmic variance for the temperature (Eq.~\eqref{sigmaM}, top left-panel), E-modes (Eq.~\eqref{sigmaE}, top right-panel), TE cross correlation (Eq.~\eqref{sigmaME}, botto left-panel) and B-modes (Eq.~\eqref{sigmaB}, bottom right-panel). 
Red curves refer to post-Born second group, blue curves to post-Born third group, orange to LSS corrections third group and green curves represent the  $(2,2)$ term of $\beta^{(2)}$. Dashed lines are negative values and the black lines trace the sum of all the terms. 
}
\label{fig:CosmicVariance}
\end{figure}
\begin{figure}[ht!]
\centering
\includegraphics[width=8cm]{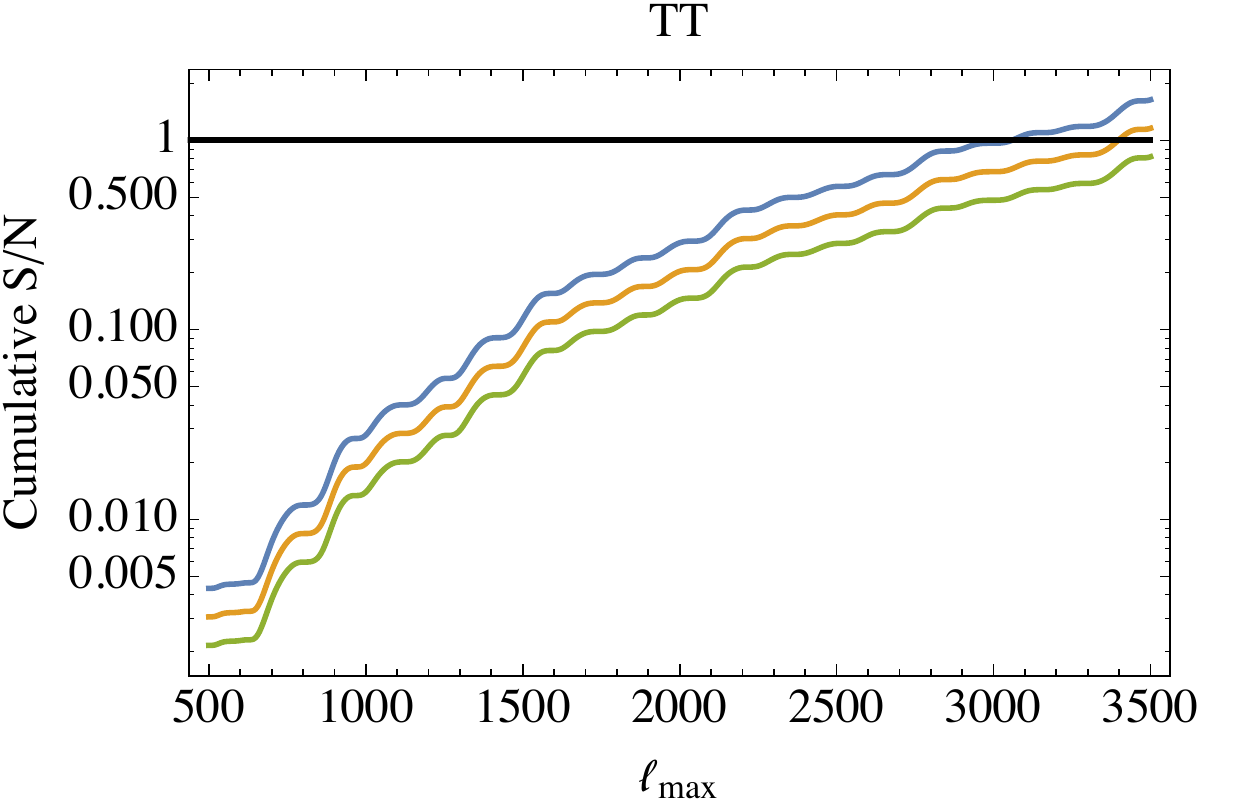}~~~~~~~~~
\includegraphics[width=8cm]{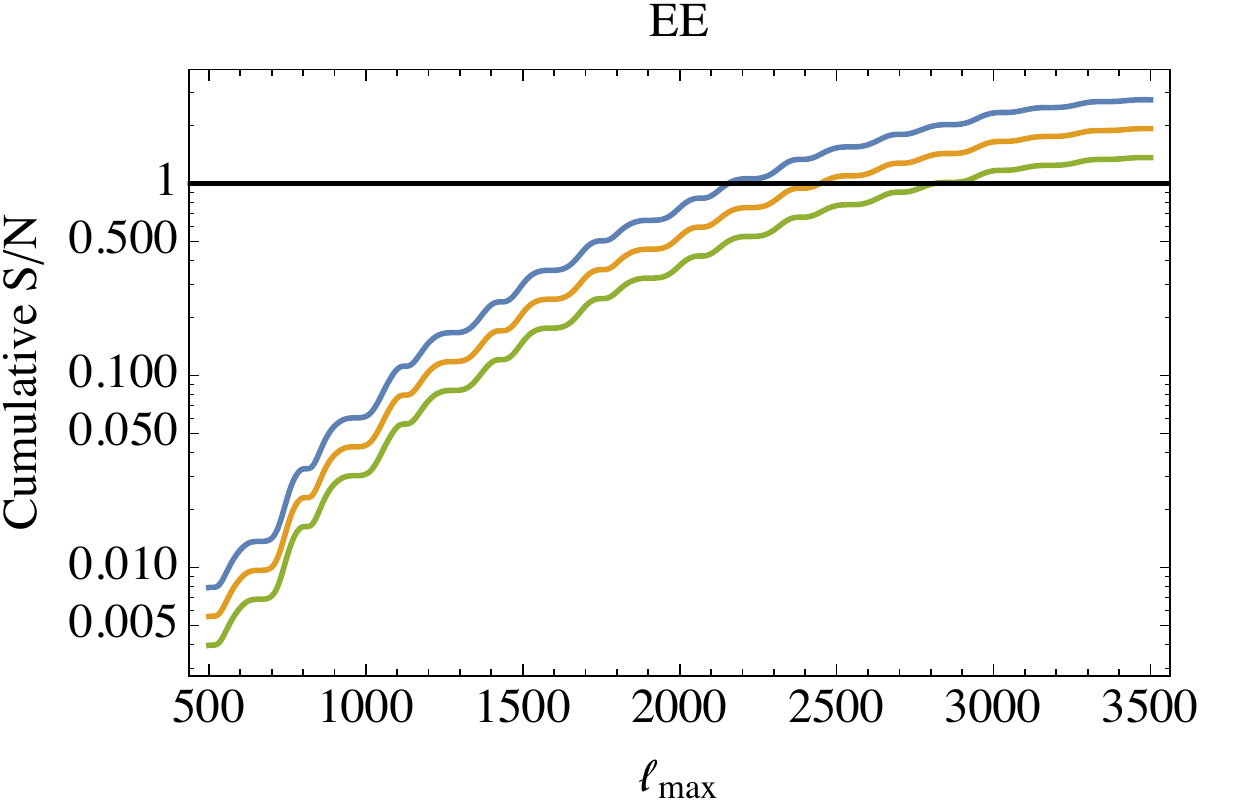}\\ \vspace{3mm}  
\includegraphics[width=8cm]{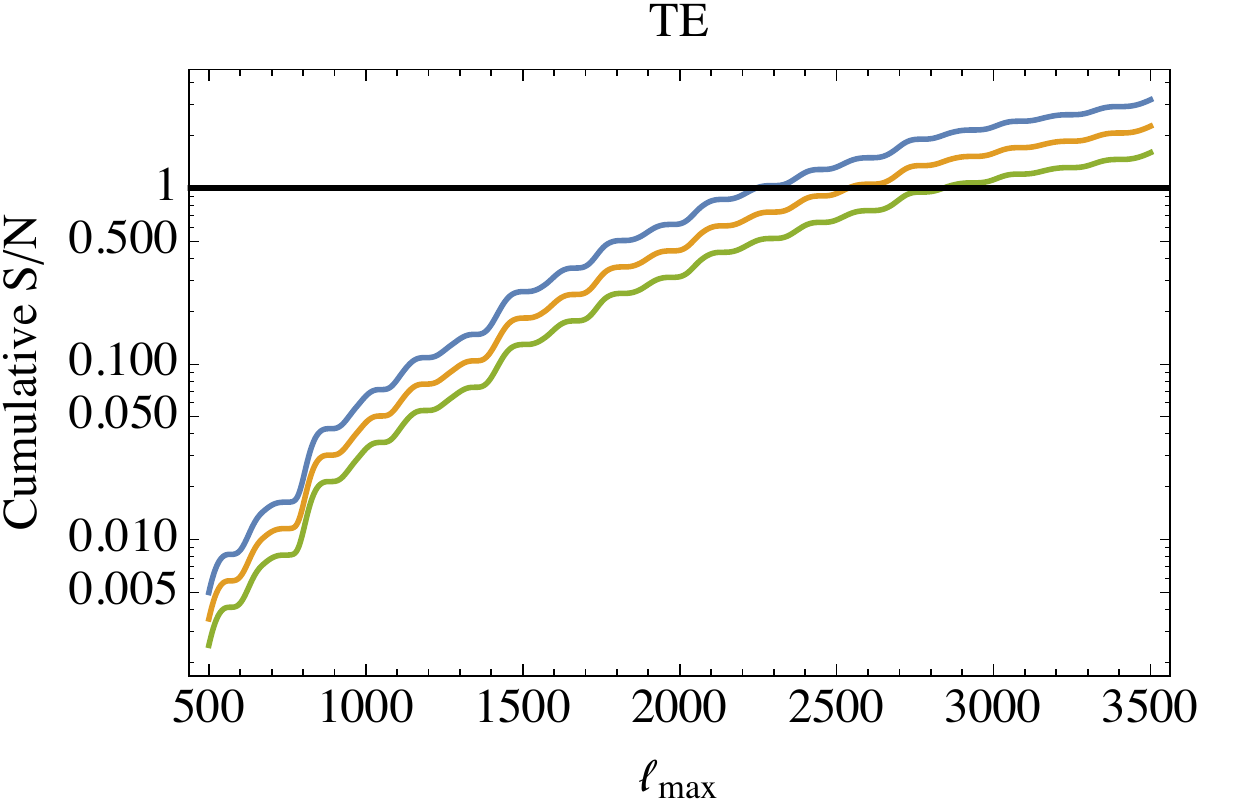}~~~~~~~~~
\includegraphics[width=8cm]{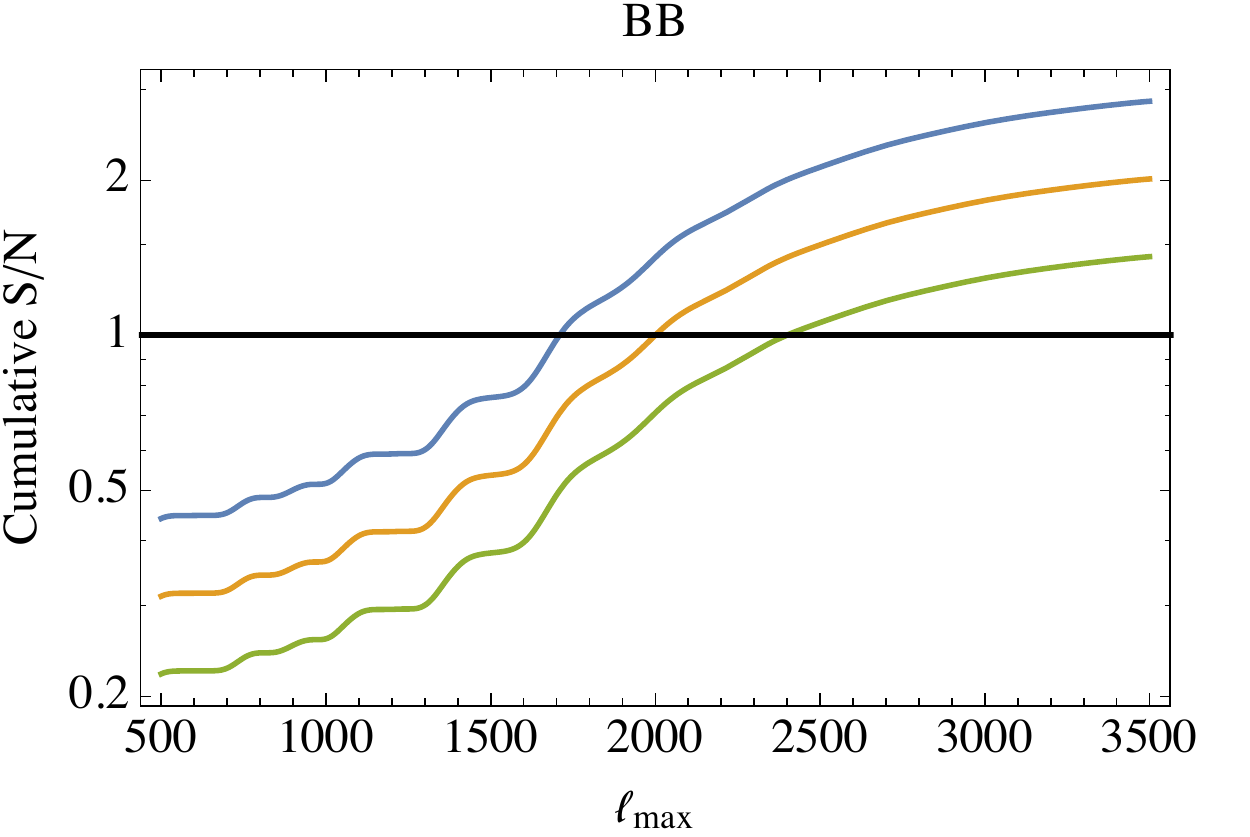}
\caption{The signal-to-noise estimates of the total next to leading order effects for different sky coverage ($f_\text{sky}=0.25$, green curves, $f_\text{sky}=0.5$, orange curves and $f_\text{sky}=1$, blue curves) are shown as functions of 
$\mathcal{\ell}_\text{max}$.
We consider the specifications of CMB S4~\cite{Abazajian:2016yjj}: $1 \ \mu K\times\text{arcmin}$ noise for temperature and $\sqrt{2} \mu K \times\text{arcmin}$ for polarization with an angular resolution of $1\  \text{arcmin}$.
}
\label{fig:SN}
\end{figure}
\end{widetext}

\section{Conclusions}
\label{Sec8}
\setcounter{equation}{0}

In this paper we have computed all the next-to-leading order corrections to the CMB power spectra of temperature  and polarization anisotropies from gravitational lensing of the photons along their path from the last scattering surface into our telescopes. We have found that most terms apart from those already taken into account in present codes~\cite{Lewis:2006fu,Blas:2011rf,Lesgourgues:2011re} are smaller than cosmic variance for a single $\ell$ mode. The only exception to  this rule are the B-mode corrections at very high $\ell$. This can be understood from the fact that cosmic variance is proportional to the amplitude of the signal which is by far smallest for the B-modes. Nevertheless by considering the lensed B-modes as Gaussian, we may underestimate their variance~\cite{Smith:2006nk}.

Several of the terms calculated in this paper have already been determined before~\cite{Pratten:2016dsm,Marozzi:2016uob,Lewis:2016tuj} and our results are in good qualitative agreement, where comparable,  with previous findings. This is a non-trivial consistency check, especially for \cite{Pratten:2016dsm,Lewis:2016tuj} which use quite different methods. Apart from rotation, the only other difference between our results and \cite{Lewis:2016tuj} comes from  the second group which has been neglected in  \cite{Lewis:2016tuj} . This leads to quite relevant differences for temperature at small scales ($\ell>3000$) and for the B-modes spectrum on all scales, whereas it does not change EE and TE spectra. The largest correction to the B-modes comes, however, from the rotation of the polarization direction which is new. It is very remarkable that our analytical results, including rotation, have been confirmed recently by N-body simulations with multiple-lens raytracing technique~\cite{Fabbian:2017wfp,Takahashi:2017hjr}. Considering the different procedures, the level of agreement between the results is impressive.

It will be interesting to investigate whether these corrections are observable. Even though for an individual value $\ell$, the corrections are below cosmic variance, this is no longer so for sufficiently large bins of $\ell$'s,  as we have shown in Fig. \ref{fig:SN}.  Let us only note here, that the rotation of the polarisation is due to the vector-degree of freedom of the gravitational field, an effect like frame dragging. Its detection would therefore  
represent a highly non trivial test of general relativity, testing its elusive spin-1 sector.
Recently, it has been proposed to measure this rotation with radio cosmic shear surveys~\cite{Thomas:2016xhb}.

But also the other terms are not negligible if a precision of 0.1\% wants to be achieved as announced in Ref.~\cite{Lesgourgues:2013bra}. For example, for $\ell$ between 2000 and 2100, cosmic variance amounts to about 2.2\%. Hence, as one easily infers from Figs.~\ref{fig:corrections} and \ref{fig:CosmicVariance}, our corrections with respect to the unlensed spectra are up to  0.1\% for the E-polarization spectrum and for the T-E cross correlation, while they are at most 0.04\% for the temperature anisotropy. For the B-polarization spectrum the correction is close to 0.5\%.

It is clear that a systematic change even below cosmic variance can affect cosmological parameters and it has to be studied whether next-to-leading order corrections from lensing can indeed influence CMB parameter estimation in the future, this is the topic of an accompanying letter~\cite{Marozzi:2016und}. While, it is unlikely that the tiny corrections of the temperature will be relevant alone, parameters depending strongly on polarization can be affected.
 Indeed, in \cite{Marozzi:2016und} we show how neglecting higher order lensing terms can lead to misinterpreting these corrections as a primordial tensor-to-scalar ratio of about $\mathcal{O}(10^{-3})$, and leads to 
a non-negligible shift of the estimated value of the effective number of relativistic species.

The fact that $\omega^{(2)}$ can significantly affect the CMB spectra has important  consequences for delensing and lensing reconstruction. Those techniques, indeed, rely on the fact that lensing is mainly sourced by a scalar lensing potential, such that an (almost) exact remapping can be done between the intrinsic CMB maps at the last scattering surface and the lensed ones nowadays. However, if $\omega^{(2)}$ contributes significantly, new estimators for lensing reconstruction have to be developed. This task is highly non-trivial and requires a proper analysis.  We shall postpone this investigation to future work.

However, independent of parameter estimation, detecting higher order corrections from CMB lensing would be extremely interesting and allow not only a handle on non-linear corrections to the gravitational potential, but also new tests of General Relativity on cosmological scales.

\section*{Acknowledgements}
We are grateful to many colleagues  for helpful discussions especially on the problem of rotation. We thank especially Camille Bonvin, Anthony Challinor, Chris Clarkson, Giulio Fabbian, Pierre Fleury, Alex Hall, Antony Lewis,  Roy Maartens and Gabriele Veneziano.
GM wishes to thank CNPq and INFN under the program TAsP (Theoretical Astroparticle Physics) for financial support.  
GF is supported by a Consolidator
Grant of the European Research Council (ERC-2015-CoG
grant 680886). 
ED is supported by the ERC Starting Grant cosmoIGM and by INFN/PD51 INDARK grant.
RD acknowledges support from the Swiss National Science Foundation.

\appendix

\begin{widetext}

\section{$\Dcal^{(i....)}(\vl{})$ terms}
\label{a:Dl}

In $\vl{}$ space, and starting from the result of \cite{Marozzi:2016uob} and of Sect. \ref{Sec5}, we obtain the corresponding expressions to evaluate the lensing corrections to the CMB polarization anisotropies up to forth order:

\bea
\Dcal^{(1)}(\vl{})&=&\frac{1}{2\pi}\int d^2 x\,\T{a}{1}\nabla_a\Pcal\,e^{i\vl{}\cdot{\bf x}}\nonumber\\
&=&-\frac{1}{\pi}\INT{2}\left[\left(\vl{}-\vl{2}\right)\cdot\vl{2}\right]
\int_{0}^{r_s}dr \frac{r_s-r}{r_s\,r}\,
\Phi_W(r,\vl{}-\vl{2})
\left[\Ecal(r_s,\vl{2})+i \Bcal(r_s,\vl{2})\right] e^{-2 i \varphi_{\ell_2}} \,,
\\
& & \nonumber \\
\Dcal^{(2)}(\vl{})&=&\frac{1}{2\pi}\int d^2 x\,\T{a}{2}\nabla_a\Pcal\,e^{i\vl{}\cdot{\bf x}}\nonumber\\
&=&
- \frac{1}{\pi}\INT{2}\left[\left(\vl{}-\vl{2}\right)\cdot\vl{2}\right]
\int_{0}^{r_s}dr \frac{r_s-r}{r_s\,r}\,
\Phi_W^{(2)}(r,\vl{}-\vl{2})
 \left[\Ecal(r_s,\vl{2})+i \Bcal(r_s,\vl{2})\right] e^{-2 i \varphi_{\ell_2}}  
\nonumber \\ & &
+
\frac{1}{\pi^2}\INT{2}\INT{3}
\,\left[\left(\vl{}+\vl{2}-\vl{3}\right)\cdot\vl{3}\right]\,
\left[\left(\vl{}+\vl{2}-\vl{3}\right)\cdot\vl{2}\right]\nonumber\\
&&\times\int_{0}^{r_s}dr\frac{r_s-r}{r_s\,r}\int_{0}^{r}dr' \frac{r-r'}{r\,r'}\,
\Phi_W(r,\vl{}+\vl{2}-\vl{3})
\bar\Phi_W(r',\vl{2})
\left[\Ecal(r_s,\vl{3})+i \Bcal(r_s,\vl{3})\right] e^{-2 i \varphi_{\ell_3}}\,,
\\
\Dcal^{(11)}(\vl{})&=&\frac{1}{2\pi}\int d^2 x\frac{1}{2}\T{a}{1}\T{b}{1}\nabla_a\nabla_b\Pcal e^{i\vl{}\cdot {\bf x}}\nonumber\\
&=&\frac{1}{2}\frac{1}{\pi^2}\INT{2}\INT{3}\,\left[\left(\vl{}+\vl{2}-\vl{3}\right)\cdot \vl{3}\right]\,\left(\vl{2}\cdot \vl{3}\right)
\int_{0}^{r_s}dr \frac{r_s-r}{r_s\,r}\int_{0}^{r_s}dr' \frac{r_s-r'}{r_s\,r'}
 \nonumber\\  && \qquad \times\Phi_W(r,\vl{}+\vl{2}-\vl{3})
\bar\Phi_W(r',\vl{2})\left[\Ecal(r_s,\vl{3})+i \Bcal(r_s,\vl{3})\right] e^{-2 i \varphi_{\ell_3}}\,,
\eea
\bea
& & \nonumber \\
\Dcal^{(3)}(\vl{})
&=&\frac{1}{2\pi}\int d^2 x\,\T{a}{3}\nabla_a\Pcal\,e^{i\vl{}\cdot{\bf x}}\nonumber\\
&=&
- \frac{1}{\pi}\INT{2}\left[\left(\vl{}-\vl{2}\right)\cdot\vl{2}\right]
\int_{0}^{r_s}dr \frac{r_s-r}{r_s\,r}\,
\Phi_W^{(3)}(r,\vl{}-\vl{2})
 \left[\Ecal(r_s,\vl{2})+i \Bcal(r_s,\vl{2})\right] e^{-2 i \varphi_{\ell_2}}  
\nonumber
\\
& & +\frac{1}{\pi^2}\INT{2}\INT{3}
\,\left[\left(\vl{}+\vl{2}-\vl{3}\right)\cdot\vl{3}\right]\,
\left[\left(\vl{}+\vl{2}-\vl{3}\right)\cdot\vl{2}\right]\int_{0}^{r_s}dr'\frac{r_s-r}{r_s\,r}
\int_{0}^{r}dr' \frac{r-r'}{r\,r'}\,
\nonumber\\
&&\times
\left[\Phi_W(r,\vl{}+\vl{2}-\vl{3})
\bar\Phi_W^{(2)}(r',\vl{2})
+\Phi_W^{(2)}(r,\vl{}+\vl{2}-\vl{3})
\bar\Phi_W(r',\vl{2})\right]
 \left[\Ecal(r_s,\vl{3})+i \Bcal(r_s,\vl{3})\right] e^{-2 i \varphi_{\ell_3}}  
\nonumber \\ &&
-\frac{1}{\pi^3}\INT{2}\INT{3}\INT{4}\left\{ \left[\left(\vl{}-\vl{2}-\vl{3}-\vl{4}\right)\cdot\vl{4}\right]\,\left[\left(\vl{}-\vl{2}-\vl{3}-\vl{4}\right)\cdot\vl{2}\right]\right.\nonumber \\
&&\left. \times \left(\vl{2}\cdot\vl{3}\right)
\int_{0}^{r_s}dr\frac{r_s-r}{r_s\,r}\int_{0}^{r}dr'\frac{r-r'}{r\,r'}
\int_{0}^{r'}dr'' \frac{r'-r''}{r'\,r''}\right.\nonumber\\
&& \times\,\Phi_W(r,\vl{}-\vl{2}-\vl{3}-\vl{4})\Phi_W(r',\vl{2})\Phi_W(r'',\vl{3})
\left[\Ecal(r_s,\vl{4})+i \Bcal(r_s,\vl{4})\right] e^{-2 i \varphi_{\ell_4}}
\nonumber\\
&&\left.+\frac{1}{2}\left[\left(\vl{}-\vl{2}-\vl{3}-\vl{4}\right)\cdot\vl{4}\right]\,\left[\left(\vl{}-\vl{2}-\vl{3}-\vl{4}\right)\cdot\vl{2}\right]\,
\left[\left(\vl{}-\vl{2}-\vl{3}-\vl{4}\right)\cdot\vl{3}\right]\right.\nonumber\\
&&\left.\times\int_{0}^{r_s}dr\frac{r_s-r}{r_s\,r}
\int_{0}^{r}dr' \frac{r-r'}{r\,r'}\,
\int_{0}^{r}dr'' \frac{r-r''}{r\,r''}\right.\nonumber\\
&& \left.\times\,\Phi_W(r,\vl{}-\vl{2}-\vl{3}-\vl{4})\Phi_W(r',\vl{2})\Phi_W(r'',\vl{3})
\left[\Ecal(r_s,\vl{4})+i \Bcal(r_s,\vl{4})\right] e^{-2 i \varphi_{\ell_4}}\right\}
\,,\\
& & \nonumber \\
\Dcal^{(12)}(\vl{})
&=&\frac{1}{2\pi}\int d^2 x\,\T{a}{1}\T{b}{2}\nabla_a\nabla_b\Pcal\,e^{i\vl\cdot{\bf x}}\nonumber\\
&=&
\frac{1}{\pi^2}\INT{2}\INT{3}\,\left[\left(\vl{}+\vl{2}-\vl{3}\right)\cdot \vl{3}\right]\,\left(\vl{2}\cdot \vl{3}\right)\int_{0}^{r_s}dr \frac{r_s-r}{r_s\,r}\nonumber\\
&&\times\int_{0}^{r_s}dr' \frac{r_s-r'}{r_s\,r'}
\Phi_W(r,\vl{}+\vl{2}-\vl{3})
\bar{\Phi}^{(2)}_W(r',\vl{2})\,
 \left[\Ecal(r_s,\vl{3})+i \Bcal(r_s,\vl{3})\right] e^{-2 i \varphi_{\ell_3}}  
\nonumber \\ & &
-\frac{1}{\pi^3}
\INT{2}\INT{3}\INT{4}\left[\left(\vl{}-\vl{2}-\vl{3}-\vl{4}\right)\cdot\vl{4}\right]\,
\left(\vl{4}\cdot\vl{2}\right)\,
\left(\vl{3}\cdot\vl{2}\right)\nonumber\\
&&\times\int_{0}^{r_s}dr \frac{r_s-r}{r_s\,r}\int_{0}^{r_s}dr'\frac{r_s-r'}{r_s\,r'}
\int_{0}^{r'}dr'' \frac{r'-r''}{r'\,r''}\,\Phi_W(r,\vl{}-\vl{2}-\vl{3}-\vl{4}) \,\nonumber
\\
&&\times \Phi_W(r',\vl{2})\Phi_W(r'',\vl{3})
\left[\Ecal(r_s,\vl{4})+i \Bcal(r_s,\vl{4})\right] e^{-2 i \varphi_{\ell_4}}
 \\
\Dcal^{(111)}(\vl{})
&=&\frac{1}{2\pi}\int d^2 x\frac{1}{6}\,\T{a}{1}\T{b}{1}\T{c}{1}\nabla_a\nabla_b\nabla_c\Pcal\,e^{i\vl{}\cdot{\bf x}}\nonumber\\
&=&
-\frac{1}{6}\frac{1}{\pi^3}\INT{2}\INT{3}\INT{4}\left[\left(\vl{}-\vl{2}-\vl{3}-\vl{4}\right)\cdot\vl{4}\right]\,
\left(\vl{2}\cdot\vl{4}\right)\,
\left(\vl{3}\cdot\vl{4}\right)
\nonumber\\
&&\times
\int_{0}^{r_s}dr \frac{r_s-r}{r_s\,r}\,\int_{0}^{r_s}dr' \frac{r_s-r'}{r_s\,r'}\,
\int_{0}^{r_s}dr'' \frac{r_s-r''}{r_s\,r''}\,\Phi_W(r,\vl{}-\vl{2}-\vl{3}-\vl{4})\nonumber\\
&&\times\Phi_W(r',\vl{2})\Phi_W(r'',\vl{3})
\left[\Ecal(r_s,\vl{4})+i \Bcal(r_s,\vl{4})\right] e^{-2 i \varphi_{\ell_4}}
\,,  
\eea
\bea
&& \nonumber \\
\Dcal^{(22)}(\vl{})
&=&\frac{1}{2\pi}\int d^2 x\,\frac{1}{2}\,\T{a}{2}\T{b}{2}\nabla_a\nabla_b\Pcal\,e^{i \vl\cdot{\bf x}}\nonumber\\
&=&
\frac{1}{2}\frac{1}{\pi^2}\INT{2}\INT{3}\,\left[\left(\vl{}+\vl{2}-\vl{3}\right)\cdot \vl{3}\right]\,\left(\vl{2}\cdot \vl{3}\right)\int_{0}^{r_s}dr \frac{r_s-r}{r_s\,r}\int_{0}^{r_s}dr' \frac{r_s-r'}{r_s\,r'}
\nonumber\\
&&\times
\Phi_W^{(2)}(r,\vl{}+\vl{2}-\vl{3})
\bar\Phi_W^{(2)}(r',\vl{2})
 \left[\Ecal(r_s,\vl{3})+i \Bcal(r_s,\vl{3})\right] e^{-2 i \varphi_{\ell_3}}  
\nonumber \\
&& -
\frac{1}{\pi^3}
\INT{2}\INT{3}\INT{4}\left[\left(\vl{}-\vl{2}-\vl{3}-\vl{4}\right)\cdot\vl{4}\right]\,
\left(\vl{4}\cdot\vl{2}\right)\,
\left(\vl{3}\cdot\vl{2}\right)
\int_{0}^{r_s}dr \frac{r_s-r}{r_s\,r}\int_{0}^{r_s}dr'\frac{r_s-r'}{r_s\,r'}
,\nonumber
\\
&&\times\,
\int_{0}^{r'}dr'' \frac{r'-r''}{r'\,r''}\Phi_W^{(2)}(r,\vl{}-\vl{2}-\vl{3}-\vl{4})\Phi_W(r',\vl{2})\Phi_W(r'',\vl{3})
 \left[\Ecal(r_s,\vl{4})+i \Bcal(r_s,\vl{4})\right] e^{-2 i \varphi_{\ell_4}}  
\nonumber \\ && 
-\frac{1}{2}\frac{1}{\pi^4}\INT{2}\INT{3}\INT{4}\INT{5}
\left[\left(\vl{}-\vl{2}-\vl{3}-\vl{4}-\vl{5}\right)\cdot\vl{5}\right]\nonumber\\
&&\times\left[\left(\vl{}-\vl{2}-\vl{3}-\vl{4}-\vl{5}\right)\cdot\vl{2}\right]
\left(\vl{5}\cdot\vl{3}\right)\,
\left(\vl{3}\cdot\vl{4}\right)
\int_{0}^{r_s}dr\frac{r_s-r}{r_s\,r}\int_{0}^{r}dr' \frac{r-r'}{r\,r'}\,
\int_{0}^{r_s}dr''\frac{r_s-r''}{r_s\,r''}
\int_{0}^{r''}dr''' \frac{r''-r'''}{r''\,r'''}
\nonumber\\
&&\times\,
\Phi_W(r,\vl{}-\vl{2}-\vl{3}-\vl{4}-\vl{5})\Phi_W(r',\vl{2})\Phi_W(r'',\vl{3})\Phi_W(r''',\vl{4})
\left[\Ecal(r_s,\vl{5})+i \Bcal(r_s,\vl{5})\right] e^{-2 i \varphi_{\ell_5}}
\,,
\\
\nonumber\\
\Dcal^{(13)}(\vl{})
&=&\frac{1}{2\pi}\int d^2 x\,\T{a}{1}\T{b}{3}\nabla_a\nabla_b\Pcal\,e^{i\vl{}\cdot{\bf x}}\nonumber\\
&=&
\frac{1}{\pi^2}\INT{2}\INT{3}\,\left[\left(\vl{}+\vl{2}-\vl{3}\right)\cdot \vl{3}\right]\,\left(\vl{2}\cdot \vl{3}\right)\int_{0}^{r_s}dr \frac{r_s-r}{r_s\,r}\int_{0}^{r_s}dr' \frac{r_s-r'}{r_s\,r'}
\nonumber\\
&&\times
\Phi_W(r,\vl{}+\vl{2}-\vl{3})
\bar{\Phi}^{(3)}_W(r',\vl{3})
\left[\Ecal(r_s,\vl{3})+i \Bcal(r_s,\vl{3})\right] e^{-2 i \varphi_{\ell_3}}  
\nonumber \\
&& 
-\frac{1}{\pi^3}
\INT{2}\INT{3}\INT{4}\left[\left(\vl{}-\vl{2}-\vl{3}-\vl{4}\right)\cdot\vl{4}\right]\,
\left(\vl{4}\cdot\vl{2}\right)\,
\left(\vl{3}\cdot\vl{2}\right)\nonumber\\
&&\times\int_{0}^{r_s}dr \frac{r_s-r}{r_s\,r}\int_{0}^{r_s}dr'\frac{r_s-r'}{r_s\,r'}
\int_{0}^{r'}dr'' \frac{r'-r''}{r'\,r''}\,
\Phi_W(r,\vl{}-\vl{2}-\vl{3}-\vl{4})
\left[\Phi_W(r',\vl{2})\Phi^{(2)}_W(r'',\vl{3})
\right. \nonumber \\ && \left.
+\Phi_W^{(2)}(r',\vl{2})\Phi_W(r'',\vl{3})\right]
 \left[\Ecal(r_s,\vl{4})+i \Bcal(r_s,\vl{4})\right] e^{-2 i \varphi_{\ell_4}}  
\nonumber \\ &&
-\frac{1}{\pi^4}
\INT{2}\INT{3}\INT{4}\INT{5}
\left\{\left[\left(\vl{}-\vl{2}-\vl{3}-\vl{4}-\vl{5}\right)\cdot\vl{5}\right]\,
\left(\vl{2}\cdot\vl{5}\right)\right.
\left(\vl{2}\cdot\vl{3}\right)
\left(\vl{3}\cdot\vl{4}\right) 
\nonumber\\
&&\times
\int_{0}^{r_s}dr \frac{r_s-r}{r_s\,r}\int_{0}^{r_s}dr'\frac{r_s-r'}{r_s\,r'}
\int_{0}^{r'}dr''\frac{r'-r''}{r'\,r''}
\int_{0}^{r''}dr''' \frac{r''-r'''}{r''\,r'''}\Phi_W(r,\vl{}-\vl{2}-\vl{3}-\vl{4}-\vl{5})\Phi_W(r',\vl{2})\nonumber\\
& & \times \Phi_W(r'',\vl{3})\Phi_W(r''',\vl{4})
\left[\Ecal(r_s,\vl{5})+i \Bcal(r_s,\vl{5})\right] e^{-2 i \varphi_{\ell_5}}
\nonumber \\
&&+\frac{1}{2}\left[\left(\vl{}-\vl{2}-\vl{3}-\vl{4}-\vl{5}\right)\cdot\vl{5}\right]\,
\left(\vl{2}\cdot\vl{5}\right)
\left(\vl{2}\cdot\vl{3}\right)\,
\left(\vl{2}\cdot\vl{4}\right)\int_{0}^{r_s}dr \frac{r_s-r}{r_s\,r}\,
\int_{0}^{r_s}dr'\frac{r_s-r'}{r_s\,r'}
\int_{0}^{r'}dr'' \frac{r'-r''}{r'\,r''}\,
\nonumber\\
&&\left.\times\,
\int_{0}^{r'}dr''' \frac{r'-r'''}{r'\,r'''}
\Phi_W(r,\vl{}-\vl{2}-\vl{3}-\vl{4}-\vl{5})\Phi_W(r',\vl{2})\Phi_W(r'',\vl{3})\Phi_W(r''',\vl{4})
\right.
\nonumber\\
&&\left.\times\,
\left[\Ecal(r_s,\vl{5})+i \Bcal(r_s,\vl{5})\right] e^{-2 i \varphi_{\ell_5}}
\right\}\,,
\\
\nonumber\\
\Dcal^{(1111)}(\vl{})
&=&\frac{1}{2\pi}\int d^2 x\,\frac{1}{24}\,\T{a}{1}\T{b}{1}\T{c}{1}\T{d}{1}\nabla_a\nabla_b\nabla_c\nabla_d\Pcal\,e^{i\vl{}\cdot{\bf x}}\nonumber\\
&=&
-\frac{1}{24}\frac{1}{\pi^4}
\INT{2}\INT{3}\INT{4}\INT{5}\left[\left(\vl{}-\vl{2}-\vl{3}-\vl{4}-\vl{5}\right)\cdot\vl{5}\right]
\nonumber\\
&&
\times 
\left(\vl{2}\cdot\vl{5}\right)\,
\left(\vl{3}\cdot\vl{5}\right)\,
\left(\vl{4}\cdot\vl{5}\right)
\int_{0}^{r_s}dr \frac{r_s-r}{r_s\,r}\int_{0}^{r_s}dr' \frac{r_s-r'}{r_s\,r'}\,
\int_{0}^{r_s}dr'' \frac{r_s-r''}{r_s\,r''}
\int_{0}^{r_s}dr'''\frac{r_s-r'''}{r_s\,r'''}
\nonumber\\
&&
\times
\Phi_W(r,\vl{}-\vl{2}-\vl{3}-\vl{4}-\vl{5})\Phi_W(r',\vl{2})\Phi_W(r'',\vl{3})
\Phi_W(r''',\vl{4})
\left[\Ecal(r_s,\vl{5})+i \Bcal(r_s,\vl{5})\right] e^{-2 i \varphi_{\ell_5}}
\,.
\eea
We do not write  the terms $\Dcal^{(4)}$ and $\Dcal^{(112)}$ because the  associated  contributions to the angular power spectra of lensed polarization tensor vanish as a consequence of statistical isotropy 
(see Sect. \ref{Sec3}).
\end{widetext}

\newpage

\section{Lensed angular power spectra for polarization} \label{AppFC}
\setcounter{equation}{0}
Following Sect. \ref{Sec3} and \cite{Marozzi:2016uob} we now present the  evaluation of the next-to-leading order corrections to E- and B-mode polarization spectra. More details are given in Ref.~\cite{Marozzi:2016uob}, where we compute, however, only the temperature anisotropy spectrum. Therefore, for  completeness, we repeat the procedure here for the polarization spectra and for the temperature polarization cross-correlation.

\subsection{Results $\tilde{C}_\ell^{\Ecal \Mcal}$}
\label{FC.1}

Let us begin by evaluating the lensed cross-correlation, $\tilde{C}_\ell^{\Ecal \Mcal}$. Up to next to next-to-leading order,
we have 
\bea
&& 
\hspace{-1cm}
-e^{2 i \varphi_{\ell}}
\langle \tilde{\Pcal}(\vl{})\bar{\tilde{\Mcal}}(\vl{}')\rangle
 = \delta(\vl{}-\vl{}') \tilde{C}_\ell^{\Ecal \Mcal} \nonumber \\
&=& \delta(\vl{}-\vl{}') {C}_\ell^{\Ecal \Mcal} 
-e^{2 i \varphi_{\ell}}\langle \Dcal (\vl{}) \bar{\Acal}(\vl{}\,')\rangle
 \,,\eea
where $\Acal(\vl{})$ is given in Eq. (\ref{AcalTotal}) and we introduce
\bea
\Dcal(\vl{})&=&\Dcal^{(0)}(\vl{})+
\sum_{i=1}^4\Dcal^{(i)}(\vl{})
+\sum_{\substack{i+j\le 4 \\  1\le i\le j}}
\Dcal^{(ij)}(\vl{})
\nonumber \\
&&
+\sum_{\substack{i+j+k\le 4 \\ 1\le i\le j\le k}} \Dcal^{(ijk)}(\vl{})+\Dcal^{(1111)}(\vl{})\,,
\eea
the 2d Fourier transforms of $\Dcal(x^a)$ defined in Eq.~\eqref{eq:expansion-compact_x}.
We now introduce the expectation values $\hat F_\ell^{(i\ldots)}$ and $\hat F_\ell^{(i\ldots ,\, j\ldots)}$ by
\bea
\delta\left( \vl{}-\vl{}\,' \right) \hat{F}_\ell^{(ij\ldots,ij\ldots)} &=&\langle \Dcal^{(ij\ldots)} (\vl{}) 
\bar{\Acal}^{(ij\ldots)}(\vl{}\,')\rangle\,,
\nonumber \\
\delta\left( \vl{}-\vl{}\,' \right) \hat{F}_\ell^{(ij\ldots,i'j'\ldots)} &=&
\langle \Dcal^{(ij\ldots)} (\vl{}) 
\bar{\Acal}^{(i'j'\ldots)}(\vl{}\,')\rangle
\nonumber \\ &+&
\langle \Dcal^{(i'j'\ldots)} (\vl{}) 
\bar{\Acal}^{(ij\ldots)}(\vl{}\,')\rangle\,, \ \ \
\eea
where the last definition applies when the coefficients $(ij\ldots)$ and $(i'j'\ldots)$ are not identical. 
The Dirac delta function $\delta\left( \vl{}-\vl{}\,' \right)$ is a consequence of  statistical isotropy.
By omitting terms of higher  than  fourth order in the Weyl potential and 
terms that vanish as a consequence of  Wick's theorem (odd number of Weyl potentials), we obtain 
\bea 
\tilde{C}^{\Ecal \Mcal}_\ell&=&C^{{\Ecal \Mcal}}_\ell+F_\ell^{(0,2)}+
F_\ell^{(0,11)}+F_\ell^{(1, 1)}+F_\ell^{(0,4)}+F_\ell^{(0,13)}
\nonumber \\ &+&
F_\ell^{(0,22)}+F_\ell^{(0,112)}+F_\ell^{(0,1111)}
 +F_\ell^{(1, 3)}+
F_\ell^{(2, 2)}
\nonumber \\
&  +&F_\ell^{(1, 12)}+F_\ell^{(1, 111)}+F_\ell^{(2, 11)}+F_\ell^{(11, 11)}\,,
\eea
where ${F}_\ell^{(i\ldots,j\ldots)}=-e^{2 i \varphi_{\ell}} \hat{F}_\ell^{(i\ldots,j\ldots)}$.

As the terms $\Dcal^{(i\ldots)}$ are simply related to the $\Acal^{(i\ldots)}$ terms, also the terms $\hat{F}_\ell^{(i\ldots ,\, j\ldots)}$  
can be easily evaluated from the  $C_\ell^{(i\ldots ,\, j\ldots)}$.
In fact, using Eq.~(\ref{Plspace}) and the results for the $\Dcal^{(i\ldots)}$ and $\Acal^{(i\ldots)}$ terms 
(see Sect.~\ref{Sec5}, Appendix \ref{a:Dl} and \cite{Marozzi:2016uob}), one finds 
 that the $\hat{F}_\ell^{(i\ldots ,\, j\ldots)}$
are given by the $C_\ell^{(i\ldots ,\, j\ldots)}$ simply by 
 substituting
\be 
 C_\ell^{\Mcal}(z_s) \quad  \rightarrow \quad  - C_\ell^{\Ecal\Mcal}(z_s) e^{-2 i \varphi_{\ell}}  \,.
 \ee
The substitution is performed for any $C_\ell^{\Mcal}(z_s)$ inside and outside the integrals.

\subsection{Results $\tilde{C}_\ell^{\Ecal}+\tilde{C}_\ell^{\Bcal}$}
\label{FC.2}
Let us also evaluate $\tilde{C}_\ell^\Ecal+\tilde{C}_\ell^\Bcal$.
Proceeding as in the previous subsection we have 
\bea
\langle \tilde{\Pcal}(\vl{})\bar{\tilde{\Pcal}}(\vl{}')\rangle
&=& 
\delta(\vl{}-\vl{}') \left[\tilde{C}_\ell^{\Ecal}+\tilde{C}_\ell^{\Bcal}\right] \nonumber \\
&=& \delta(\vl{}-\vl{}') \left[{C}_\ell^{\Ecal}+{C}_\ell^{\Bcal}\right] 
+\langle \Dcal (\vl{}) \bar{\Dcal}(\vl{}\,')\rangle \,. \ \ \ \
\eea
We now introduce $M_\ell^{(i\ldots)}$ and $M_\ell^{(i\ldots ,\, j\ldots)}$ given by 
\bea
\delta\left( \vl{}-\vl{}\,' \right) M_\ell^{(ij\ldots,ij\ldots)} &=&\langle \Dcal^{(ij\ldots)} (\vl{}) 
\bar{\Dcal}^{(ij\ldots)}(\vl{}\,')\rangle\,,
\nonumber \\
\delta\left( \vl{}-\vl{}\,' \right) M_\ell^{(ij\ldots,i'j'\ldots)} &=&
\langle \Dcal^{(ij\ldots)} (\vl{}) 
\bar{\Dcal}^{(i'j'\ldots)}(\vl{}\,')\rangle
\nonumber \\ &+&
\langle \Dcal^{(i'j'\ldots)} (\vl{}) 
\bar{\Dcal}^{(ij\ldots)}(\vl{}\,')\rangle, \ \ \ 
\eea
where again the last definition applies when the coefficients $(ij\ldots)$ and $(i'j'\ldots)$ are not identical. 
The delta Dirac function $\delta\left( \vl{}-\vl{}\,' \right)$ is a consequence of  statistical isotropy.
As before, by omitting terms of higher  than  fourth order in the Weyl potential and 
terms that vanish as a consequence of  Wick's theorem, we obtain 
\bea 
 \left[\tilde{C}_\ell^{\Ecal}+\tilde{C}_\ell^{\Bcal}\right] &=& \left[C_\ell^{\Ecal}+C_\ell^{\Bcal}\right]\!+
M_\ell^{(0,11)}\!+M_\ell^{(1, 1)}\!+M_\ell^{(0,2)}\!
\nonumber \\
&+&
M_\ell^{(0,13)}\!+M_\ell^{(0,22)}\!+M_\ell^{(0,112)}
\!\!+M_\ell^{(0,1111)}\!
\nonumber \\
& + &
\!M_\ell^{(1, 3)}\!+\!
M_\ell^{(2, 2)}\!+\!M_\ell^{(1, 12)}\!+\!M_\ell^{(1, 111)}\!
\nonumber \\
& + &
\!M_\ell^{(2, 11)}\!+\!M_\ell^{(11, 11)}\,.
\eea
As for the case of the $F_\ell^{(i\ldots ,\, j\ldots)}$ terms,
also in this case we can obtain the  $M_\ell^{(i\ldots ,\, j\ldots)}$ terms
starting from the results for the $C_\ell^{(i\ldots ,\, j\ldots)}$.
These will be obtained 
by the $C_\ell^{(i\ldots ,\, j\ldots)}$ via the substitution
\be 
 C_\ell^{\Mcal}(z_s)  \quad \rightarrow  \quad C_\ell^{\Ecal}(z_s) + C_\ell^{\Bcal}(z_s) \,,
   \ee
 performed for any $C_\ell^{\Mcal}(z_s)$ inside and outside the integrals.

\subsection{Results $\tilde{C}_\ell^{\Ecal}-\tilde{C}_\ell^{\Bcal}$}
\label{FC.3}

Let us finally move to the evaluation of $\tilde{C}_\ell^\Ecal-\tilde{C}_\ell^\Bcal$.
Proceeding as in the previous subsections we have 
\bea
\langle \tilde{\Pcal}(\vl{}){\tilde{\Pcal}}(\vl{}')\rangle 
 &=& 
\delta(\vl{}+\vl{}') \left[\tilde{C}_\ell^{\Ecal}-\tilde{C}_\ell^{\Bcal}\right]e^{-4 i \varphi_\ell}\nonumber \\
&=& 
\delta(\vl{}+\vl{}') \left[{C}_\ell^{\Ecal}-{C}_\ell^{\Bcal}\right] e^{-4 i \varphi_\ell}
+\langle \Dcal (\vl{}) {\Dcal}(\vl{}\,')\rangle \,. \nonumber \\
\eea
We now introduce $\hat{N}_\ell^{(i\ldots ,\, j\ldots)}$ defined as follows
\bea
\delta\left( \vl{}+\vl{}\,' \right) \hat{N}_\ell^{(ij\ldots,ij\ldots)} &=&\langle \Dcal^{(ij\ldots)} (\vl{}) 
{\Dcal}^{(ij\ldots)}(\vl{}\,')\rangle\,,
\nonumber \\
\delta\left( \vl{}+\vl{}\,' \right) \hat{N}_\ell^{(ij\ldots,i'j'\ldots)} &=&
\langle \Dcal^{(ij\ldots)} (\vl{}) 
{\Dcal}^{(i'j'\ldots)}(\vl{}\,')\rangle
\nonumber \\
& + &
\langle \Dcal^{(i'j'\ldots)} (\vl{}) 
{\Dcal}^{(ij\ldots)}(\vl{}\,')\rangle\ \qquad
\eea
where the last definition applies when the coefficients $(ij\ldots)$ and $(i'j'\ldots)$ are different. 
The $\delta\left( \vl{}+\vl{}\,' \right)$ is a consequence of  statistical isotropy and of the fact that in general 
$A(\vl{})=\bar{A}(-\vl{})$.
As before, by omitting terms of higher  than  fourth order in the Weyl potential and 
terms that vanish as a consequence of  Wick's theorem, we obtain 
\bea 
 \left[\tilde{C}_\ell^{\Ecal}-\tilde{C}_\ell^{\Bcal}\right] &=& \left[C_\ell^{\Ecal}-C_\ell^{\Bcal}\right]\!+N_\ell^{(0,2)}\! +
N_\ell^{(0,11)}\!+N_\ell^{(1, 1)}\!
\nonumber \\
& + &
N_\ell^{(0,4)}\!+N_\ell^{(0,13)}\!+N_\ell^{(0,22)}\!+N_\ell^{(0,112)}
\nonumber \\
&  & \!+N_\ell^{(0,1111)}+N_\ell^{(1, 3)}+
N_\ell^{(2, 2)}+N_\ell^{(1, 12)}
\nonumber \\
& + &N_\ell^{(1, 111)}+N_\ell^{(2, 11)}+N_\ell^{(11, 11)}\,,
\eea
where ${N}_\ell^{(i\ldots ,\, j\ldots)}=e^{4 i \varphi_\ell} \hat{N}_\ell^{(i\ldots ,\, j\ldots)}$

Like for the other terms, we can obtain the $\hat{N}_\ell^{(i\ldots ,\, j\ldots)}$ terms
starting from the results for the $C_\ell^{(i\ldots ,\, j\ldots)}$ by  substituting
\be 
 C_\ell^{\Mcal}(z_s)  \quad \rightarrow \quad \left[C_\ell^{\Ecal}(z_s) - C_\ell^{\Bcal}(z_s)\right] e^{-4 \varphi_\ell} \,,
   \ee
for any $C_\ell^{\Mcal}(z_s)$ inside and outside the integrals.

Using these results  we obtain the corrections to the different polarization power spectra.
The general rules to follow are  specified in Eqs. (\ref{subET})-(\ref{subB}).


\section{Rotation angle using the Sachs formalism}
\label{AppA}
In this Appendix we determine the rotation angle of the Sachs basis described in the main text, and show that the result obtained is equivalent to the rotation angle of the amplification matrix (the Jacobian of the lens map). 

For this purpose, we work in GLC coordinates \cite{Gasperini:2011us} where photon directions are fixed and given by the direction of the incoming photons at the observer.
GLC coordinates consist of a timelike  coordinate $\tau$  (which can always be identified with the proper time in the synchronous 
gauge \cite{BenDayan:2012pp}),
a null coordinate $w$, and two angular coordinates $\tilde{\theta}^a$ ($a=1,2$). 
The GLC line-element  depends on six arbitrary functions ($\Ups, U^a, \ga_{ab}=\ga_{ba} $), 
and takes the form
\beq
ds^2\! = \!\Ups^2 dw^2\!-\!2\Ups dw d\tau+ \! \gamma_{ab}(d\tilde\theta^a\!- \! U^a dw)(d\tilde\theta^b\!-U^b dw)
\label{4}
 \eeq
 with $a,b=1,2$, 
where $\gamma_{ab}$ and its inverse  $\gamma^{ab}$ lower and  raise  two-dimensional indices. 
In GLC coordinates the past light-cone of a given observer is defined by  $w = w_o =$ constant, and 
null geodesics stay at fixed values of the angular coordinates $\tilde{\theta}^a= \tilde{\theta}^a_o =$ constant
(with $\tilde{\theta}^a_o$ specifying the direction of observation). In these coordinates, photon geodesics are given by $k_\mu=\pa_\mu w$, or, equivalently $k^\mu=\U^{-1}\delta^\mu_\tau$. On the one hand,  $w$ represent the fully non-linear potential for the photon four-momentum $k_\mu$. On the other hand, the fact that $\tilde\theta^a$ remain constant along the photon path implies that they can be identified, up to some internal degrees of freedom\footnote{These internal degrees of freedom can lead to some misalignement with the observed angles if not properly addressed \cite{Mitsou:2017ynv}. However, this misalignement can just appear as some corrections at the observer position and these are completely sub-leading with respect to the lensing terms here considered.} \cite{Fanizza:2013doa,Fleury:2016htl}, with the incoming photon directions, i.e. the observed direction of the source. This fact ensures that observables evaluated in GLC coordinates are already functions of the observed angles, as required.

To clarify the geometric meaning of these variables, let us consider the limiting case  of  a spatially flat Friedmann-Lema\^itre-Robertson-Walker (FLRW) Universe with scale factor $a(t)$. 
In this case the geodesic light-cone variables are
\bea
&&
w= r+\eta,\qquad \tau=t,\qquad  \Ups = a(t), \qquad U^a=0,
\nonumber \\ &&
\gamma_{ab}\,d \tilde\theta^a d\tilde\theta^b = a^2(t)\,r^2 (d \theta^2 +\sin^2 \theta d\phi^2),
\label{FR}
\eea
where $\eta$ is the conformal time of the FLRW metric: $d\eta= dt/a$.

Let us now introduce the so-called Sachs basis $\{\tilde{s}_A^\mu\}$ \cite{Sachs:1961zz,Seitz:1994xf}, 
namely the two  4-vectors  $\tilde{s}_A^\mu$ ($A = 1,2$)  defined by the conditions 
\cite{Fleury:2013sna,Pitrou:2012ge}:
\begin{align}
\label{eq:SachsBasis1}
g_{\mu\nu}\tilde s_A^\mu \tilde s_B^\nu&=\delta_{AB}  \,,\\
\label{eq:SachsBasis2}
\tilde s_A^\mu u_\mu =0  ,& \qquad
\tilde s_A^\mu k_\mu =0,\\
\label{eq:SachsBasis4}
\Pi^\mu_\nu k^\la\nabla_{\la} \tilde s_A^\nu=0 
\\
\text{with} \qquad & \Pi^\mu_\nu = \delta^\mu_\nu - \frac{k^\mu k_\nu}{(u^\alpha k_\alpha)^2} - \frac{k^\mu u_\nu + u^\mu k_\nu}{u^\alpha k_\alpha} \, ,
\end{align}
where $\Pi^\mu_\nu$ is a projector on the two-dimensional space orthogonal to 
the four velocity $u_\mu$ and to the spatial photon direction  $n_\mu = u_\mu + (u^\alpha k_\alpha)^{-1} k_\mu$ with $n^\alpha n_\alpha = 1$ and $n^\alpha u_\alpha = 0$.

Following \cite{Fanizza:2013doa}, it can then be shown that in GLC coordinates the screen space, normal to incoming photon geodesics and 
the observers worldline, is simply given by the 2-dimensional subspace spanned by the angles $\tilde\theta^a$. 
We can then restrict the discussion to the angular part of the Sachs basis, which is determined up to a global rotation by the equations~\cite{Fanizza:2013doa}
\beq
\gamma_{ab}\,\tilde s^a_A\tilde s^b_B=\delta_{AB}\qquad,\qquad  k^\mu\nabla_\mu \tilde s^a_A = \nabla_\tau \tilde s^a_A=0  \, .
\label{eq:Sachsproperties}
\eeq
Let us underline that this implies that the angular part of the Sachs basis is parallel transported in GLC gauge. This is a property of the GLC coordinates and is a consequence of the way in which the angles are defined
in this gauge.

The second condition of \eqref{eq:Sachsproperties} can be rewritten  as $\epsilon^{AB}\pa_\tau{\tilde s^a_A} \tilde s_{aB}=0$,  where $\epsilon^{AB}$ is the Levi-Civita symbol in flat space. Note  that an arbitrary orthonormal basis of the screen allows a residual freedom of   rotation given by  $\mathcal R \in SO(2)$. Indeed, if $s^a_A$ is a solution of $\gamma_{ab} s^a_A s^b_B=\delta_{AB}$, $ \tilde s^a_A={\mathcal R_A}^B s^a_B$ is also a solution, where
\beq
{\mathcal R_A}^B=\left(
\begin{array}{cc}
\cos\beta&\sin\beta\\
-\sin\beta&\cos\beta
\end{array}
\right)\,,
\label{eq:R}
\eeq
with an arbitrary rotation angle $\beta$.
Therefore, the expression of the time-dependent rotation 
angle $\beta$ is uniquely given by the second condition in Eqs.~\eqref{eq:Sachsproperties}. Starting from a generic orthonormal zweibein 
$(s_B)$, in order to satisfy also the second condition of \eqref{eq:Sachsproperties} we choose the rotation $\mathcal R$ such that the rotated zweibein is parallel transported along lightlike geodesics. To achieve this the rotation angle  $\beta$ has to satisfy the relation
\beq
\pa_\tau{\beta}=\frac{1}{2}\epsilon^{AB}\pa_\tau s^a_A s_{aB}\,,
\label{eq:betacondition}
\eeq
see also Appendix A of \cite{Fanizza:2013doa}.
In \cite{Fanizza:2014baa}, an exact expression for $\beta$ is obtained in this context (see Eqs.~(A.3)-(A.4)).
Let us underline that the value of $\beta$ is gauge invariant. Even though we are performing the calculation in GLC gauge, Eq.~(\ref{eq:betacondition}) was obtained from the covariant 
Eq.~(\ref{eq:SachsBasis4}). This covariant equation will always result in the same rotation angle $\beta$ to lowest non-vanishing order, irrespective of the  gauge used.
In fact, as a consequence of the higher order Stewart-Walker lemma~\cite{Bruni:1996im,Yoo:2017svj} $\beta^{(2)}$ is gauge invariant since both  $\beta^{(1)}$  and $\beta^{(0)}$ vanish.

Here we are interested in solving \eqref{eq:Sachsproperties} up to second order in perturbation theory. In doing this we make use of Poisson gauge, in particular we follow the approach of~\cite{Fanizza:2015swa} where Poisson gauge quantities are written in terms of the GLC coordinates.
Having this in mind, let us define the background Sachs basis by
\bea
\left(\begin{matrix}\bar s^{a}_1\\ \bar s^{a}_2\end{matrix}\right)=\left[a(\tau)\,r(\tau,w)\right]^{-1}
\left(
\begin{matrix}
1&0\\
0&\sin^{-1}\tilde\theta^1
\end{matrix}\right)\,,
\eea
and to zeroth order
\bea 
\left(\ga^{(0)}_{ab}\right) = a^2(\tau)\,r^2(\tau,w)\left(\begin{matrix}
1&0\\
0&\sin^{2}\tilde\theta^1
\end{matrix}\right)\,.
\eea
We decompose the perturbed Sachs basis $\tilde s^a_A$ uniquely into a symmetric part and a rotation as follows,
\begin{equation}
\tilde s_{aA}=\chi_{ab}\,\bar s^b_B\,\Rcal^B_A= s_{a B}\,\Rcal^B_A\,,
\label{guess}
\end{equation}
where $\chi_{ab}$ is symmetric and $\Rcal^B_A$ is the two dimensional rotation matrix defined above. 
The matrix $\chi_{ab}$ is chosen to ensure $\ga_{ab}s^a_As^b_B=\delta_{AB}$. Moreover, this decomposition is very helpful because, as long as we expand $\chi_{ab}$ and $\beta$ up to the desired order, their degrees of freedom decouple, and we obtain $\chi_{ab}$ and $\beta$ respectively from the first and second conditions in Eqs. \eqref{eq:Sachsproperties}.
In this way, we obtain, to zeroth order
\bea
s_{aA}^{(0)}=\ga_{ab}^{(0)}\bar s^b_A
\eea
where $\Rcal^B_A$ can be fixed equal to $\delta^B_A$.
Due to the factorization of the time dependence,  we have that $\pa_\tau{ (s^a_A)^{(0)}}\propto (s^a_A)^{(0)}$ and $\pa_\tau{ \ga_{ab}^{(0)}}\propto \ga_{ab}^{(0)}$.
At first order,  $\gamma_{ab}=\gamma^{(0)}_{ab}+\gamma^{(1)}_{ab}$ and $s^a_A=(s^{a}_A)^{(0)}+(s^{a}_A)^{(1)}$, the normalisation condition yields
\beq
(s^{c}_A)^{(1)}+\gamma^{(0)}_{ab}(s^{a}_A)^{(0)}(s^{b}_B)^{(1)}(s^{c}_B)^{(0)}=-\gamma_{(0)}^{cb}\gamma^{(1)}_{ba}(s^{a}_A)^{(0)}\,.
\eeq
From this equation, after some algebra, by expand $\chi_{ab}$ and $\beta$ in Eq. \eqref{guess}  to first order, we uniquely obtain
\beq
\chi_{ab}^{(1)}=\gamma^{(1)}_{ab}/2.
\label{first}
\eeq

For our purpose, we expand $\beta$ in Eq.~\eqref{eq:R} up to fourth order,  since in principle we require the rotation of the Sachs basis  up to fourth order to compute all the contributions to the next-to-leading order of the polarization spectra, i.e. $\beta=\beta^{(0)}+\beta^{(1)}+\beta^{(2)}+\beta^{(3)}+\beta^{(4)}$. Since the background is isotropic and first order perturbations are purely scalar perturbations which do not induce rotation,  $\beta^{(0)}$ and $\beta^{(1)}$ do not induce a local rotation of the basis and can be set to zero.  For completeness, we show this explicitly below. Therefore,  we can write the rotation matrix up to fourth order as
\bea
{\mathcal {R}_A}^B=\left[ 1-\frac{\left( \beta^{(2)} \right)^2}{2} \right]\delta^B_A+\left( \beta^{(2)} + \beta^{(3)} + \beta^{(4)}\right){\epsilon_A}^B\,. \nonumber \\
&&
\eea
\begin{widetext}
Hence the  parallel transported Sachs basis is
\bea
\tilde  s^a_A={\mathcal {R}_A}^B s^a_B
&=&\left\{ \left[ 1-\frac{\left( \beta^{(2)} \right)^2}{2} \right]\delta^B_A+\left( \beta^{(2)} + \beta^{(3)} + \beta^{(4)}\right){\epsilon_A}^B \right\}\nonumber\\
&&\times \left[\s{a}{B}{0}+\s{a}{B}{1}+( s^{a}_{B})^{(2)}+(s^{a}_{B})^{(3)}+(s^{a}_{B})^{(4)}\right]\nonumber\\
&=& s^a_A
-\frac{\left( \beta^{(2)} \right)^2}{2}\s{a}{A}{0}
+\beta^{(2)}{\epsilon_A}^B\left[ \s{a}{B}{0} + \s{a}{B}{1} + ( s^{a}_{B})^{(2)}\right]\nonumber\\
&&+\beta^{(3)}{\epsilon_A}^B\left[ \s{a}{B}{0} + \s{a}{B}{1} \right]
+\beta^{(4)}{\epsilon_A}^B \s{a}{B}{0} \,,
\label{eq:RotatedSachs}
\eea
\end{widetext}
where $(s^{a}_{\;B})$ is an arbitrary ortho-normal zweibein on the screen and we have used that up to first order, $( s^{a}_{\;B})$ can be chosen such that there is no rotation, hence  $\tilde s^{a}_{\;B}= s^{a}_{\;B}$.
In the main text we note that  $\beta^{(3)}$ and $\beta^{(4)}$ do not contribute at next to leading order for reasons of statistical isotropy, we can thus just focus on determining $\beta^{(2)}$.

Before that, we prove that  the solution \eqref{first}  
combined with Eq.~\eqref{eq:betacondition} implies $\beta^{(1)}=$ constant. Of course $\beta^{(0)}$ is constant since our background is isotropic.
Indeed, Eq.~\eqref{eq:betacondition} for the background yields
\bea
\pa_\tau \beta^{(0)}&=&\frac{1}{2}\epsilon^{AB}\pa_\tau \s{a}{A}{0}(s_{aB})^{(0)}
\nonumber \\
&\propto& \epsilon^{AB}\s{a}{A}{0}(s_{aB})^{(0)}=\epsilon^{AB}\delta_{AB}=0\,, \qquad
\eea
because $\epsilon^{AB}$ is antisymmetric whereas $\delta_{AB}$ is symmetric.
With a global rotation we can choose $\beta^{(0)}=0$, so ${\Rcal^B_A}^{(0)}=\delta^B_A$, as we already said above.
 In the same way, we can show that $\pa_\tau {\beta^{(1)}}$ vanishes. We have that
\bea
\pa_\tau \beta^{(1)}
&=&-\frac{1}{4}\epsilon^{AB}\pa_\tau \gamma_{(0)}^{ab}\gamma^{(1)}_{bc}(s^{c}_A)^{(0)}(s_{aB})^{(0)}
\nonumber\\
&&
-\frac{1}{4}\epsilon^{AB}\gamma_{(0)}^{ab}\pa_\tau\gamma^{(1)}_{bc}(s^{c}_A)^{(0)}(s_{aB})^{(0)}\nonumber\\
&&-\frac{1}{4}\epsilon^{AB} \gamma_{(0)}^{ab}\gamma^{(1)}_{bc}\pa_\tau(s^{c}_A)^{(0)}(s_{aB})^{(0)}
\nonumber\\
&&
+\frac{1}{4}\epsilon^{AB}\pa_\tau \s{a}{A}{0}\gamma^{(1)}_{ab}(s^{b}_B)^{(0)}\,.
\eea
Considering that the last two terms cancel and using $\epsilon^{AB}\s{a}{A}{0}\s{b}{B}{0}\propto \epsilon^{ab}$ \cite{Fanizza:2014baa}, we obtain
\be
\pa_\tau \beta^{(1)} =-F\,\frac{1}{4}\epsilon^{cd}\pa_\tau \gamma_{(0)}^{ab}\gamma^{(1)}_{bc}\ga_{da}^{(0)}
-G\,\frac{1}{4}\epsilon^{cb}\pa_\tau\gamma^{(1)}_{bc}\,,
\ee
which vanish separately  for arbitrary functions $F$ and $G$ as in both cases the epsilon tensor is contracted with a symmetric expression. This means that also $\beta^{(0)}+\beta^{(1)}$ can be set equal to zero,  ${\Rcal^B_A}^{(0+1)}=\delta^B_A$, and
\beq
\tilde{s}_{aA}^{(1)}=\frac{1}{2}\gamma^{(1)}_{ab}(s^{b}_A)^{(0)}\,,
\eeq
or
\beq
(\tilde{s}^{c}_A)^{(1)}=-\frac{1}{2}\gamma_{(0)}^{cb}\gamma^{(1)}_{ba}(s^{a}_A)^{(0)}
\label{eq:firstorderSachs}\,.
\eeq

Let us now determine the second-order contribution to the Sachs basis. The orthogonality condition 
at the second order is
\bea
&&( s^{c}_{A})^{(2)}
+\ga^{(0)}_{ab}\s{a}{A}{0}( s^{b}_{B})^{(2)}\s{c}{B}{0}
=
\nonumber \\
&=&\frac{3}{4}\s{a}{A}{0}\ga^{(1)}_{ab}\ga_{(0)}^{bd}\ga^{(1)}_{de}\ga_{(0)}^{ec}
-\s{a}{A}{0}\ga^{(2)}_{ab}\ga_{(0)}^{bc} \qquad
\eea
which gives
\beq
\chi^{(2)}_{ab}=\frac{1}{2}\ga^{(2)}_{ab}
-\frac{1}{8}\ga^{(1)}_{ac}\ga_0^{cd}\ga^{(1)}_{db}
\eeq
so that
\bea
\left(\tilde s_{A\,a}\right)^{(2)}
&=& \left( s_{A\,a}\right)^{(2)}+\beta^{(2)} \left( s_{aB} \right)^{(0)}\epsilon^B_A\nonumber\\
&=&\left(\frac{1}{2}\ga^{(2)}_{ad}
-\frac{1}{8}\ga^{(1)}_{ab}\ga_0^{bc}\ga^{(1)}_{cd}\right)\s{d}{A}{0}
\nonumber \\
&+&\beta^{(2)} \left( s_{aB} \right)^{(0)}\epsilon^B_A\,.
\eea
We now compute the rotation angle using
Eq.~\eqref{eq:betacondition}. At second order it yields
\bea
\pa_\tau{\beta^{(2)}}&=&
\frac{1}{2}\epsilon^{AB}\left[ \pa_\tau{( s^{a}_{A})^{(2)}}s_{aB}^{(0)}
\right.
\nonumber\\
&& \left.
+\pa_\tau{\s{a}{A}{0}} s_{aB}^{(2)}+\pa_\tau{\s{a}{A}{1}}s_{aB}^{(1)}\right] \,.
\label{eq:rotsecond}
\eea
It is easy to verify that first and second term on the rhs of Eq.~\eqref{eq:rotsecond} cancel just as  for the first order rotation angle. We focus on the remaining term:
\bea
&& \hspace{-0.6cm}\epsilon^{AB}\pa_\tau{\s{a}{A}{1}}s_{aB}^{(1)}
\!=\!-\frac{1}{4}\epsilon^{AB}\pa_\tau \ga_{(0)}^{ab}\ga^{(1)}_{bc}\s{c}{A}{0} \ga^{(1)}_{ad}\s{d}{B}{0}
\nonumber\\
&&\qquad-\frac{1}{4}\epsilon^{AB} \ga_{(0)}^{ab}\pa_\tau\ga^{(1)}_{bc}\s{c}{A}{0} \ga^{(1)}_{ad}\s{d}{B}{0}\nonumber\\
&&\qquad-\frac{1}{4}\epsilon^{AB} \ga_{(0)}^{ab}\ga^{(1)}_{bc}\pa_\tau\s{c}{A}{0} \ga^{(1)}_{ad}\s{d}{B}{0}.  \label{e:C18}
\eea
Using the identities $\epsilon^{AB}\s{a}{A}{0}\s{b}{B}{0}=\ga_{(0)}^{-1/2} \epsilon^{ab}$, with $\det{\ga^{(0)}_{ab}}\equiv \ga_{(0)}$ and $\pa_\tau \s{a}{A}{0}=-\frac{1}{4}\frac{\pa_\tau\ga_{(0)}}{\ga_{(0)}} \s{a}{A}{0}$, as well as the  antisymmetry of $\epsilon^{cd}$,
Eq. \eqref{e:C18} simplifies to
\bea
\epsilon^{AB}\pa_\tau{\s{a}{A}{1}}s_{aB}^{(1)}
&=&-\frac{1}{4}\ga_{(0)}^{-1/2} \ga_{(0)}^{ab}\pa_\tau\ga^{(1)}_{bc}\epsilon^{cd}  \ga^{(1)}_{da}\,.
\nonumber \\
&&
\eea
Hence
\beq
\pa_\tau\beta^{(2)}=-\frac{1}{8}\ga_{(0)}^{-1/2} \ga_{(0)}^{ab}\pa_\tau\ga^{(1)}_{bc}\epsilon^{cd}  \ga^{(1)}_{da}\,.
\eeq
The first order perturbations of the angular part of the metric, $\ga^{(1)}_{ab}$  can be expressed in terms of the first order deflection angle in Poisson gauge as follows (see \cite{Fanizza:2015swa})
\beq
\gamma^{(1)}_{ab}=\gamma^{(0)}_{ac}\pa_b\theta^{c(1)}+\gamma^{(0)}_{cb}\pa_a\theta^{c(1)}.
\eeq
Using also  $\pa_\tau \ga^{(0)}_{ab}=\frac{1}{2}\frac{\pa_\tau\ga_{(0)}}{\ga_{(0)}}\ga^{(0)}_{ab}$, we obtain the second order rotation in terms of first order deflection angles,
\bea
\pa_\tau\beta^{(2)}
&=&-\frac{1}{8}\ga_{(0)}^{-1/2} \pa_c\pa_\tau\theta^{a(1)}\epsilon^{cd}  \ga^{(1)}_{da}
\nonumber\\
&-&
\frac{1}{8}\ga_{(0)}^{-1/2} \ga_{(0)}^{ab}\gamma^{(0)}_{ce}\pa_b\pa_\tau\theta^{e(1)}\epsilon^{cd}  \ga^{(1)}_{da}\nonumber\\
&&-\frac{1}{16}\frac{\pa_\tau\ga_{(0)}}{\ga_{(0)}^{3/2}} \pa_c\theta^{a(1)}\epsilon^{cd}  \ga^{(1)}_{da}
\nonumber\\
&-&
\frac{1}{16}\frac{\pa_\tau\ga_{(0)}}{\ga_{(0)}^{3/2}} \ga_{(0)}^{ab}\ga^{(0)}_{ce}\pa_b\theta^{e(1)}\epsilon^{cd}  \ga^{(1)}_{da}.\eea

We finally  express the rotation angle in term of the  Weyl potential. Using the expression for the deflection angle given in the main text, Eq.~\eqref{TheOrd1}, we obtain 
\begin{widetext}
\bea
\pa_\tau\beta^{(2)}
&=&a^2\ga_{(0)}^{-1/2}\,\epsilon^{ab} \gamma^{(0)}_{bc}\ga_{(0)}^{de}\pa_d\int_{\eta}^{\eta_o}d\eta_1a^2(\eta_1)\ga_{(0)}^{cf}(\eta_1)\int_{\eta_1}^{\eta_o}d\eta_2\pa_f\Phi_W(\eta_2)\,\pa_e\pa_a \int_{\eta}^{\eta_o}d\eta_3\Phi_W(\eta_3)\nonumber\\
&&+\,a^2\ga_{(0)}^{-1/2}\,\epsilon^{ab} \pa_b\int_{\eta}^{\eta_o}d\eta_1a^2(\eta_1)\ga_{(0)}^{cd}(\eta_1)\int_{\eta_1}^{\eta_o}d\eta_2\pa_d\Phi_W(\eta_2) \pa_c\pa_a\int_{\eta}^{\eta_o}d\eta_3\Phi_W(\eta_3) \,.\nonumber\\
\eea
\end{widetext}
Note that here $\Phi_W(\eta_i) \equiv \Phi_W(\eta_i, \bn(\eta_o-\eta_i)$ where $\eta_o$ is present time and $\bn$ is the directions of the geodesic given by $\tilde\theta^a$.
This expression can be further simplified using $r_i\equiv\eta_o-\eta_i$ and
 $\ga_{(0)}^{ab}=\left[a(\tau)\,r(\tau,w)\right]^{-2}\hat\ga_{(0)}^{ab}=\left[a(\eta)r\right]^{-2}\hat\ga_{(0)}^{ab}$.  We then find\footnote{Hereafter we move between the proper time in GLC and the conformal time $\eta$ in Poisson gauge simply considering the background relation $\pa_\tau= a^{-1} \pa_\eta$. In theory we should go from 
 the $\tau$ variable to the background variables corresponding to our observed redshift, but the effect of neglecting this is always subleading in the number of angular derivatives.}
\bea
\pa_\eta\beta^{(2)}
&=&2\,\frac{\hat\ga_{(0)}^{-1/2}}{r^2}\,\epsilon^{ab} \hat\ga_{(0)}^{cd}\int_0^{r}\frac{dr_1}{r_1^2}\int_{0}^{r_1}dr_2\pa_b\pa_d\Phi_W(r_2)
\nonumber\\
&&\qquad \times
\int_{0}^{r}dr_3\pa_a\pa_c\Phi_W(r_3)\,.
\eea
Here we have used 
$\Phi_W(r_i) = \Phi_W(\eta_o-r_i, \bn r_i)$.
This result can be  integrated to yield (we use $\int_{\eta_s}^{\eta_o}d\eta = \int_0^{r_s} dr$ and adopt the boundary condition $\beta^{(2)}(\eta_o)=0$)
\bea
\beta^{(2)}(r_s)
&=&2\,\epsilon^{ab}\,\int_{0}^{r_s}\frac{dr}{r^2}\,\int_{0}^{r}\frac{dr_1}{r_1^2}\int_{0}^{r_1}dr_2\nabla_b\nabla_c\Phi_W(r_2)
\nonumber\\
&&\qquad \times
\int_{0}^{r}dr_3\nabla_a\nabla^c\Phi_W(r_3)\,,
\label{eq:rotation}
\eea
where, in going from partial to covariant derivatives, we go from standard angular derivatives to normalized angular
derivatives (e.g. $\partial_{\tilde{\varphi}} \rightarrow (1/\sin \tilde{\theta}) \partial_{\tilde{\varphi}}$).

Of course a global (time independent) rotation is irrelevant, what has  physical meaning is just the difference of this angle between the source and the observer position, namely $\Delta \beta=\beta(\eta_s)-\beta(\eta_o)$. Therefore, the choice $\beta^{(0)}=\beta^{(1)}=0$ is irrelevant.

We now show that $\beta^{(2)}$ agrees with the rotation angle in the amplification matrix, 
which is of the form, see e.g.  Eq.~(2.9) of \cite{Marozzi:2016uob},
\bea
\left(\Acal_{b}^{a}\right)  = \left(\frac{\partial\theta^{a}_s}{\partial\theta^{b}_o}\right)  &=& 
 \left(\begin{array}{cc} 1-\kappa & 0 \\  0 & 1-\kappa\end{array}\right)\, +\, \left(\begin{array}{cc} -\gamma_1 &~ 
 -\gamma_2 \\ -\gamma_2 & ~ \gamma_1\end{array}\right)\, 
 \nonumber \\
 &+&
  \left(\begin{array}{cc} 0 &~ -\omega \\ \omega &~ 0\end{array}\right) \,.
\nonumber
\eea 
For scalar perturbation $\om$ vanishes (at first order). At second order, scalar perturbations induce non-vanishing vector and tensor perturbations and therefore also a non-vanishing $\om^{(2)}$.
In order to compute $\om^{(2)}$ it we insert the expression for $\Psi_{ab}^{(2)}$ given in Eq. (2.15) of Ref.~\cite{Marozzi:2016uob}
\begin{widetext}
\bea
\omega^{(2)}&=&-\frac{1}{2}\hat\ga_{(0)}^{-1/2}\epsilon^{ab}\Psi_{ab}^{(2)}\nonumber\\
&=&2\,\epsilon^{ab}\int_{0}^{r_s}dr\,\frac{r_s-r}{r_s\,r}\left[\nabla_a\nabla^c\Phi_W(r)\int_0^rdr_1\,\frac{r-r_1}{r\,r_1}\nabla_b\nabla_c\Phi_W(r_1)\right]\nonumber\\
&=&2\,\epsilon^{ab}\int_{0}^{r_s}\frac{dr}{r^2}\,\int_{0}^{r}dr_1\,\left[\nabla_a\nabla^c\Phi_W(r_1)\int_{0}^{r_1}\frac{dr_2}{r_2^2}\,\int_{0}^{r_2}dr_3\,\nabla_b\nabla_c\Phi_W(r_3)\right],
\label{intermediate}
\eea
where we have used the  relation
\beq
\int_{0}^{r_s}dr\,\frac{r_s-r}{r_s\,r}\,f(r)=\int_{0}^{r_s}\frac{dr}{r^2}\int_{0}^{r}dr_1\,f(r_1) 
-\lim_{r\rightarrow 0}\left[ \frac{r_s-r}{r_s\,r}\int_0^r dr_1 f(r_1) \right]\,,
\eeq
for both  inner and outer integrals. The third line of Eq.~\eqref{intermediate} can be further transformed as follows
\bea
\omega^{(2)}&=&2\,\epsilon^{ab}\int_{0}^{r_s}\frac{dr}{r^2}\,\int_{0}^{r}dr_1\,\left[
\frac{d}{dr_1}\left( \int_{0}^{r_1}dr_4\nabla_a\nabla^c\Phi_W(r_4)\right)
\int_{0}^{r_1}\frac{dr_2}{r_2^2}\,\int_{0}^{r_2}dr_3\,\nabla_b\nabla_c\Phi_W(r_3)\right]\nonumber\\
&=&
2\,\epsilon^{ab}\int_{0}^{r_s}\frac{dr}{r^2}\,
\int_{0}^{r}dr_1\nabla_a\nabla^c\Phi_W(r_1)\int_{0}^{r}\frac{dr_2}{r_2^2}\,\int_{0}^{r_2}dr_3\,\nabla_b\nabla_c\Phi_W(r_3)\nonumber\\
&&
-2\,\epsilon^{ab}\int_{0}^{r_s}\frac{dr}{r^2}\,\int_{0}^{r}dr_1\,\left[
\int_{0}^{r_1}dr_4\nabla_a\nabla^c\Phi_W(r_4)
\frac{d}{dr_1}\left( \int_{0}^{r_1}\frac{dr_2}{r_2^2}\,\int_{0}^{r_2}dr_3\,\nabla_b\nabla_c\Phi_W(r_3)\right)\right]\nonumber\\
&=&
2\,\epsilon^{ab}\int_{0}^{r_s}\frac{dr}{r^2}\,\int_{0}^{r}dr_1
\nabla_a\nabla^c\Phi_W(r_1)\int_{0}^{r}\frac{dr_2}{r_2^2}\,\int_{0}^{r_2}dr_3\,\nabla_b\nabla_c\Phi_W(r_3)\nonumber\\
&&-2\,\epsilon^{ab}\int_{0}^{r_s}\frac{dr}{r^2}\,\int_{0}^{r} \frac{dr_1}{r_1^2}\,\left[
\int_{0}^{r_1}dr_4\nabla_a\nabla^c\Phi_W(r_4)
\,\int_0^{r_1}dr_3\,\nabla_b\nabla_c\Phi_W(r_3)\right] \qquad\label{eq:cornerstone1}\\
&=&
\beta^{(2)}-2\,\epsilon^{ab}\int_{0}^{r_s}\frac{dr}{r^2}\,\int_{0}^r\frac{dr_1}{r_1^2}\,\left[
\int_{0}^{r_1}dr_4\nabla_a\nabla^c\Phi_W(r_4)
 \,\int_0^{r_1}dr_3\,\nabla_b\nabla_c\Phi_W(r_3)\right] \,.  \qquad
\label{eq:cornerstone}
\eea
\end{widetext}
To obtain \eqref{eq:cornerstone1},  we have performed an integration by part in the first and second lines of the previous expression. The last term in Eq.~\eqref{eq:cornerstone} vanishes: indeed, the antisymmetric tensor 
$\epsilon^{ab}$ multiplies a symmetric expression. This proves the equivalence of the rotation angles $\om^{(2)}$ and $\beta^{(2)}$. 

This is not surprising. While the lens map really describes the change of the position in the sky due to lensing by foreground structures, the amplification matrix gives the variation of this change as function of direction. On the other hand, the geodesic deviation equation, which is solved to obtain the rotation of the Sachs basis, yields to change of the distance vector between neighbouring geodesics projected onto the screen. If these maps contain a non-trivial rotation, to lowest non-vanishing order these rotations do agree.

We finally express $\beta^{(2)}$ in $\boldsymbol \ell$ space. Using the flat sky approximation we expand the Weyl potential in Fourier space,
\beq
\Phi_W(z,\boldsymbol{x})=\frac{1}{2\pi}\int d^2\ell \,\Phi_W(z,\boldsymbol{\ell})\,e^{-i\, \boldsymbol{\ell}\cdot\boldsymbol{x}} \,.
\eeq
As in the main text, to each redshift $z$ there corresponds a comoving distance $r(z)$.
Inserting this expansion in Eq.~\eqref{eq:rotation} we find
\begin{widetext}
\bea
\beta^{(2)}&=&\frac{2\,\epsilon^{ab}}{(2\pi)^2}\int_{0}^{r_s}dr\,\frac{r_s-r}{r_s\,r}
\int d^2\ell_1\,\ell_{1a}\ell_1^c\,\Phi_W(z,\boldsymbol{\ell}_1)\,e^{-i\, \boldsymbol{\ell}_1\cdot\boldsymbol{x}}
\int_0^{r}dr_1\,\frac{r-r_1}{r\,r_1}
\int d^2\ell_2 \,\ell_{2b}\ell_{2c}\,\Phi_W(z_1,\boldsymbol{\ell}_2)\,e^{-i\, \boldsymbol{\ell}_2\cdot\boldsymbol{x}}\nonumber\\
&=&\frac{2}{(2\pi)^2}\int_{0}^{r_s}dr\,\frac{r_s-r}{r_s\,r}\int_0^{r}dr_1\,\frac{r-r_1}{r\,r_1}
\int d^2\ell_1\,\int d^2\ell_2
\epsilon^{ab}\ell_{1a}\,\ell_{2b}\left(\boldsymbol{\ell}_1\cdot\boldsymbol{\ell}_2\right)\,\Phi_W(z,\boldsymbol{\ell}_1)\Phi_W(z_1,\boldsymbol{\ell}_2)\,
e^{-i\, \left(\boldsymbol{\ell}_1+\boldsymbol{\ell}_2\right)\cdot\boldsymbol{x}}\nonumber\\
&=&\frac{2}{(2\pi)^2}\int_{0}^{r_s}dr\,\frac{r_s-r}{r_sr}\int_0^{r}dr_1\,\frac{r-r_1}{rr_1}
\int d^2\ell_1\,\int d^2\ell_2
\boldsymbol{n}\cdot\left( \boldsymbol{\ell}_2\land\boldsymbol{\ell}_1 \right)\left(\boldsymbol{\ell}_1\cdot\boldsymbol{\ell}_2\right)\,\Phi_W(z,\boldsymbol{\ell}_1)\Phi_W(z_1,\boldsymbol{\ell}_2)\,
e^{-i\, \left(\boldsymbol{\ell}_1+\boldsymbol{\ell}_2\right)\cdot\boldsymbol{x}}. \qquad
\eea
Here, we remember, $\boldsymbol{n}$ is the direction of the light ray, orthogonal to the plane containing the $\boldsymbol{\ell}$ vectors. 
Then, by applying Limber approximation and using Eqs.~\eqref{limber_weyl} and \eqref{limber_weyl2}, we obtain
\bea \label{beta_square}
\langle (\beta^{(2)})^2 \rangle &=& \int_0^{r_s} \frac{dr}{r^2} \int_0^r \frac{dr_1}{r_1^2} \int \frac{d\ell_1 d\ell_2}{32 \left(2 \pi \right)^2} \ell_1^5 \ell_2^5\left( \frac{r - r_1}{rr_1} \right)^2 \left( \frac{r_s - r}{r_s r} \right)^2 P_R \left( \frac{\ell_1 +1/2}{r} \right) P_R \left( \frac{\ell_2 +1/2}{r_1} \right)
\nonumber \\
&& \qquad
 \left[ T_{\Phi + \Psi} \left(  \frac{\ell_1 +1/2}{r} ,z \right)  T_{\Phi + \Psi} \left(  \frac{\ell_2 +1/2}{r_1} ,z_1 \right) \right]^2
\eea
\end{widetext}


\section{Fisher Analysis}
\label{app:Fisher}
We briefly summarise the Fisher formalism adopted in this work to estimate the theoretical bias introduced by neglecting next-to-leading order lensing. In the ideal case of a cosmic variance limited survey, the Fisher matrix is defined by
\be
F_{\alpha \beta} = \sum_\ell \sum_{X,Y} \frac{\partial C_\ell^X}{\partial q_\alpha}\frac{\partial C_\ell^Y}{\partial q_{\beta}} \text{Cov}^{-1}_{\ell [ X, Y]}\,,
\ee
where 
$X$ and $Y$ denote the corresponding power spectra $\left( \Mcal, \Ecal,\Ecal\Mcal,\Bcal \right)$, $q_\alpha$ are the cosmological parameters and the covariance matrix is~\cite{Verde:2009tu}
\begin{widetext}
\be
\text{Cov}_{\ell} = \frac{2}{2 \ell +1}\left( \begin{array}{cccc}
\left( C^{\Mcal}_\ell \right)^2 & \left( C^{\Ecal\Mcal}_\ell \right)^2 & C^{\Mcal}_\ell C^{\Ecal\Mcal}_\ell &0 \\
 \left( C^{\Ecal\Mcal}_\ell \right)^2 & \left( C^{\Ecal}_\ell \right)^2  & C^{\Ecal}_\ell C^{\Ecal\Mcal}_\ell &0  \\
  C^{\Mcal}_\ell C^{\Ecal\Mcal}_\ell &  C^{\Ecal}_\ell C^{\Ecal\Mcal}_\ell & \frac{1}{2} \left( \left( C_\ell^{\Ecal\Mcal} \right)^2 + C_\ell^{\Mcal} C_\ell^{\Ecal} \right) &0 \\
  0 & 0 & 0 & \left( C_\ell^{\Bcal} \right)^2 
\end{array}\right) \, .
\ee
\end{widetext}
To estimate the impact on the cosmological parameter estimation induced by neglecting a correction $\Delta C_\ell$ on the leading contribution $C_\ell$ we follow the formalism introduced in Refs.~\cite{Knox:1998fp,Heavens:2007ka,Kitching:2008eq}. Therefore the shift of the best-fit is determined by
\be
\Delta_{q_\alpha} = \sum_\beta \left[ F^{-1}\right]_{\alpha \beta} B_\beta\,,
\ee
with
\be
B_\beta = \sum_\ell \sum_{X,Y} \Delta C_\ell^X \frac{ \partial C_\ell^Y}{\partial q_\beta} \text{Cov}^{-1}_{\ell \left[ X,Y \right]} \, .
\ee

Strictly speaking, a Fisher matrix analysis applies only for Gaussian distributions which is not the case of cosmological parameters in general and even less for higher order corrections. But to lowest order in the deviation from the best-fit value every statistic is Gaussian, and hence for the tiny deviations which we find a Fisher analysis is expected to be sufficient.
The impact of deviation from Gaussian statistics of the lensed power spectra has been studied in~\cite{Smith:2006nk}, concluding that the errors induced on the $\left( \Mcal, \Ecal,\Ecal\Mcal\right)$
lensed power spectra are negligible, while on B-modes the Gaussian approximation may underestimate the variance.

\bibliographystyle{utcaps}
\bibliography{biblio_CMBlensing}

\end{document}